\documentclass[10pt]{elsarticle}
\usepackage[left=2.50cm, right=2.50cm, top=2.50cm, bottom=2.50cm]{geometry}
\usepackage{helvet}
\usepackage{amsmath, amsfonts, amssymb}
\usepackage[english]{babel}
\usepackage{romannum}
\usepackage{array}

\usepackage{graphicx}
\usepackage{caption}
\usepackage{subcaption}
\usepackage{float}
\usepackage{placeins}
\usepackage{lineno}
\usepackage{color}

\usepackage{setspace}
\usepackage{url}
\usepackage{bm}
\usepackage{multirow}
\usepackage{indentfirst}
\usepackage{booktabs}
\usepackage{amsthm}
\theoremstyle{remark}
\usepackage{hyperref}
\hypersetup{colorlinks, unicode,
    pdfauthor={Hao Hong, Yanhui Zhong, Bei Zhang, Shuo Xu, Minchuan Tan, Quansheng Zang, Hussein Rappel, St\'ephane P. A. Bordas},
    pdftitle={Dynamic response of tunnels and buried conduits under moving loads by coupling nonmatching isogeometric and scaled-boundary discretizations},
    linkcolor=blue, citecolor=blue, urlcolor=blue}

\usepackage{silence}
\usepackage{cleveref}
\crefname{figure}{Fig.}{Figs.}
\Crefname{figure}{Fig.}{Figs.}

\makeatletter
\def\frontmattersuperscript#1{{\m@th\ensuremath{^{\mbox{\fontsize\sf@size\sf@size#1}}}}}
\AtBeginEnvironment{frontmatter}{\let\@textsuperscript\frontmattersuperscript}

\renewcommand*\env@matrix[1][\arraystretch]{%
  \edef\arraystretch{#1}%
  \hskip -\arraycolsep
  \let\@ifnextchar\new@ifnextchar
  \array{*\c@MaxMatrixCols c}}
\makeatother

\newcommand{\ii}{\mathrm{i}}
\newcommand{\dd}{\mathrm{d}}
\newcommand{\bs}[1]{\bm{#1}}
\newcommand{\mat}[1]{\mathbf{#1}}

\begin{document}
\captionsetup[figure]{labelfont={bf},labelformat={default},labelsep=period,name={Fig.}}

\begin{frontmatter}

\title{Dynamic response of tunnels and buried conduits under moving loads by coupling nonmatching isogeometric and scaled-boundary discretizations}

\author[china1,lux,china2]{Hao~Hong}
\author[china1,china2]{Yanhui~Zhong}
\author[china1,china2]{Bei~Zhang}
\author[china1,china2]{Shuo~Xu}
\author[china1,china2]{Minchuan~Tan}
\author[china1,china2]{Quansheng~Zang\fnref{ind1}\corref{cor1}}
\author[Exeter]{Hussein~Rappel}
\author[lux]{St\'ephane~P.~A.~Bordas}

\address[china1]{School of Water Conservancy and Transportation, Zhengzhou University, Zhengzhou 450001, China}
\address[lux]{Team Legato, Department of Engineering, Faculty of Science, Technology and Medicine, University of Luxembourg, Luxembourg}
\address[china2]{International Joint Research Center of Trenchless Technology for Underground Infrastructures, Zhengzhou 450001, China}
\address[Exeter]{Department of Engineering, Faculty of Environment, Science and Economy, University of Exeter, Exeter, UK}

\cortext[cor1]{Corresponding author}

\fntext[ind1]{E-mail: qszang1991@zzu.edu.cn}

\date{(Dated: \today)}
\pagenumbering{arabic}

\begin{abstract}
Moving-load analysis of tunnels and buried conduits requires resolving local structural response together with wave propagation and radiation in the surrounding soil. We develop a nonmatching coupling formulation within a Fourier-based dimensional reduction framework (2.5D), allowing independent discretization of the near-field and propagation domains. An $L^2$ projection on a common partition of the physical interface transfers displacements between the traces, and its Hermitian adjoint transfers interface forces so the reduced system preserves virtual work. The near field uses a geometry-independent field approximation with truncated hierarchical B-splines, permitting local refinement on fixed CAD geometry. The surrounding soil uses quadtree scaled-boundary isogeometric analysis with unified B\'ezier trace extraction and connects to the homogeneous unbounded exterior through a scaled-boundary impedance. Homogeneous half-space comparisons and an analytical cylindrical-cavity solution verify the components and the coupled system. In the cavity benchmark, hierarchical refinement improves near-field accuracy with the interface trace and surrounding discretization held fixed. A layered pavement with a buried conduit and twin modified-D tunnels demonstrate coupling across open and closed interfaces, including disconnected components. Conduit cover mainly changes peak radial and hoop stress magnitudes. Transverse load offset also shifts their governing sectors. For the twin tunnels, increasing the centre spacing from $1.75$ to $3.00$ times the excavation width reduces the unloaded tunnel's peak mean cross-sectional displacement by $25.5\%$ and its peak tangential lining stress by $66.6\%$.
\end{abstract}


\end{frontmatter}

% \linenumbers

\section{Introduction}
\label{sec:introduction}

Tunnels and buried conduits subjected to moving loads respond over markedly different spatial scales. Displacement and stress gradients are concentrated in linings, conduit walls, pavement layers and the soil immediately around the structure, while the three-component wave field transmitted through the ground can propagate far from the structure and must be represented together with radiation into the unbounded soil. Under moving wheel loads, the pavement layers shape the subsurface stress field \citep{ZhangLiu-66,Wu2020PavementMovingLoad}, and the transmitted loading induces stresses and displacements in buried pipes \citep{ZhangWong2024BuriedPipeTruck,Xu2017BuriedPipelineTraffic}. For adjacent tunnels, vibration travels through the intervening soil from the loaded tunnel to its neighbour \citep{Kuo2011TwinTunnel,He2019TwinTunnelInteraction}. Dynamic analysis of an underground structure under moving loads therefore has to resolve curved structural geometry, material interfaces and local stresses near the structure, and wave propagation and radiation in the surrounding soil. These regions differ in their approximation requirements, mesh organization, geometric representation and computational cost, which motivates independently designed regional discretizations and a means of coupling them.

For linear dynamic problems whose geometry and material distribution are invariant along the travelling direction, Fourier reduction transforms the three-dimensional propagation problem into a family of two-and-a-half-dimensional (2.5D) cross-sectional problems retaining all three displacement components. This reduction complements the two- and three-dimensional time-domain finite-element models that have been compared for train-induced vibration in tunnels \citep{Real2015TunnelFEM2D3D}. The framework is established in finite/infinite-element formulations for moving loads \citep{Yang2003MovingLoadFEMIEM}, coupled 2.5D finite-element/boundary-element models have been used for train-induced tunnel vibration and validated against measurements \citep{JinThompson2018}, and the same class of formulations has been applied to tunnels in poroelastic half-spaces \citep{He2018PoroelasticTunnel25D}. Numerical models for vibration from underground railways have also been compared with one another \citep{Gupta2007UndergroundRailwayModels}. Related methods address the interaction of bounded structures with layered half-spaces \citep{FrancoisSchevenels-57} and moving loads in transversely isotropic saturated media \citep{Huang2025Improved25DFEBE}. For unbounded media, the scaled boundary finite element method (SBFEM) provides a radial treatment that does not require a fundamental solution \citep{WolfSong2001SBFEM} and has been coupled to bounded finite-element regions for ground-vibration analysis \citep{YaseriBazyarJavady-XX}. \citet{YangLi2021IGASBIGA} combined NURBS-based isogeometric analysis (IGA) for the bounded domain with scaled-boundary isogeometric analysis (SBIGA) for the unbounded domain in a 2.5D railway-vibration model. The present formulation allows the structural near field and the surrounding propagation region to be discretized and refined independently, without requiring matching interface meshes or basis functions. A common-partition $L^2$ projection and its Hermitian-adjoint force transfer couple the two regions while preserving discrete interface virtual work.

Within the bounded cross-section, the near field around a tunnel lining or conduit and the surrounding soil present different approximation tasks. In conventional isoparametric IGA, the geometry and displacement field typically share the same underlying spline basis. Geometry-independent field approximation (GIFT) instead permits independent selection of the field space and CAD parameterization \citep{Atroshchenko2018GIFT}. For tunnel linings, conduit walls, pavement layers and loaded regions, this freedom permits local refinement of the displacement space on the original NURBS geometry; conventional IGA refinement can also preserve the geometry. Adaptive GIFT studies for time-harmonic acoustics include PHT-splines \citep{Videla2019AdaptiveGIFT} and enriched approximations \citep{Jansari2022AdaptiveGIFT}. Truncated hierarchical B-splines (THB-splines) organize multilevel approximation spaces through local refinement and truncation \citep{Giannelli2012THB,Giannelli2016THBIGA}, with related developments in macro-element refinement for B\'ezier projection \citep{Dijkstra2026MacroTHB} and in element-local data structures based on B\'ezier extraction \citep{Borden2011BezierExtraction}. These tools support refined near-field approximation on fixed geometry. Their application to 2.5D moving-load elastodynamics calls for a THB-GIFT dynamic operator containing all three displacement components and the longitudinal-wavenumber coupling terms. The surrounding soil, by contrast, calls for flexible spatial partitioning, coarse-to-fine transitions and connection to the unbounded exterior. Scaled-boundary analysis can use automatically generated polygonal or polyhedral cells \citep{Liu2017AutomaticPolyhedralSBFEM}; quadtree scaled-boundary discretizations can combine curved NURBS boundaries with straight internal traces \citep{Li2024NQSBFEM}, and high-order tree meshes and far-field treatments have also been used in elastodynamics \citep{Lozano2025GuidedWaveSBFEM}. These capabilities motivate quadtree-SBIGA for the bounded surrounding domain: the quadtree organizes the partition and local refinement, unified extracted boundary bases represent straight, hanging and CAD-curved edges, and the scaled-boundary radial equations determine the element-interior dynamic response from boundary unknowns. Together with the exterior impedance, this construction supplies the propagation-domain operator. THB-GIFT is therefore used where curved structural geometry, material layers and local stresses require independent refinement of the field variables, and quadtree-SBIGA where the soil must carry the transmitted waves to the radiating exterior.

Independent choices of regional analysis spaces generally produce interface traces with different nodes, element partitions or basis functions. The benefits of those choices can enter a unified dynamic system through nonmatching interface transfer. Mortar methods impose weak constraints between independent spline patches \citep{Brivadis2015IGAMortar}. Weak substitution eliminates slave degrees of freedom using a mortar-type master--slave relation \citep{Dornisch2015WeakSubstitution}; related patch coupling has also been applied to scaled-boundary isogeometric representations \citep{Chasapi2020Mortar}. Nitsche formulations have been used to couple nonmatching two- and three-dimensional NURBS patches \citep{Nguyen2014NitscheNURBS}, for locally refined multipatch isogeometric analysis \citep{Si2023AdaptiveMultipatchNitsche} and for patches with different element sizes, polynomial degrees and parametrizations \citep{Antonelli2026GapShiftedIGA}. Direct transfer operators between nonmatching meshes \citep{Dvorak2024InterfaceOperators} and coupling methods that assign different discretization roles to the regions \citep{Bai2025NonmatchingDDMD} provide further routes to connecting heterogeneous spaces. Nonconforming finite-element/boundary-element coupling has also used virtual work to relate kinematic constraints to interface-force transfer \citep{Yang2021NonconformingIGAFEBE}. In the present 2.5D frequency-domain problem, displacement coefficients and dynamic operators are complex-valued, and the traces may include polynomial, hanging-edge-restricted and rational CAD bases. A kinematic map with a well-defined physical-interface integration must therefore be paired with the corresponding adjoint force transfer. The $L^2$ projection supplies a variational trace map and its associated best-approximation property; the Hermitian-adjoint pairing supplies interface virtual-work consistency.

The main contribution is a work-consistent interface formulation that couples nonmatching THB-GIFT and quadtree-SBIGA discretizations in 2.5D moving-load analysis. Our earlier NURBS-trace infinite-element formulation \citep{Zhong2027NURBSTraceInfinite} connected the bounded and infinite-element domains through a shared interface basis and boundary degrees of freedom. The present coupling acts within the bounded domain, at the interface between the structural near field and the propagation region. Hierarchical refinement of the THB-GIFT near field improves local resolution on fixed CAD geometry, and the quadtree-SBIGA propagation region connects to the unbounded exterior through a scaled-boundary impedance. The cylindrical-cavity benchmark verifies the coupled response against an analytical solution and quantifies the improvement in near-field accuracy with the interface trace and propagation discretization held fixed. The pavement--conduit example combines an open interface for the layered pavement and a separate closed interface around the conduit and examines how cover and load offset affect conduit-wall stresses. The twin-tunnel example couples two disconnected near-field regions through the surrounding soil and evaluates the effect of spacing on the unloaded tunnel's mean cross-sectional motion and local lining stresses.

The remainder of the paper is organized as follows. \Cref{sec:problem} presents the Fourier-reduced 2.5D governing equations and the THB-GIFT near-field dynamic discretization. \Cref{sec:trace} develops the bounded quadtree-SBIGA operator using unified B\'ezier extraction and the scaled-boundary exterior impedance. \Cref{sec:coupling} gives the nonmatching $L^2$ trace projection, work-consistent system coupling and moving-load response reconstruction. \Cref{sec:numerical} evaluates numerical accuracy, regional refinement and engineering applications through verification studies, a layered pavement with a buried conduit, and twin modified-D tunnels. \Cref{sec:conclusions} summarizes the main findings.

\section{Fourier-Reduced 2.5D Problem and THB-GIFT Near Field}
\label{sec:problem}

The Fourier-reduced 2.5D THB-GIFT discretization is formulated for a linearly elastic or viscoelastic medium whose geometry, material layout, and linearized operator are invariant in the travelling direction. The surrounding quadtree-SBIGA domain and exterior impedance are developed in \Cref{sec:trace}, and the interface coupling and moving-load response are presented in \Cref{sec:coupling}.

\subsection{Domain Decomposition and Fourier-Reduced Governing Operator}
\label{subsec:fourier_problem}

The cross-sectional computational setting is decomposed as
\begin{equation}
    \Omega
    =
    \Omega_{\mathrm T}\cup\Omega_{\mathrm Q}\cup\Omega_\infty,
    \qquad
    \Gamma_{\mathrm{TQ}}=\partial\Omega_{\mathrm T}\cap\partial\Omega_{\mathrm Q},
    \qquad
    \Gamma_\infty=\partial\Omega_{\mathrm Q}\cap\partial\Omega_\infty .
    \label{eq:hybrid_domain_decomposition}
\end{equation}
Here $\Omega_{\mathrm T}$ denotes one or more, possibly disconnected, THB-GIFT near-field regions, each of which may comprise multiple NURBS patches. The bounded quadtree-SBIGA surrounding domain is $\Omega_{\mathrm Q}$, and $\Omega_\infty$ is the semi-infinite or infinite exterior represented by a scaled-boundary impedance on $\Gamma_\infty$. The cross-section is described by the transverse coordinate $x$ and the vertical coordinate $y$, and the displacement vector is $\bs{u}=[u_x,u_y,u_z]^{\mathsf T}$. The longitudinal--temporal transform pair adopts the phase convention
\begin{equation}
\begin{aligned}
    \tilde{\bs{u}}(x,y;k,\omega)
    &=
    \int_{-\infty}^{\infty}
    \int_{-\infty}^{\infty}
    \bs{u}(x,y,z,t)
    \exp\left[-\ii(\omega t-kz)\right]
    \dd z\,\dd t,
    \\
    \bs{u}(x,y,z,t)
    &=
    \frac{1}{(2\pi)^2}
    \int_{-\infty}^{\infty}
    \int_{-\infty}^{\infty}
    \tilde{\bs{u}}(x,y;k,\omega)
    \exp\left[\ii(\omega t-kz)\right]
    \dd k\,\dd\omega .
\end{aligned}
    \label{eq:fourier_u}
\end{equation}
where $k$ is the longitudinal wavenumber and $\omega$ is the circular frequency. Accordingly, $\partial/\partial z$ is replaced by $-\ii k$. This substitution retains all three displacement components in the cross-sectional analysis. The weak form, inertia, traction spectrum, and exterior impedance are evaluated in the same frequency--wavenumber domain. The transformed strain vector, ordered as $[\varepsilon_{xx},\varepsilon_{yy},\varepsilon_{zz},\gamma_{xy},\gamma_{yz},\gamma_{zx}]^{\mathsf T}$, is
\begin{equation}
    \tilde{\bs{\varepsilon}}
    =
    \begin{bmatrix}
    \partial_x & 0 & 0\\
    0 & \partial_y & 0\\
    0 & 0 & -\ii k\\
    \partial_y & \partial_x & 0\\
    0 & -\ii k & \partial_y\\
    -\ii k & 0 & \partial_x
    \end{bmatrix}
    \tilde{\bs{u}} .
    \label{eq:strain_25d}
\end{equation}
Each layer is modelled as a linear isotropic solid. Piecewise-homogeneous media assign the appropriate constitutive parameters to THB integration leaves and S-elements, with physical material interfaces aligned with the corresponding analysis partitions. For an isotropic layer, the three-dimensional constitutive matrix is
\begin{equation}
    \mat{D}=
    \begin{bmatrix}
    \lambda+2\mu & \lambda & \lambda & 0 & 0 & 0\\
    \lambda & \lambda+2\mu & \lambda & 0 & 0 & 0\\
    \lambda & \lambda & \lambda+2\mu & 0 & 0 & 0\\
    0 & 0 & 0 & \mu & 0 & 0\\
    0 & 0 & 0 & 0 & \mu & 0\\
    0 & 0 & 0 & 0 & 0 & \mu
    \end{bmatrix},
    \quad
    \lambda=\frac{E\nu}{(1+\nu)(1-2\nu)},\quad
    \mu=\frac{E}{2(1+\nu)} .
    \label{eq:isotropic_D}
\end{equation}
Linear viscoelastic response retains the same matrix form after replacing the elastic constants by complex-valued counterparts. The numerical examples represent material damping by the frequency-independent hysteretic factor
\begin{equation}
    E^\ast=E(1+2\ii\zeta),\qquad
    \lambda^\ast=\frac{E^\ast\nu}{(1+\nu)(1-2\nu)},\qquad
    \mu^\ast=\frac{E^\ast}{2(1+\nu)} ,
    \label{eq:complex_moduli}
\end{equation}
where $\zeta$ is the material damping ratio. Setting $\zeta=0$ recovers the undamped linear-elastic case. In the following derivation, $\lambda$ and $\mu$ may therefore denote either the real elastic constants or their complex counterparts, depending on the material model used in a given example. The transformed problem is then posed independently for each pair $(k,\omega)$. For each bounded material subdomain $\Omega_a\in\{\Omega_{\mathrm T},\Omega_{\mathrm Q}\}$ with prescribed traction boundary $\Gamma_t^a$, the local weak form reads
\begin{equation}
    \int_{\Omega_a}
    \delta\tilde{\bs{\varepsilon}}^{\mathrm{H}}
    \mat{D}
    \tilde{\bs{\varepsilon}}
    \dd\Omega
    -
    \omega^2
    \int_{\Omega_a}
    \rho\,
    \delta\tilde{\bs{u}}^{\mathrm{H}}
    \tilde{\bs{u}}
    \dd\Omega
    =
    \int_{\Gamma_t^a}
    \delta\tilde{\bs{u}}^{\mathrm{H}}
    \tilde{\bs{t}}
    \dd\Gamma ,
    \label{eq:weak_25d}
\end{equation}
where $(\cdot)^{\mathrm{H}}$ denotes the conjugate transpose. The minus sign in the inertia term follows the time-harmonic convention in \Cref{eq:fourier_u}. Equation~\eqref{eq:weak_25d} supplies the bounded-domain operator for both the THB-GIFT and quadtree scaled-boundary discretizations. The unbounded exterior $\Omega_\infty$ is condensed to an impedance on $\Gamma_\infty$ in \Cref{subsec:exterior_relation}.

\subsection{Geometry-Independent THB Analysis Space and Hierarchical Refinement}
\label{subsec:thb_gift_space}

Each THB-GIFT region may comprise several NURBS patches. On any one patch, the fixed geometry map is
\begin{equation}
    \bs{x}=\bs{F}(\xi,\eta)
    =
    \sum_{A=1}^{n_g}
    R_A^g(\xi,\eta)\,\bs{P}_A,
    \qquad
    (\xi,\eta)\in\widehat{\Omega},
    \label{eq:gift_geometry}
\end{equation}
where $R_A^g$ are the geometric NURBS basis functions and $\bs{P}_A$ are the CAD control points. On the same parametric domain as the fixed geometry map, the transformed displacement is approximated independently by active THB basis functions,
\begin{equation}
    \widehat{\widetilde{\bs{u}}}_h(\xi,\eta;k,\omega)
    =
    \mat{N}_{\mathrm T}(\xi,\eta)\,\tilde{\bs{d}}_{\mathrm T},
    \qquad
    \mat{N}_{\mathrm T}=
    \begin{bmatrix}
    T_1 & \cdots & T_{n_T}
    \end{bmatrix}\otimes\mat{I}_3 ,
    \label{eq:thb_field}
\end{equation}
where $\tilde{\bs{d}}_{\mathrm T}$ collects the three displacement components associated with the active THB basis functions. The physical field is the pullback $\widetilde{\bs{u}}_h(\bs{x};k,\omega)=\widehat{\widetilde{\bs{u}}}_h(\bs{F}^{-1}(\bs{x});k,\omega)$ for $\bs{x}\in\Omega_{\mathrm T}$.
In general, the geometric basis $\{R_A^g\}_{A=1}^{n_g}$ and the analysis basis $\{T_I\}_{I=1}^{n_T}$ are different, and $n_g$ need not equal $n_T$. THB refinement therefore modifies only the analysis space, while the NURBS geometry map, its control points, and the geometry Jacobian remain fixed.

\begin{figure}[!htbp]
\centering
\includegraphics[width=\linewidth]{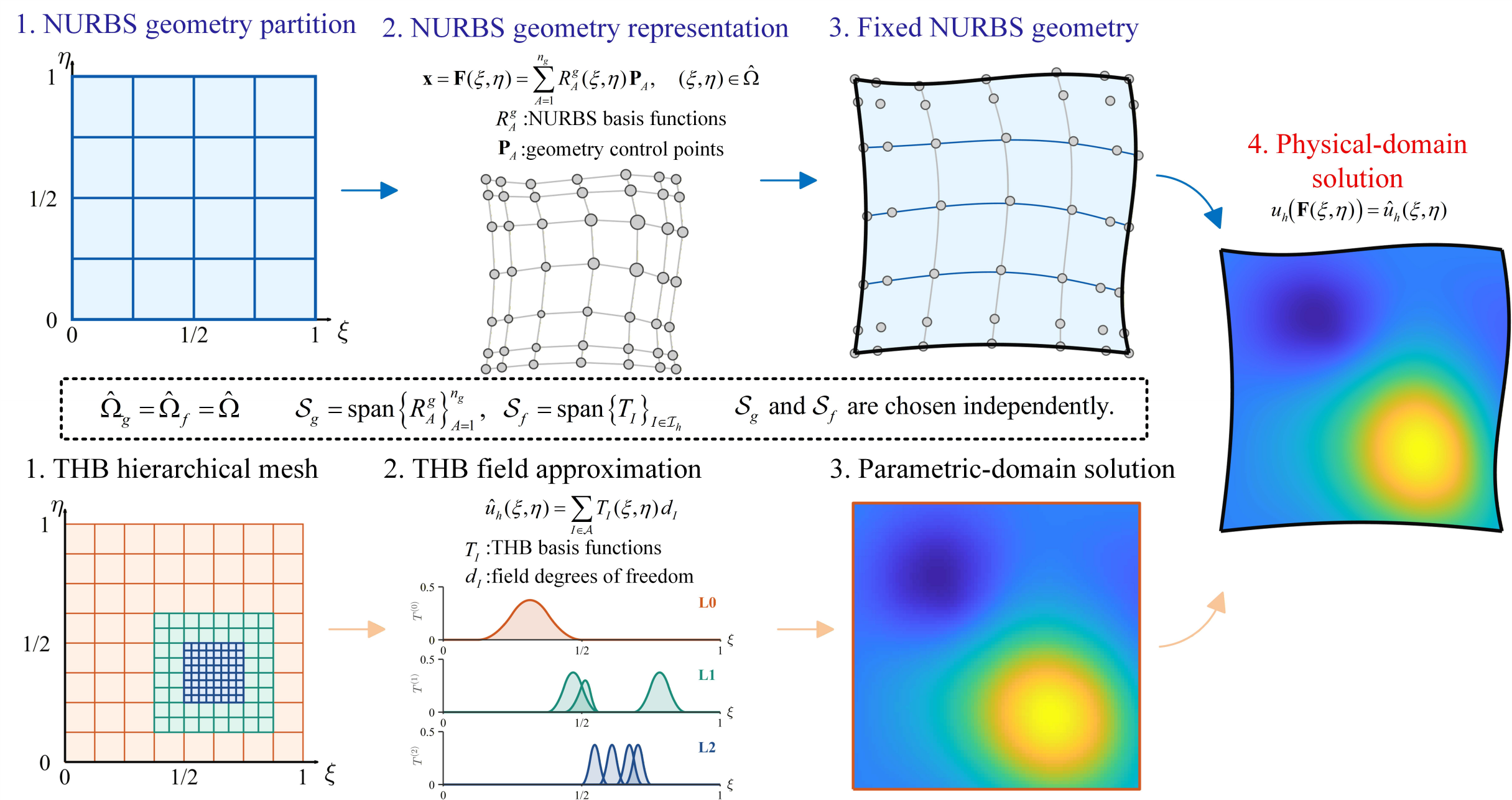}
\caption{Geometry-independent THB-GIFT representation of the near-field patch. The upper branch maps the parametric domain to the curved NURBS geometry, whose control net, weights, knot vectors, and basis remain fixed during refinement. The lower branch defines the independently refined THB field space. The hierarchical mesh and one-dimensional profiles show active basis scales along an interior cut; the solution panel shows the field in parameter and physical space. The geometry space $\mathcal{S}_g=\operatorname{span}\{R_A^g\}$ and field space $\mathcal{S}_f=\operatorname{span}\{T_I\}$ are both evaluated on the common parameter domain $\widehat{\Omega}$.}
\label{fig:thb_gift_refinement}
\end{figure}

\begin{figure}[p]
    \centering
    \captionsetup[subfigure]{
        font=scriptsize,
        justification=centering,
        singlelinecheck=true,
        skip=0.5pt
    }
    
    % ============================================================
    % Row 1: global overview
    % ============================================================
    
    \begin{subfigure}[t]{0.42\linewidth}
        \centering
        \includegraphics[width=\linewidth]
        {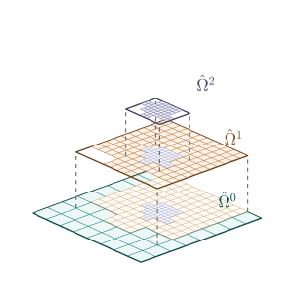}
        \caption{Nested domains}
        \label{fig:thb_leaf_mesh}
    \end{subfigure}%
    \hspace{0.025\linewidth}%
    \begin{subfigure}[t]{0.42\linewidth}
        \centering
        \includegraphics[width=\linewidth]
        {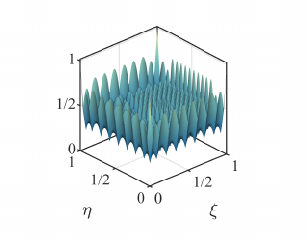}
        \caption{Final THB family $\mathcal{T}^{[2]}$}
        \label{fig:thb_final_three_level}
    \end{subfigure}
    
    \par\vspace{1.4mm}
    
    % ============================================================
    % Row 2: complete tensor-product B-spline families
    % ============================================================
    
    \begin{subfigure}[t]{0.30\linewidth}
        \centering
        \includegraphics[width=\linewidth]
        {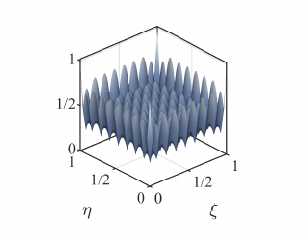}
        \caption{$\mathcal{B}^{0}$}
        \label{fig:thb_bspline_level0}
    \end{subfigure}%
    \hspace{0.008\linewidth}%
    \begin{subfigure}[t]{0.30\linewidth}
        \centering
        \includegraphics[width=\linewidth]
        {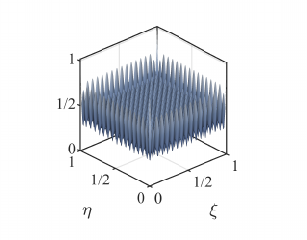}
        \caption{$\mathcal{B}^{1}$}
        \label{fig:thb_bspline_level1}
    \end{subfigure}%
    \hspace{0.008\linewidth}%
    \begin{subfigure}[t]{0.30\linewidth}
        \centering
        \includegraphics[width=\linewidth]
        {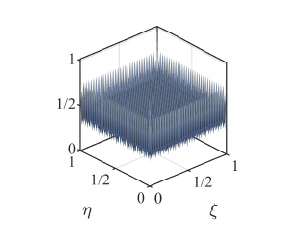}
        \caption{$\mathcal{B}^{2}$}
        \label{fig:thb_bspline_level2}
    \end{subfigure}
    
    \par\vspace{1.3mm}
    
    % ============================================================
    % Row 3: active HB contributions
    % ============================================================
    
    \begin{subfigure}[t]{0.30\linewidth}
        \centering
        \includegraphics[width=\linewidth]
        {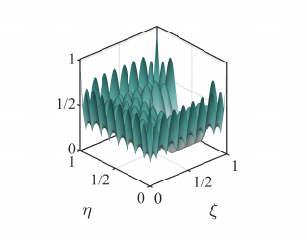}
        \caption{$\mathcal{H}_{0}^{[1]}$}
        \label{fig:thb_hb_level0}
    \end{subfigure}%
    \hspace{0.008\linewidth}%
    \begin{subfigure}[t]{0.30\linewidth}
        \centering
        \includegraphics[width=\linewidth]
        {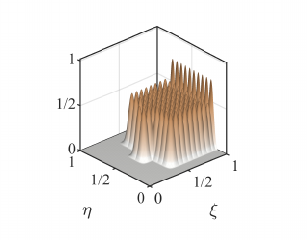}
        \caption{$\mathcal{H}_{1}^{[1]}$}
        \label{fig:thb_hb_level1}
    \end{subfigure}%
    \hspace{0.008\linewidth}%
    \begin{subfigure}[t]{0.30\linewidth}
        \centering
        \includegraphics[width=\linewidth]
        {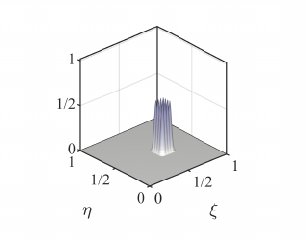}
        \caption{$\mathcal{H}_{2}^{[2]}$}
        \label{fig:thb_hb_level2}
    \end{subfigure}
    
    \par\vspace{1.3mm}
    
    % ============================================================
    % Row 4: truncation and hierarchical assembly
    % ============================================================
    
    \begin{subfigure}[t]{0.30\linewidth}
        \centering
        \includegraphics[width=\linewidth]
        {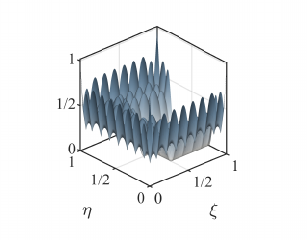}
        \caption{$\operatorname{Trunc}^{1} \!\left(\mathcal{H}_{0}^{[1]}\right)$}
        \label{fig:thb_truncated_level0}
    \end{subfigure}%
    \hspace{0.008\linewidth}%
    \begin{subfigure}[t]{0.30\linewidth}
        \centering
        \includegraphics[width=\linewidth]
        {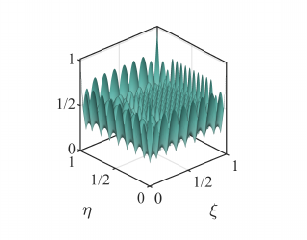}
        \caption{$\mathcal{T}^{[1]}$}
        \label{fig:thb_two_level}
    \end{subfigure}%
    \hspace{0.008\linewidth}%
    \begin{subfigure}[t]{0.30\linewidth}
        \centering
        \includegraphics[width=\linewidth]
        {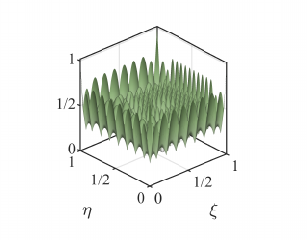}
        \caption{$\operatorname{Trunc}^{2} \!\left(\mathcal{T}^{[1]}\right)$}
        \label{fig:thb_recursive_truncation}
    \end{subfigure}
    
    \vspace{0.4mm}
    
    \caption{Three-level HB/THB construction used to define the active analysis basis. The first row shows the nested refinement domains and the final family $\mathcal{T}^{[2]}$. The remaining panels compare complete tensor-product families $\mathcal{B}^{\ell}$, stage-dependent HB contributions $\mathcal{H}_{0}^{[1]}$, $\mathcal{H}_{1}^{[1]}$ and $\mathcal{H}_{2}^{[2]}$, and the truncation steps leading to $\mathcal{T}^{[1]}$ and $\operatorname{Trunc}^{2}(\mathcal{T}^{[1]})$. Each three-dimensional panel shows the pointwise upper envelope $E_{\mathcal F}(\xi,\eta)=\max_{\phi\in\mathcal F}\phi(\xi,\eta)$ of the indicated basis-function family over the parametric domain.}
    \label{fig:thb_basis_construction}
    \end{figure}

The active basis, its support locality, and its trace regularity follow from nested spline spaces. Let
\begin{equation}
    V^0\subset V^1\subset\cdots\subset V^L
    \label{eq:nested_spline_spaces}
\end{equation}
be a nested sequence of tensor-product B-spline spaces of degree $(p_\xi,p_\eta)$, and let
\begin{equation}
    \widehat{\Omega}^0\supseteq\widehat{\Omega}^1\supseteq\cdots\supseteq\widehat{\Omega}^{L}
    \label{eq:nested_refinement_domains}
\end{equation}
be the corresponding nested refinement domains, with $\widehat{\Omega}^0=\widehat{\Omega}$. The complete tensor-product B-spline basis at level $\ell$ is $\mathcal{B}^{\ell}=\{B_i^\ell:\,i\in\mathcal{I}^{\ell}\}$.
For a construction stage with finest active level $m$, set $\widehat{\Omega}^{m+1}_{[m]}=\emptyset$ and $\widehat{\Omega}^{r}_{[m]}=\widehat{\Omega}^{r}$ for $r=0,\ldots,m$. The HB contribution from level $\ell$ at this stage is
\begin{equation}
    \mathcal{H}_{\ell}^{[m]}
    =
    \left\{
    B_i^\ell\in\mathcal{B}^{\ell}:
    \operatorname{supp}B_i^\ell\subseteq\widehat{\Omega}^{\ell}_{[m]}
    \ \mathrm{and}\
    \operatorname{supp}B_i^\ell\nsubseteq\widehat{\Omega}^{\ell+1}_{[m]}
    \right\},
    \qquad 0\le \ell\le m .
    \label{eq:hb_active_set}
\end{equation}
This stage notation is used because, in a three-level construction, $\mathcal{H}_{1}^{[1]}$ is the level-1 active contribution of the intermediate two-level hierarchy, whereas $\mathcal{H}_{1}^{[2]}$ excludes the functions replaced by level-2 refinement. Because of the nested relation in \eqref{eq:nested_spline_spaces}, every coarse basis function admits a two-scale representation on the next level,
\begin{equation}
    B_i^\ell
    =
    \sum_j c_{ij}^{\ell+1}B_j^{\ell+1}.
    \label{eq:two_scale_relation}
\end{equation}
The THB basis replaces the active hierarchical function by its truncated version,
\begin{equation}
    \operatorname{Trunc}^{\ell+1}B_i^\ell
    =
    \sum_{\operatorname{supp}B_j^{\ell+1}\nsubseteq\widehat{\Omega}^{\ell+1}}
    c_{ij}^{\ell+1}B_j^{\ell+1},
    \label{eq:thb_truncation}
\end{equation}
and applies this operation successively through finer levels. For a set of basis functions $\mathcal{F}$, truncation acts elementwise as $\operatorname{Trunc}^{\ell}(\mathcal{F})=\{\operatorname{Trunc}^{\ell}\phi:\phi\in\mathcal{F}\}$.
The two- and three-level THB families used in \Cref{fig:thb_basis_construction} are then
\begin{align}
    \mathcal{T}^{[1]}
    &=
    \operatorname{Trunc}^{1}\!\left(\mathcal{H}_{0}^{[1]}\right)
    \cup
    \mathcal{H}_{1}^{[1]},
    \label{eq:two_level_thb_family}\\
    \mathcal{T}^{[2]}
    &=
    \operatorname{Trunc}^{2}\!\left(\mathcal{T}^{[1]}\right)
    \cup
    \mathcal{H}_{2}^{[2]} .
    \label{eq:three_level_thb_family}
\end{align}
When no ambiguity is possible, the equivalent shorthand $\mathcal{T}^{0:1}:=\mathcal{T}^{[1]}$ and $\mathcal{T}^{0:2}:=\mathcal{T}^{[2]}$ is used. Truncation reduces the overlap of coarse functions inside refined regions and retains the partition-of-unity and non-negativity properties that are important for stable adaptive isogeometric analysis \citep{Giannelli2016THBIGA,VentaVinuela2026THBPhaseField}.

The continuity of the THB approximation is controlled by the underlying B-spline knot multiplicities rather than by the leaf-cell layout. Across an internal knot line of multiplicity $r$, a degree-$p$ basis is $C^{p-r}$ continuous. Thus simple internal knots give $C^{p-1}$ continuity. Material interfaces are handled by aligning the analysis partition with the physical layer breaks and, when a displacement-only continuous layer interface is required, increasing the corresponding knot multiplicity. Perfect bonding enforces displacement and traction continuity, while strains and individual stress components may remain discontinuous across material interfaces. Local THB refinement itself does not introduce finite-element-style hanging-node constraints inside a patch. The common interface partition in \Cref{subsec:layer_conforming_interface} assumes this layer-aligned trace.

\subsection{Leaf-Cell Discretization and Near-Field Dynamic Stiffness}
\label{subsec:thb_leaf_integration}

The THB basis functions are integrated over the leaf partition $\mathcal{Q}_{\mathrm T}$ of the hierarchical mesh. On a leaf cell $Q$, the geometry Jacobian is $J_F(\xi,\eta)=\det[\partial(x,y)/\partial(\xi,\eta)]$. Substitution of \eqref{eq:thb_field} into \eqref{eq:strain_25d} gives the 2.5D strain-displacement matrix $\mat{B}_{\mathrm T}(k)$. The near-field stiffness and mass matrices are
\begin{align}
    \mat{K}_{\mathrm T}(k)
    &=
    \sum_{Q\in\mathcal{Q}_{\mathrm T}}
    \int_Q
    \mat{B}_{\mathrm T}^{\mathrm{H}}(k)
    \mat{D}
    \mat{B}_{\mathrm T}(k)
    |J_F|\,\dd\xi\,\dd\eta,
    \nonumber\\
    \mat{M}_{\mathrm T}
    &=
    \sum_{Q\in\mathcal{Q}_{\mathrm T}}
    \int_Q
    \rho\,
    \mat{N}_{\mathrm T}^{\mathrm{H}}
    \mat{N}_{\mathrm T}
    |J_F|\,\dd\xi\,\dd\eta .
    \label{eq:thb_stiffness_mass}
\end{align}
The THB near-field dynamic stiffness is therefore
\begin{equation}
    \mat{S}_{\mathrm T}(k,\omega)
    =
    \mat{K}_{\mathrm T}(k)-\omega^2\mat{M}_{\mathrm T} .
    \label{eq:thb_dynamic_stiffness}
\end{equation}
For a multi-patch THB discretization with $n_p$ patches sharing at least $C^0$ seams, the global operators are assembled as
\begin{equation}
    \mat{K}_{\mathrm T}
    =
    \sum_{p=1}^{n_p}
    \left(\mat{A}_{\mathrm T}^p\right)^{\mathsf T}
    \mat{K}_{\mathrm T}^p
    \mat{A}_{\mathrm T}^p,
    \qquad
    \mat{M}_{\mathrm T}
    =
    \sum_{p=1}^{n_p}
    \left(\mat{A}_{\mathrm T}^p\right)^{\mathsf T}
    \mat{M}_{\mathrm T}^p
    \mat{A}_{\mathrm T}^p,
    \label{eq:multipatch_assembly}
\end{equation}
where $\mat{A}_{\mathrm T}^p$ is the Boolean assembly operator that merges shared seam degrees of freedom through a common global numbering. The same assembly convention applies to $\mat{S}_{\mathrm T}$, $\tilde{\bs{f}}_{\mathrm T}$, and any recovered response vector. The THB trace needed by the coupling is made explicit by
\begin{equation}
    \tilde{\bs{d}}_{\Gamma}^{\mathrm T}
    =
    \mat{R}_{\Gamma,\mathrm T}\tilde{\bs{d}}_{\mathrm T},
    \qquad
    \tilde{\bs{u}}_{\Gamma}^{\mathrm T}(s)
    =
    \mat{N}_{\mathrm T}^{\Gamma}(s)\tilde{\bs{d}}_{\Gamma}^{\mathrm T},
    \label{eq:thb_interface_trace}
\end{equation}
where the restriction matrix $\mat{R}_{\Gamma,\mathrm T}$ selects the independent nonzero THB trace degrees of freedom and $\mat{N}_{\mathrm T}^{\Gamma}$ is the corresponding physical boundary basis. Thus \Cref{eq:thb_dynamic_stiffness,eq:thb_interface_trace} provide, respectively, the near-field operator and the master trace passed to the coupling construction in \Cref{sec:coupling}. The transformed surface load is assembled consistently from
\begin{equation}
    \tilde{\bs{f}}_{\mathrm T}
    =
    \int_{\Gamma_t^{\mathrm T}}
    \mat{N}_{\mathrm T}^{\mathrm{H}}\tilde{\bs{t}}\,
    \dd\Gamma,
    \label{eq:thb_load_vector}
\end{equation}
and the THB monitoring response at $\bs{x}_{\mathrm{obs}}=\bs{F}(\xi_{\mathrm{obs}},\eta_{\mathrm{obs}})$ is $\tilde{\bs{u}}_h(\bs{x}_{\mathrm{obs}};k,\omega)=\mat{N}_{\mathrm T}(\xi_{\mathrm{obs}},\eta_{\mathrm{obs}})\tilde{\bs{d}}_{\mathrm T}$. Monitors in the surrounding domain use the corresponding SBIGA trace and radial evaluation.

\section{Quadtree-SBIGA Surrounding Domain and Exterior Impedance}
\label{sec:trace}

In the applications, the quadtree-SBIGA domain represents the soil between the structure-scale near fields and the radiating exterior. A common extracted trace representation connects the quadtree topology and CAD geometry to the surrounding-domain operator. Degree-$p$ Bernstein traces describe straight edges, de Casteljau restrictions enforce compatibility on hanging sub-edges, and rational B\'ezier extraction preserves curved NURBS boundaries. The extracted edge data generate the 2.5D scaled-boundary coefficient matrices and the bounded operator $\mat{S}_{\mathrm{dom}}(k,\omega)$. Condensation of the unbounded exterior on $\Gamma_\infty$ then yields the complete quadtree operator $\mat{S}_{\mathrm Q}(k,\omega)$ passed to the coupling formulation.

\subsection{Unified Polynomial and Rational B\'ezier Trace Representation}
\label{subsec:quadtree_trace}

Let an edge be parametrized by $\eta\in[-1,1]$. The degree-$p$ Bernstein polynomials are
\begin{equation}
    B_i^p(\eta)
    =
    \binom{p}{i}
    \left(\frac{1-\eta}{2}\right)^{p-i}
    \left(\frac{1+\eta}{2}\right)^i,
    \qquad i=0,\ldots,p ,
    \label{eq:bernstein_definition}
\end{equation}
which form a basis for the polynomial space of degree at most $p$ on the edge. Write the basis row as $\bs{B}_p(\eta)=[\,B_0^p(\eta)\ \cdots\ B_p^p(\eta)\,]$.
The scalar trace on a straight edge is represented as
\begin{equation}
    u_e(\eta)=
    \bs{B}_p(\eta)\,
    \mat{C}_e\,
    \bs{d}_e ,
    \label{eq:poly_trace_scalar}
\end{equation}
where $\bs{d}_e\in\mathbb{R}^{p+1}$ contains two vertex values and $p-1$ edge-interior trace values. The matrix $\mat{C}_e\in\mathbb{R}^{(p+1)\times(p+1)}$ maps the trace unknowns to the local Bernstein control values used in the trace interpolation. For a matching straight edge with consistent orientation, $\mat{C}_e=\mat{I}_{p+1}$; if the local edge orientation is reversed, $\mat{C}_e$ is the reversal permutation. More general choices of $\mat{C}_e$ are used below for hanging interfaces and extracted CAD segments.

Equation~\eqref{eq:poly_trace_scalar} is the one-span instance of the extraction interface. More generally, let an edge or spline trace be partitioned into $n_{\mathrm{span}}(e)$ nonzero B\'ezier spans $e_s$. On each span,
\begin{equation}
    \left.u_e\right|_{e_s}(\eta_s)
    =
    \bs{B}_{p_s}(\eta_s)
    \mat{C}_{e}^{s}
    \bs{d}_e,
    \qquad
    s=1,\ldots,n_{\mathrm{span}}(e),
    \label{eq:spanwise_trace_extraction}
\end{equation}
where $\mat{C}_{e}^{s}$ contains the B-spline/B\'ezier extraction, orientation, and any restriction to the active edge variables. A straight polynomial edge has one span. Knot-span refinement increases $n_{\mathrm{span}}(e)$, $p$-refinement changes the local Bernstein degree, and a hanging restriction is composed with $\mat{C}_{e}^{s}$. Polynomial, spline, and restricted traces therefore use their respective span data within the same integration and assembly procedure.

On a straight edge, Bernstein and Lagrange polynomials of the same degree span the same space; Bernstein polynomials are adopted because they are the local basis of B\'ezier extraction~\citep{Borden2011BezierExtraction}. The same basis supports de Casteljau subdivision, degree elevation, and rational B\'ezier functions with weights on CAD-curved edges.

For a polygonal S-element with $n_e$ sides, the scalar trace contains $n_e$ vertex values and $(p-1)n_e$ edge-interior values, giving $pn_e$ local scalar trace values. The vertex values are shared by all incident elements, and edge-interior values are shared by the two elements adjacent to the same edge. On a mesh with $n_v$ vertices and $n_{\mathrm{edge}}$ unique global edges, the total number of scalar trace unknowns is $n_{\text{scalar}}=n_v+(p-1)n_{\mathrm{edge}}$.
Global $C^0$ continuity on matching edges is obtained by sharing these trace unknowns directly. For $p=1$, the edge-interior term vanishes and the trace contains only vertex values. The Bernstein basis then reduces exactly to the standard two-node edge interpolation of low-order SBFEM.

\begin{figure}[!htbp]
    \centering
    \begin{subfigure}[t]{0.32\textwidth}
        \centering
        \includegraphics[width=\linewidth]{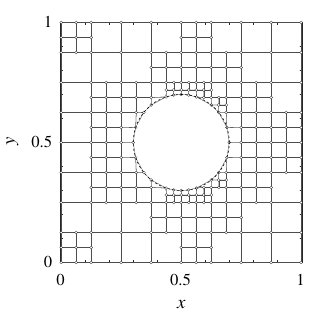}
        \caption{Two-node linear SBFEM edges}
        \label{fig:mesh_traditional_sbfem}
    \end{subfigure}
    \hfill
    \begin{subfigure}[t]{0.32\textwidth}
        \centering
        \includegraphics[width=\linewidth]{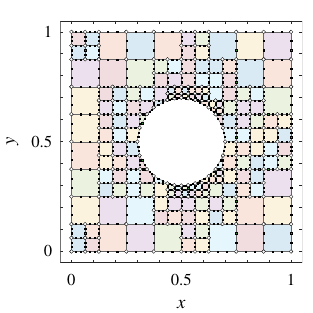}
        \caption{Chordal Bernstein trace}
        \label{fig:mesh_bernstein_p3}
    \end{subfigure}
    \hfill
    \begin{subfigure}[t]{0.32\textwidth}
        \centering
        \includegraphics[width=\linewidth]{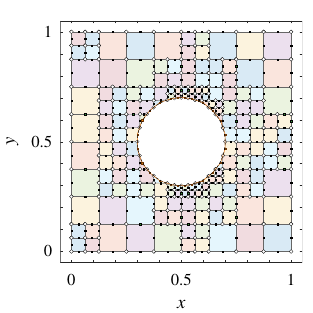}
        \caption{CAD-NURBS boundary trace}
        \label{fig:mesh_nurbs_bernstein_p3}
    \end{subfigure}
    \caption{Boundary-trace representations used to separate approximation order from geometric representation. The two-node SBFEM and high-order Bernstein models retain a chordal boundary approximation, whereas the CAD-NURBS model uses a globally shared curved-boundary trace with local rational B\'ezier extraction.}
    \label{fig:trace_discretization_evolution}
\end{figure}

\Cref{fig:trace_discretization_evolution} separates trace-order enrichment on a fixed chordal skeleton from the rational CAD boundary representation defined below.

\begin{figure}[!htbp]
\centering
\includegraphics[width=0.75\linewidth]{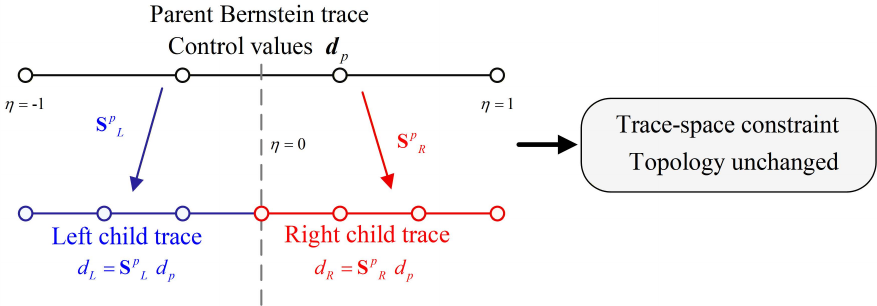} 
\caption{B\'ezier subdivision constraint for a hanging quadtree interface. The child-edge trace values are obtained from the parent Bernstein trace by subdivision matrices, preserving degree-$p$ compatibility in trace space while leaving the S-element topology unchanged.}
\label{fig:bezier_hanging}
\end{figure}

A balanced quadtree may place a long edge of a coarse cell next to two shorter edges of refined cells. The high-order extraction enforces compatibility without changing the coarse-cell topology: de Casteljau subdivision constrains the child traces to the parent Bernstein trace, as illustrated in \Cref{fig:bezier_hanging}. Let a parent edge be split at its midpoint $\eta=0$. The Bernstein control values of the left and right half-edges, denoted $\bs{d}_L,\bs{d}_R\in\mathbb{R}^{p+1}$, follow from the parent control values $\bs{d}_P$ as
\begin{equation}
    \bs{d}_L = \mat{S}_L^p\,\bs{d}_P, \qquad
    \bs{d}_R = \mat{S}_R^p\,\bs{d}_P,
    \label{eq:decasteljau_subdivision}
\end{equation}
where $\mat{S}_L^p,\mat{S}_R^p\in\mathbb{R}^{(p+1)\times(p+1)}$ are the standard degree-$p$ midpoint de Casteljau subdivision operators~\citep{Borden2011BezierExtraction}. Each row sums to unity, so the constrained sub-edge traces preserve the partition-of-unity property of the parent trace.

In the element extraction used here, the independent trace variables are the parent trace variables, and a child edge receives its local Bernstein control values through a local restriction row:
\begin{equation}
    \bs{a}_{c}
    =
    \mat{C}_{e}^{\mathrm{hang}}\bs{d}_{P},
    \qquad
    \mat{C}_{e}^{\mathrm{hang}}
    \in
    \{\mat{S}_{L}^{p},\mat{S}_{R}^{p}\}
    \quad
    \text{up to edge orientation.}
    \label{eq:hang_constraint}
\end{equation}
Embedding \eqref{eq:hang_constraint} in the local edge extraction enforces compatibility before global assembly. If the trace topology is already split at the hanging point, a matching sub-edge identity or orientation permutation provides the same restriction.

Curved traces use the rational counterpart of the same local B\'ezier representation.
\label{subsec:nurbs_trace}

A CAD boundary curve is represented as a NURBS of order $q$ and degree $p_c=q-1$. To distinguish the CAD-curve parameter from the scaled-boundary radial coordinate used later, the curve parameter is denoted by $\chi$ in this subsection. With knot vector $\Xi=[\chi_0,\ldots,\chi_m]$, control points $\bs{P}_i$, and weights $w_i$, the physical point at parameter $\chi$ is
\begin{equation}
    \bs{c}(\chi) = \frac{\sum_{i=0}^{n} N_{i,q}(\chi)\,w_i\,\bs{P}_i}
                          {\sum_{i=0}^{n} N_{i,q}(\chi)\,w_i},
    \label{eq:nurbs_curve}
\end{equation}
where $N_{i,q}(\chi)$ are the B-spline basis functions defined by the Cox--de Boor recurrence.

\begin{figure}[!htbp]
\centering
\includegraphics[width=\linewidth]{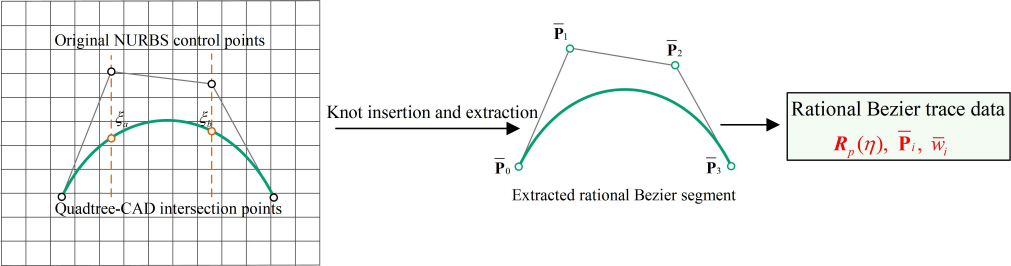}
\caption{CAD-NURBS boundary segment represented as local rational trace data. Knot insertion isolates the quadtree boundary segment while preserving the CAD curve exactly; the extracted rational B\'ezier segment provides the basis values, control points, and weights used in scaled-boundary integration.}
\label{fig:nurbs_bezier_extraction}
\end{figure}

The physical endpoints of a quadtree edge on the CAD boundary are projected onto $\bs{c}(\chi)$ to obtain $\chi_a$ and $\chi_b$ before knot insertion (\Cref{fig:nurbs_bezier_extraction}).

The endpoint parameters are inserted to full multiplicity, and internal knots subdivide $[\chi_a,\chi_b]$ into its nonzero B\'ezier spans. If the analysis trace degree $p$ exceeds the CAD degree $p_c$, spanwise rational degree elevation preserves the geometry. If $p<p_c$, exact CAD representation instead requires separate geometry and analysis degrees. Subject to this degree condition, the extracted rational B\'ezier spans reproduce the NURBS boundary within numerical precision and enter the same spanwise interface as polynomial edges. On a span of degree $p$, the rational Bernstein basis is
\begin{equation}
    R_i^p(\eta)=
    \frac{B_i^p(\eta)\,\bar{w}_i}
    {\sum_{j=0}^{p}B_j^p(\eta)\,\bar{w}_j},
    \qquad i=0,\ldots,p ,
    \label{eq:rational_bezier}
\end{equation}
and replaces the polynomial basis in~\eqref{eq:poly_trace_scalar}, yielding
\begin{equation}
    \left.\tilde{\bs{u}}_e\right|_{e_s}(\eta_s)
    =
    \left[
    \bs{R}_p(\eta_s)\,\mat{C}_e^s\otimes\mat{I}_3
    \right]
    \tilde{\bs{d}}_e .
    \label{eq:rational_trace}
\end{equation}
The span extraction matrix $\mat{C}_e^s$ has the same role as on a straight edge. Setting all weights to unity recovers $\bs{R}_p(\eta)=\bs{B}_p(\eta)$. Polynomial and rational edges thus enter the same trace interface through their span data and weights. The SBFEM coefficient matrices~\eqref{eq:coefficient_matrices} are evaluated span by span from the same boundary-integral expressions, with the rational basis and its derivatives substituted for the polynomial ones.

\begin{figure}[!htbp]
    \centering

    \begin{subfigure}[t]{0.32\textwidth}
        \centering
        \includegraphics[width=\linewidth]{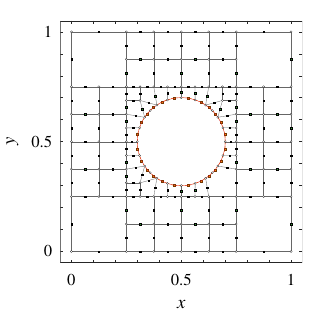}
        \caption{$\ell_{\max}=4$, $h_{\min}=1/16$, $p=2$}
        \label{fig:mesh_level4_p2}
    \end{subfigure}
    \hfill
    \begin{subfigure}[t]{0.32\textwidth}
        \centering
        \includegraphics[width=\linewidth]{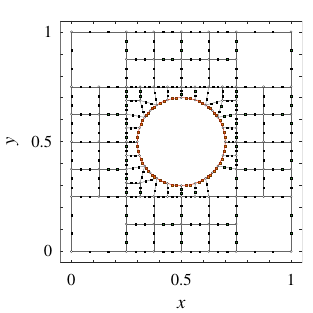}
        \caption{$\ell_{\max}=4$, $h_{\min}=1/16$, $p=3$}
        \label{fig:mesh_level4_p3_order}
    \end{subfigure}
    \hfill
    \begin{subfigure}[t]{0.32\textwidth}
        \centering
        \includegraphics[width=\linewidth]{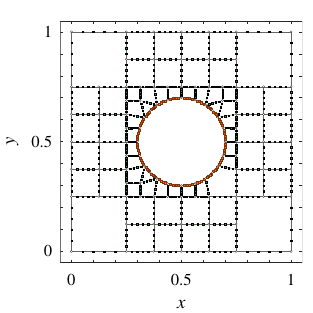}
        \caption{$\ell_{\max}=4$, $h_{\min}=1/16$, $p=5$}
        \label{fig:mesh_level4_p5}
    \end{subfigure}

    \vspace{0.8em}

    \begin{subfigure}[t]{0.32\textwidth}
        \centering
        \includegraphics[width=\linewidth]{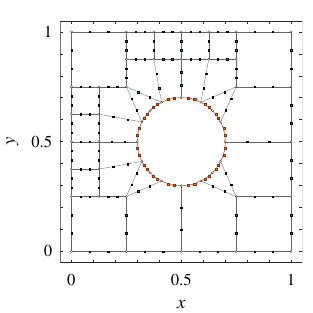}
        \caption{$\ell_{\max}=3$, $h_{\min}=1/8$, $p=3$}
        \label{fig:mesh_level3_p3}
    \end{subfigure}
    \hfill
    \begin{subfigure}[t]{0.32\textwidth}
        \centering
        \includegraphics[width=\linewidth]{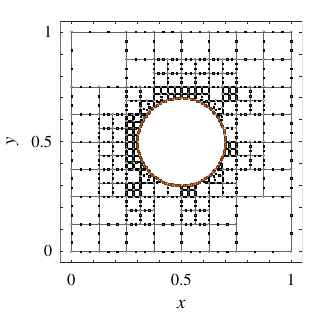}
        \caption{$\ell_{\max}=5$, $h_{\min}=1/32$, $p=3$}
        \label{fig:mesh_level5_p3}
    \end{subfigure}
    \hfill
    \begin{subfigure}[t]{0.32\textwidth}
        \centering
        \includegraphics[width=\linewidth]{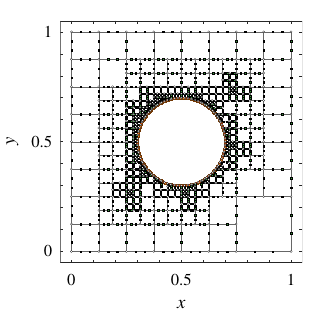}
        \caption{$\ell_{\max}=6$, $h_{\min}=1/64$, $p=3$}
        \label{fig:mesh_level6_p3}
    \end{subfigure}

    \caption{Polynomial-order elevation and adaptive quadtree refinement for the CAD-NURBS boundary trace. The upper row varies the trace degree on the fixed adaptive mesh with $\ell_{\max}=4$. The lower row varies the local refinement depth at fixed trace degree $p=3$. For the unit-square root cell, the finest admissible leaf-cell size is $h_{\min}=2^{-\ell_{\max}}$.}
    \label{fig:s04_mesh_and_order_refinement}
\end{figure}

\Cref{fig:s04_mesh_and_order_refinement} illustrates trace-degree elevation on a fixed quadtree and local quadtree refinement at fixed degree. Split or hanging sub-edges use the restriction in \Cref{eq:hang_constraint}, and the curved boundary retains the local rational B\'ezier extraction of the global CAD-NURBS trace.

Several disconnected CAD boundaries, such as voids, inclusions, or material interfaces, are handled by associating each CAD edge with a curve index and endpoint parameters; the coefficient expressions are unchanged.

The local trace construction is examined independently through a two-dimensional plate-with-a-circular-hole elasticity benchmark. The unit-square domain contains a centred circular hole of radius $0.2$ and uses plane-stress elasticity with $E=10$ and $\nu=0.3$. The left boundary is fixed and a unit horizontal traction acts on the right boundary. The strain energy $\mathcal{U}=\tfrac12\bs{u}^{\mathsf T}\mat{K}\bs{u}$ simultaneously exercises straight Bernstein traces, exact rational B\'ezier geometry, local quadtree refinement, and hanging interfaces. \Cref{tab:trace_verification_hole} reports polynomial-order and local-level refinement. The differences $\delta_p$ and $\delta_h$ are measured against the highest order for the same boundary representation and against $\ell_{\max}=7$, respectively.

\begin{table}[!htbp]
\centering
\caption{Numerical verification of the extraction-based quadtree trace construction for the plate-with-a-circular-hole benchmark.}
\label{tab:trace_verification_hole}
\footnotesize
\setlength{\tabcolsep}{4.5pt}
\begin{tabular}{llrrr}
\toprule
\multicolumn{5}{l}{\textbf{Panel A: fixed manual mesh, polynomial-order refinement}}\\
Boundary/trace representation & $p$ & DOFs & $\mathcal{U}$ & Relative difference $\delta_p$\\
\midrule
Two-node SBFEM baseline & 1 & 446 & 0.073827847131 & --\\
Chordal Bernstein trace & 2 & 1236 & 0.074511890704 & $3.37\times10^{-4}$\\
Chordal Bernstein trace & 3 & 2026 & 0.074532727935 & $5.74\times10^{-5}$\\
Chordal Bernstein trace & 5 & 3606 & 0.074536168556 & $1.12\times10^{-5}$\\
Chordal Bernstein trace & 7 & 5186 & 0.074536805382 & $2.69\times10^{-6}$\\
Chordal Bernstein trace & 9 & 6766 & 0.074537005605 & 0\\
Local rational B\'ezier trace & 2 & 1236 & 0.074842850584 & $1.60\times10^{-4}$\\
Local rational B\'ezier trace & 3 & 2026 & 0.074853168845 & $2.20\times10^{-5}$\\
Local rational B\'ezier trace & 5 & 3606 & 0.074854469428 & $4.64\times10^{-6}$\\
Local rational B\'ezier trace & 7 & 5186 & 0.074854729892 & $1.16\times10^{-6}$\\
Local rational B\'ezier trace & 9 & 6766 & 0.074854816709 & 0\\
\bottomrule
\end{tabular}

\vspace{2mm}

\begin{tabular}{lrrrrr}
\toprule
\multicolumn{6}{l}{\textbf{Panel B: adaptive quadtree with global CAD-NURBS trace}}\\
Trace representation & $p$ & $\ell_{\max}$ & $h_{\min}$ & $\mathcal{U}$ & Relative difference $\delta_h$\\
\midrule
CAD-NURBS trace & 9 & 3 & 1/8   & 0.074854189071 & $6.25\times10^{-6}$\\
CAD-NURBS trace & 9 & 4 & 1/16  & 0.074854197106 & $6.14\times10^{-6}$\\
CAD-NURBS trace & 9 & 5 & 1/32  & 0.074854426831 & $3.07\times10^{-6}$\\
CAD-NURBS trace & 9 & 6 & 1/64  & 0.074854656553 & $2.01\times10^{-12}$\\
CAD-NURBS trace & 9 & 7 & 1/128 & 0.074854656553 & 0\\
\bottomrule
\end{tabular}

\begin{flushleft}
\footnotesize
Relative differences are taken with respect to the $p=9$ result of the same boundary representation (Panel A) and the $\ell_{\max}=7$ result (Panel B).
\end{flushleft}
\end{table}

On the fixed manual mesh of Panel A, the strain energy increases monotonically with $p$ for both trace families. The chordal and locally extracted rational traces give strain energies that differ by about $0.4\%$ at the highest order.

The rational representation preserves the circular geometry exactly under the stated degree condition. Panel B fixes $p=9$ and refines the quadtree locally around the circular boundary, which is represented by a globally shared CAD-NURBS trace. The relative difference in strain energy from the finest-mesh result stays below $10^{-5}$ at all tested levels. The lower row of \Cref{fig:s04_mesh_and_order_refinement} illustrates selected refinement levels at $p=3$. The same extracted boundary operators form the 2.5D scaled-boundary coefficient matrices below.

\subsection{Extracted 2.5D SBIGA Bounded-Domain Operator}
\label{subsec:sbiga25d}
\label{subsec:scaled_boundary_approx}
\label{subsec:coefficient_matrices}
\label{subsec:dynamic_stiffness}

At element level, polynomial, hanging, and rational B\'ezier edge traces enter one extraction operator. Let $\bs{a}_{\Omega_e}$ collect the local Bernstein or rational-B\'ezier control values on an S-element, and let $\bs{d}_{\Omega_e}$ collect its unique scalar trace unknowns after shared vertices and edges have been merged. The element extraction relation is
\begin{equation}
    \bs{a}_{\Omega_e}
    =
    \mat{C}_{\Omega_e}^{s}\,
    \bs{d}_{\Omega_e}.
    \label{eq:element_extraction}
\end{equation}
The rows of $\mat{C}_{\Omega_e}^{s}$ are assembled from the edge-span maps introduced in \Cref{subsec:quadtree_trace}. Matching straight edges contribute an identity or reversal permutation, hanging child edges contribute $\mat{S}_L^p$ or $\mat{S}_R^p$, and spline or rational CAD edges contribute one block per nonzero B\'ezier span. Shared vertex rows are merged at element level. For the three-component 2.5D displacement, the vector extraction is
\begin{equation}
    \mat{C}_{\Omega_e}^{(3)}
    =
    \mat{C}_{\Omega_e}^{s}\otimes\mat{I}_3 .
    \label{eq:vector_extraction}
\end{equation}
The operator $\mat{C}_{\Omega_e}^{(3)}$ transfers the element dynamic stiffness from the local extracted trace to the global trace variables in \eqref{eq:Sdom_assembly}. Thus, scaled-boundary integration uses the same form of boundary interpolation for each edge category.

For a scaled-boundary element with scaling centre $\bs{x}_0$, any point inside the element is represented by a radial coordinate $\xi$ and a boundary coordinate $\eta$. In this section $\xi$ always denotes the scaled-boundary radius; it is unrelated to the first THB parametric coordinate used in \Cref{subsec:thb_gift_space}. The physical coordinates are
\begin{equation}
    \bs{x}(\xi,\eta)=\bs{x}_0+\xi\left[\bs{x}_b(\eta)-\bs{x}_0\right],
    \qquad 0\le \xi \le 1 ,
    \label{eq:scaled_mapping}
\end{equation}
where $\xi$ is the radial coordinate and $\eta$ is the boundary coordinate. In what follows, $x_b$ and $y_b$ denote the boundary coordinates measured from the scaling centre, and
\begin{equation}
    J_b=x_b y_{b,\eta}-y_b x_{b,\eta}
    \label{eq:boundary_jacobian}
\end{equation}
is the scaled-boundary boundary Jacobian. The sign of $J_b$ is fixed by the counter-clockwise ordering of the element boundary, and a positive $J_b$ ensures that the scaling centre sees the whole boundary without radial self-intersection.

The transformed displacement inside the S-element is separated into an extracted boundary interpolation and an unknown radial function,
\begin{equation}
    \tilde{\bs{u}}(\xi,\eta)
    =
    \mat{N}(\eta)\,
    \tilde{\bs{u}}_{\xi}(\xi),
    \qquad
    \mat{N}(\eta)=\bs{N}(\eta)\otimes\mat{I}_3 ,
    \label{eq:sbiga_displacement}
\end{equation}
where $\bs{N}$ is the extracted trace basis defined in \Cref{sec:trace}. The scaled-boundary ordinary differential equation governs the radial variation $\tilde{\bs{u}}_{\xi}(\xi)$ with boundary-only trace unknowns and no radial Lagrange layers.

Using \Cref{eq:strain_25d} and the scaled-boundary mapping, the strain field separates into three contributions:
\begin{equation}
    \tilde{\bs{\varepsilon}}
    =
    \mat{B}_1(\eta)\,
    \tilde{\bs{u}}_{\xi,\xi}
    +
    \frac{1}{\xi}
    \mat{B}_2(\eta)\,
    \tilde{\bs{u}}_{\xi}
    +
    \mat{B}_3(\eta;k)\,
    \tilde{\bs{u}}_{\xi}.
    \label{eq:strain_decomposition}
\end{equation}
The three terms contain the radial derivative, the boundary derivative, and the longitudinal-wavenumber coupling, respectively. The corresponding strain-displacement blocks are
\begin{align}
    \mat{B}_1 &= \mat{b}_1\mat{N}, &
    \mat{B}_2 &= \mat{b}_2\mat{N}_{,\eta}, &
    \mat{B}_3 &= \mat{b}_3(k)\mat{N},
    \label{eq:B_blocks}
\end{align}
The matrices $\mat{b}_1$ and $\mat{b}_2$ are the conventional two-dimensional SBFEM geometric derivative operators associated with \eqref{eq:scaled_mapping}~\citep{WolfSong2001SBFEM,Song2018SBFEMBook}. The additional 2.5D coupling is explicit in
\begin{equation}
    \mat{b}_3(k)
    =
    \begin{bmatrix}
    0 & 0 & 0\\
    0 & 0 & 0\\
    0 & 0 & -\ii k\\
    0 & 0 & 0\\
    0 & -\ii k & 0\\
    -\ii k & 0 & 0
    \end{bmatrix}.
    \label{eq:b3}
\end{equation}
The decomposition in \eqref{eq:strain_decomposition} extends the conventional SBFEM coefficient structure to the Fourier-reduced setting. The matrices $\mat{E}_0$, $\mat{E}_1$, and $\mat{E}_2$ retain their two-dimensional roles, whereas $\mat{E}_3$, $\mat{E}_4$, and $\mat{E}_5$ collect the longitudinal-wavenumber contributions. Substitution into the weak form gives
\begin{align}
    \mat{E}_0 &= \int_{\Gamma_e} \mat{B}_1^{\mathrm{H}}\mat{D}\mat{B}_1 J_b\,\dd\eta, &
    \mat{E}_1 &= \int_{\Gamma_e} \mat{B}_2^{\mathrm{H}}\mat{D}\mat{B}_1 J_b\,\dd\eta, \nonumber\\
    \mat{E}_2 &= \int_{\Gamma_e} \mat{B}_2^{\mathrm{H}}\mat{D}\mat{B}_2 J_b\,\dd\eta, &
    \mat{E}_3 &= \int_{\Gamma_e} \mat{B}_3^{\mathrm{H}}\mat{D}\mat{B}_1 J_b\,\dd\eta, \nonumber\\
    \mat{E}_4 &= \int_{\Gamma_e} \mat{B}_3^{\mathrm{H}}\mat{D}\mat{B}_2 J_b\,\dd\eta, &
    \mat{E}_5 &= \int_{\Gamma_e} \mat{B}_3^{\mathrm{H}}\mat{D}\mat{B}_3 J_b\,\dd\eta, \nonumber\\
    \mat{M}_0 &= \int_{\Gamma_e} \rho\,\mat{N}^{\mathrm{H}}\mat{N}J_b\,\dd\eta .
    \label{eq:coefficient_matrices}
\end{align}
All integrals are evaluated edge by edge with Gaussian quadrature. The same formulas apply to polynomial and rational edges because the basis evaluation and the extraction matrix have already been absorbed into $\mat{N}$. When $k=0$, the matrices $\mat{E}_3$, $\mat{E}_4$, and $\mat{E}_5$ vanish except for numerical round-off, and the formulation reduces to the usual two-dimensional SBFEM coefficient form.

Let $\mat{S}_{\mathrm b}(\xi)$ map radial boundary displacement to the outward nodal force of a bounded scaled-boundary element. With $\tau=\ln\xi$, its dynamic stiffness satisfies
\begin{equation}
    \frac{\dd\mat{S}_{\mathrm b}}{\dd\tau}
    =-\left(\mat{S}_{\mathrm b}-\mat{E}_1-\xi\mat{E}_3\right)
    \mat{E}_0^{-1}\left(\mat{S}_{\mathrm b}-\mat{E}_1^{\mathsf T}+\xi\mat{E}_3^{\mathsf T}\right)
    +\mat{E}_2+\xi\left(\mat{E}_4-\mat{E}_4^{\mathsf T}\right)
    +\xi^2\mat{E}_5-\omega^2\xi^2\mat{M}_0 .
    \label{eq:riccati_domain}
\end{equation}
The initial stiffness near the scaling centre is obtained from the invariant subspace of the static SBFEM eigenproblem associated with modes that remain finite as $\xi\rightarrow0$. Equivalently, the starting matrix $\mat{K}_0$ satisfies
\begin{equation}
    \left(\mat{E}_1-\mat{K}_0\right)
    \mat{E}_0^{-1}
    \left(\mat{K}_0-\mat{E}_1^{\mathsf T}\right)
    +
    \mat{E}_2
    =
    \mat{0}.
    \label{eq:static_initial_stiffness}
\end{equation}
The initial condition is $\mat{S}_{\mathrm b}(0)=\mat{K}_0$ in the limiting sense. The element dynamic stiffness is $\mat{S}_e=\mat{S}_{\mathrm b}(1)$. It is computed in the local extracted trace space and transferred to the global trace unknowns through \Cref{eq:vector_extraction}. The bounded-domain global stiffness is assembled as
\begin{equation}
    \mat{S}_{\mathrm{dom}}
    =
    \sum_e
    \mat{A}_e^{\mathsf T}
    \left[
    \mat{C}_{\Omega_e}^{(3)}
    \right]^{\mathsf T}
    \mat{S}_e
    \mat{C}_{\Omega_e}^{(3)}
    \mat{A}_e ,
    \label{eq:Sdom_assembly}
\end{equation}
where $\mat{A}_e$ is the Boolean assembly operator. Because $\mat{A}_e$ and $\mat{C}_{\Omega_e}^{(3)}$ are real, the algebraic transpose is used; the complex conjugation of the sesquilinear weak form is already contained in $\mat{S}_e$.

\subsection{Exterior Impedance and Complete Quadtree Operator}
\label{subsec:exterior_relation}

An impedance relation on the artificial boundary $\Gamma_\infty$ couples the outer quadtree boundary to the semi-infinite medium. The exterior scaled-boundary solution is condensed onto this finite trace; it is distinct from a finite-thickness absorbing layer. Let $\tilde{\bs{u}}_{\Gamma}$ be the transformed displacement trace on $\Gamma_\infty$, and let $\tilde{\bs{q}}_{\Gamma}$ be the nodal force applied by the bounded quadtree domain to the exterior medium. The exterior contribution is represented by
\begin{equation}
    \tilde{\bs{q}}_{\Gamma}
    =
    \mat{S}_{\infty}(k,\omega)
    \tilde{\bs{u}}_{\Gamma}.
    \label{eq:exterior_impedance_relation}
\end{equation}
In a scaled-boundary exterior element, $\xi$ points from the artificial boundary into the unbounded medium. With the trace extraction of \Cref{subsec:quadtree_trace}, its impedance satisfies
\begin{equation}
    \frac{\dd\mat{S}_{\infty}}{\dd\tau}
    =\left(\mat{S}_{\infty}+\mat{E}_1+\xi\mat{E}_3\right)
    \mat{E}_0^{-1}\left(\mat{S}_{\infty}+\mat{E}_1^{\mathsf T}-\xi\mat{E}_3^{\mathsf T}\right)
    -\mat{E}_2-\xi\left(\mat{E}_4-\mat{E}_4^{\mathsf T}\right)
    -\xi^2\mat{E}_5+\omega^2\xi^2\mat{M}_0 .
    \label{eq:riccati_exterior}
\end{equation}

Under the Fourier convention in \Cref{eq:fourier_u}, the complementary longitudinal mixed terms satisfy
\begin{equation}
    \int_{\Gamma_e}\mat{B}_1^{\mathrm{H}}\mat{D}\mat{B}_3 J_b\,\dd\eta
    =
    -\mat{E}_3^{\mathsf T},
    \qquad
    \int_{\Gamma_e}\mat{B}_2^{\mathrm{H}}\mat{D}\mat{B}_3 J_b\,\dd\eta
    =
    -\mat{E}_4^{\mathsf T},
    \label{eq:mixed_transpose_convention}
\end{equation}
where $(\cdot)^{\mathsf T}$ denotes the algebraic transpose. The complex conjugation has already entered the definitions of $\mat{E}_3$ and $\mat{E}_4$ through $\mat{B}_3^{\mathrm{H}}$. The complementary terms therefore use the algebraic transpose; otherwise the sign of the $k$-dependent coupling is changed.

The outgoing-wave condition is enforced through the large-$\xi$ behaviour of the exterior dynamic stiffness. The high-$\xi$ expansion is written as
\begin{equation}
    \mat{S}_{\infty}(\xi)
    =
    \xi\mat{A}_1+\mat{A}_0+O(\xi^{-1}),
    \qquad \xi\rightarrow\infty .
    \label{eq:exterior_asymptotic}
\end{equation}
Substitution into the exterior Riccati equation determines the two coefficients from
\begin{align}
    &\left(\mat{A}_1+\mat{E}_3\right)
    \mat{E}_0^{-1}
    \left(\mat{A}_1-\mat{E}_3^{\mathsf T}\right)
    -\mat{E}_5+\omega^2\mat{M}_0
    =\mat{0},
    \label{eq:A1_exterior}\\
    &\left(\mat{A}_1+\mat{E}_3\right)
    \mat{E}_0^{-1}
    \left(\mat{A}_0+\mat{E}_1^{\mathsf T}\right)
    +
    \left(\mat{A}_0+\mat{E}_1\right)
    \mat{E}_0^{-1}
    \left(\mat{A}_1-\mat{E}_3^{\mathsf T}\right)
    \nonumber\\
    &\hspace{28mm}
    -\left(\mat{E}_4-\mat{E}_4^{\mathsf T}\right)
    -\mat{A}_1
    =\mat{0}.
    \label{eq:A0_exterior}
\end{align}
The outgoing or exponentially decaying invariant subspace selects the admissible solution for $\mat{A}_1$, and $\mat{A}_0$ supplies the first finite-radius correction. The Riccati equation is initialized at a sufficiently large $\xi_h$ by
\begin{equation}
    \mat{S}_{\infty}(\xi_h)
    =
    \xi_h\mat{A}_1+\mat{A}_0 .
    \label{eq:exterior_initial}
\end{equation}
Integration in $\tau=\ln\xi$ proceeds from $\ln\xi_h$ to zero, and convergence of $\mat{S}_\infty(1)$ is checked with respect to $\xi_h$ and the radial-integration tolerance. The sign in \eqref{eq:exterior_impedance_relation} defines $\tilde{\bs{q}}_{\Gamma}$ as the nodal force applied by the bounded quadtree domain to the exterior medium.

Let $\mat{A}_{\infty}$ map the compact exterior trace vector on the selected artificial-boundary edges to the global quadtree trace ordering. The top free surface and the THB--quadtree interface are excluded from this map; only the exterior artificial boundary contributes to $\mat{S}_{\infty}$. The complete quadtree-domain dynamic stiffness is then
\begin{equation}
    \mat{S}_{\mathrm Q}(k,\omega)
    =
    \mat{S}_{\mathrm{dom}}(k,\omega)
    +
    \mat{A}_{\infty}^{\mathsf T}
    \mat{S}_{\infty}(k,\omega)
    \mat{A}_{\infty}.
    \label{eq:quadtree_total_stiffness}
\end{equation}
The operator $\mat{S}_{\mathrm Q}$ already includes the local trace extraction, hanging-edge restrictions, exterior insertion, and the global trace ordering used by the coupling in \Cref{sec:coupling}.

\section{Projection-Coupled THB-GIFT/Quadtree-SBIGA Formulation}
\label{sec:coupling}

The independently assembled operators $\mat{S}_{\mathrm T}$ and $\mat{S}_{\mathrm Q}$ meet on nonmatching traces. In the applications, these traces bound the near field around a conduit or tunnel lining and lie within homogeneous soil, so they are numerical coupling interfaces rather than physical material interfaces. A projection $\boldsymbol{\Pi}$ maps the retained THB trace to the quadtree trace over a common physical partition. The global transformation $\mat{G}$ embeds this map in a work-consistent reduced system, from which the moving-load response is reconstructed.

\subsection{Common-Partition Projection for Nonmatching Traces}
\label{subsec:layer_conforming_interface}
\label{subsec:trace_projection}

The coupling interface may comprise open, closed, or disconnected components. Its general topology is
\begin{equation}
    \Gamma_{\mathrm{TQ}}
    =
    \bigcup_{j=1}^{n_I}\Gamma_j,
    \qquad
    \Gamma_i\cap\Gamma_j=\emptyset\quad(i\ne j).
    \label{eq:general_interface}
\end{equation}
Each connected component is partitioned independently. THB knot breaks, the endpoints of the SBIGA edges and of their nonzero B\'ezier spans, and intersections with material boundaries are combined into non-overlapping integration segments whose union defines the common partition $\mathcal{P}_\Gamma$. Every $g\in\mathcal{P}_\Gamma$ lies within one THB trace span, one nonzero B\'ezier span of a quadtree edge or sub-edge, and one material region. Common-refinement methods provide the broader precedent for integrating transfer relations over intersections of nonmatching interface meshes \citep{JiaoHeath2004CommonRefinement}; here $\mathcal{P}_\Gamma$ supplies the physical integration segments for the THB-to-SBIGA projection matrices while both subdomains retain their independent meshes. Disconnected slave traces share no degrees of freedom. Open components share adjacent endpoint variables, whereas a closed component merges its coincident first and last endpoint once.

The scalar interface trace spaces are
\begin{equation}
    V_{\mathrm T}^\Gamma
    =
    \operatorname{span}\{N_{\mathrm T,I}^{\Gamma}\},
    \qquad
    V_{\mathrm Q}^\Gamma
    =
    \operatorname{span}\{N_{\mathrm Q,J}^{\Gamma}\}.
    \label{eq:interface_trace_spaces}
\end{equation}
The vector-valued displacement spaces are $V_{\mathrm T}^\Gamma\otimes\mathbb{R}^3$ and $V_{\mathrm Q}^\Gamma\otimes\mathbb{R}^3$. The retained THB trace is the master, and the quadtree interface variables form the slave trace. Both basis rows are evaluated at the same physical points on each common segment. Their componentwise integrals define the global $L^2$ projection through a mortar-type weak orthogonality relation:
\begin{align}
    \mat{M}_{\mathrm{QQ}}
    &=
    \sum_{g\in\mathcal{P}_{\Gamma}}
    \int_g
    \left(\bs{N}_{\mathrm Q}^{g}\right)^{\mathsf T}
    \bs{N}_{\mathrm Q}^{g}\,\dd\Gamma,
    \nonumber\\
    \mat{M}_{\mathrm{QT}}
    &=
    \sum_{g\in\mathcal{P}_{\Gamma}}
    \int_g
    \left(\bs{N}_{\mathrm Q}^{g}\right)^{\mathsf T}
    \bs{N}_{\mathrm T}^{g}\,\dd\Gamma,
    \qquad
    \mat{M}_{\mathrm{QQ}}\boldsymbol{\Pi}_{s}
    =
    \mat{M}_{\mathrm{QT}}.
    \label{eq:interface_l2_projection}
\end{align}
Here $\boldsymbol{\Pi}_{s}$ maps scalar THB coefficients to scalar quadtree coefficients. After corner and periodic endpoint variables have been merged, suppose that the slave basis is linearly independent, the quadrature weights and physical Jacobians are positive, and each common segment contains at least $p_{\mathrm Q}+1$ distinct Gauss points, where $p_{\mathrm Q}$ is the slave trace degree. On a polynomial span, a degree-$p_{\mathrm Q}$ trace that vanishes at these points vanishes identically; on a rational B\'ezier span, the same conclusion follows for its numerator when the weight function is positive. Consequently, $\bs{c}^{\mathsf T}\mat{M}_{\mathrm{QQ}}\bs{c}=\sum_g\sum_q w_qJ_q(\bs{N}_{\mathrm Q}^{g}\bs{c})^2>0$ for every nonzero real $\bs{c}$. The discrete Gram matrix is symmetric positive definite, and the projection is uniquely determined. Its entries are real and depend on the geometry, trace spaces, common partition and quadrature, but not on $k$ or $\omega$. The projection can therefore be assembled once and reused across spectral samples while these interface data remain fixed.

If a master trace $\bs{N}_{\mathrm T}^{\Gamma}\bs{d}_{\mathrm T}^{\Gamma}$ is also representable as $\bs{N}_{\mathrm Q}^{\Gamma}\bs{c}$, evaluating both matrices in \eqref{eq:interface_l2_projection} with the same points and weights gives $\mat{M}_{\mathrm{QT}}\bs{d}_{\mathrm T}^{\Gamma}=\mat{M}_{\mathrm{QQ}}\bs{c}$. Thus $\boldsymbol{\Pi}_{s}\bs{d}_{\mathrm T}^{\Gamma}=\bs{c}$ even when the integrals have quadrature error. The THB, Bernstein and rational Bernstein bases form partitions of unity, and the hanging-edge restriction rows sum to one; constant traces and uniform translations are therefore reproduced exactly. For other master traces, the map is the best approximation in the slave space in the quadrature-induced $L^2$ inner product. The corresponding three-component projection is
\begin{equation}
    \boldsymbol{\Pi}
    =
    \boldsymbol{\Pi}_{s}\otimes\mat{I}_3,
    \qquad
    \tilde{\bs{d}}_{\Gamma}^{\mathrm Q}
    =
    \boldsymbol{\Pi}\tilde{\bs{d}}_{\Gamma}^{\mathrm T}.
    \label{eq:vector_interface_projection}
\end{equation}
The projection relates coefficient vectors; the resulting traces in distinct approximation spaces need not be equal pointwise. Matching traces reduce \eqref{eq:interface_l2_projection} to the identity or an orientation permutation. Reversed element sides are handled by the edge extraction of \Cref{subsec:quadtree_trace}, and segmentwise quadrature prevents orientation or material changes from crossing an integration interval. The assembled $\boldsymbol{\Pi}$ applies without additional operators to open, closed, and disconnected interfaces.

\begin{figure}[!htbp]
\centering
\begin{subfigure}[t]{0.35\linewidth}
    \centering
    \includegraphics[width=\linewidth]{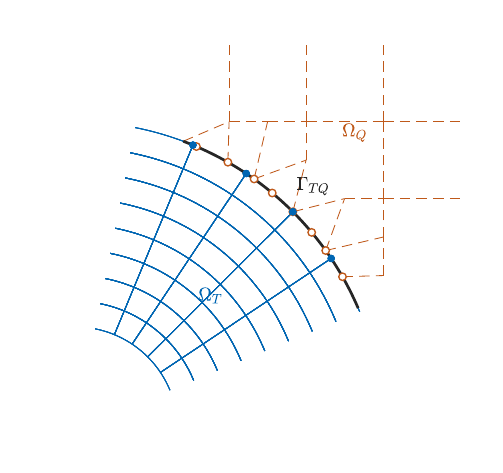}
    \caption{}
    \label{fig:fig07a_coupled_discretization}
\end{subfigure}
\begin{subfigure}[t]{0.35\linewidth}
    \centering
    \includegraphics[width=\linewidth]{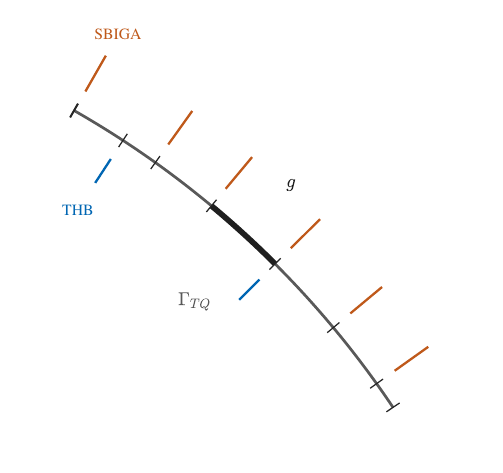}
    \caption{}
    \label{fig:fig07b_common_partition}
\end{subfigure}
\caption{(a) Coupled THB-GIFT/SBIGA discretization on the physical interface. (b) Local nonmatching trace partitions and their common integration partition: blue, THB master trace partition; orange, SBIGA slave trace partition; black, common integration partition; $g$, representative common segment. Projection integrals are evaluated segmentwise on this common partition.}
\label{fig:interface_common_partition}
\end{figure}

\subsection{Work-Consistent Coupling and Reduced System}
\label{subsec:hybrid_global_system}

\begin{figure}[!htbp]
\centering
\includegraphics[width=\linewidth]{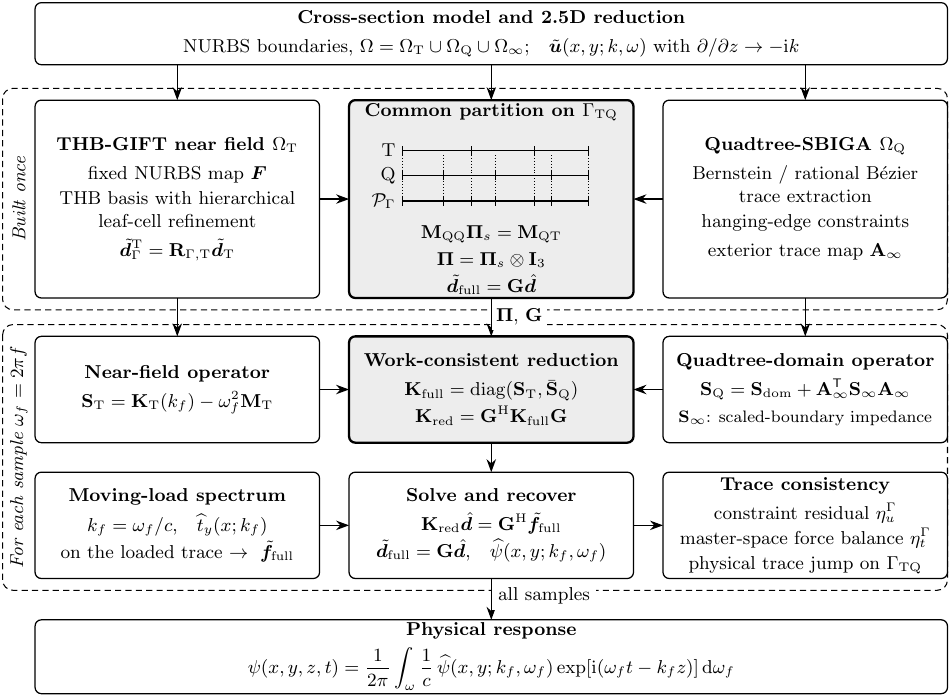}
\caption{Main line of the projection-coupled THB-GIFT/quadtree-SBIGA formulation. The THB-GIFT near field and the quadtree-SBIGA domain are discretized independently, and their traces meet only on the common partition of $\Gamma_{\mathrm{TQ}}$, where the projection $\boldsymbol{\Pi}$ and the transformation $\mat{G}$ are formed once. For each frequency sample on the moving-load line, the regional operators are reduced by the Hermitian congruence $\mat{G}^{\mathrm H}(\cdot)\mat{G}$; the recovered spectra are then transformed back to the physical response.}
\label{fig:coupled_reduction_workflow}
\end{figure}

Let $\mat{R}_{\Gamma,\mathrm T}$ restrict the full THB vector to its assembled master-interface degrees of freedom. The quadtree vector is ordered as $\tilde{\bs{d}}_{\mathrm Q}=[(\tilde{\bs{d}}_\Gamma^{\mathrm Q})^{\mathsf T},(\tilde{\bs{d}}_F^{\mathrm Q})^{\mathsf T}]^{\mathsf T}$, where $\tilde{\bs{d}}_F^{\mathrm Q}$ contains all remaining quadtree degrees of freedom, including the outer trace on which $\mat{S}_{\mathrm Q}$ acts. The full vectors before interface elimination are
\begin{equation}
    \tilde{\bs{d}}_{\mathrm{full}}
    =
    \begin{bmatrix}
    \tilde{\bs{d}}_{\mathrm T}\\
    \tilde{\bs{d}}_\Gamma^{\mathrm Q}\\
    \tilde{\bs{d}}_F^{\mathrm Q}
    \end{bmatrix},
    \qquad
    \tilde{\bs{f}}_{\mathrm{full}}
    =
    \begin{bmatrix}
    \tilde{\bs{f}}_{\mathrm T}\\
    \tilde{\bs{f}}_\Gamma^{\mathrm Q}\\
    \tilde{\bs{f}}_F^{\mathrm Q}
    \end{bmatrix}.
    \label{eq:hybrid_full_vectors}
\end{equation}
External loading contributes to $\tilde{\bs{f}}_{\mathrm T}$ or $\tilde{\bs{f}}_F^{\mathrm Q}$ according to its physical support. Interface entries represent internal forces, while the exterior impedance is already contained in $\mat{S}_{\mathrm Q}$.

With the ordering in \eqref{eq:hybrid_full_vectors}, the constraint \eqref{eq:vector_interface_projection} gives
\begin{equation}
    \tilde{\bs{d}}_{\mathrm{full}}
    =
    \begin{bmatrix}
    \tilde{\bs{d}}_{\mathrm T}\\
    \tilde{\bs{d}}_\Gamma^{\mathrm Q}\\
    \tilde{\bs{d}}_F^{\mathrm Q}
    \end{bmatrix}
    =
    \underbrace{
    \begin{bmatrix}
    \mat{I} & \mat{0}\\
    \boldsymbol{\Pi}\mat{R}_{\Gamma,\mathrm T} & \mat{0}\\
    \mat{0} & \mat{I}
    \end{bmatrix}}_{\mat{G}}
    \begin{bmatrix}
    \tilde{\bs{d}}_{\mathrm T}\\
    \tilde{\bs{d}}_F^{\mathrm Q}
    \end{bmatrix}.
    \label{eq:hybrid_transformation}
\end{equation}
The unreduced block stiffness is
\begin{equation}
    \mat{K}_{\mathrm{full}}
    =
    \begin{bmatrix}
    \mat{S}_{\mathrm T} & \mat{0}\\
    \mat{0} & \bar{\mat{S}}_{\mathrm Q}
    \end{bmatrix},
    \label{eq:hybrid_full_stiffness}
\end{equation}
where $\bar{\mat{S}}_{\mathrm Q}$ denotes \eqref{eq:quadtree_total_stiffness} after the same quadtree permutation used in \eqref{eq:hybrid_transformation}, with all disconnected interface variables placed before the remaining quadtree variables. The full displacement is recovered as $\tilde{\bs{d}}_{\mathrm{full}}=\mat{G}\hat{\bs{d}}$.

Conservative load transfer on nonmatching curved interfaces and virtual-work-based nonconforming finite-element/boundary-element coupling provide related precedents for pairing kinematic and force transfer \citep{Jaiman2006ConservativeTransfer,Yang2021NonconformingIGAFEBE}. In the present frequency-domain setting, the work-consistent force map follows from the sesquilinear virtual variation of the slave trace:
\begin{equation}
    \delta\tilde{\bs{d}}_{\Gamma}^{\mathrm Q}
    =
    \boldsymbol{\Pi}\delta\tilde{\bs{d}}_{\Gamma}^{\mathrm T},
    \qquad
    \left(\delta\tilde{\bs{d}}_{\Gamma}^{\mathrm Q}\right)^{\mathrm{H}}
    \bs{f}_{\Gamma}^{\mathrm Q}
    =
    \left(\delta\tilde{\bs{d}}_{\Gamma}^{\mathrm T}\right)^{\mathrm{H}}
    \boldsymbol{\Pi}^{\mathrm{H}}\bs{f}_{\Gamma}^{\mathrm Q}.
    \label{eq:interface_virtual_work}
\end{equation}
In the sesquilinear weak form \eqref{eq:weak_25d}, the quadtree interface forces contribute to the retained THB equations through the adjoint map $\boldsymbol{\Pi}^{\mathrm{H}}$. For disconnected components, the assembled projection transfers forces componentwise without cross-component slave terms. The final coupled frequency-domain system is
\begin{equation}
    \mat{K}_{\mathrm{red}}\hat{\bs{d}}
    =
    \hat{\bs{f}},
    \qquad
    \mat{K}_{\mathrm{red}}
    =
    \mat{G}^{\mathrm{H}}
    \mat{K}_{\mathrm{full}}
    \mat{G},
    \qquad
    \hat{\bs{f}}
    =
    \mat{G}^{\mathrm{H}}
    \tilde{\bs{f}}_{\mathrm{full}} .
    \label{eq:global_system}
\end{equation}
The reduced unknown $\hat{\bs{d}}=[\tilde{\bs{d}}_{\mathrm T}^{\mathsf T},(\tilde{\bs{d}}_F^{\mathrm Q})^{\mathsf T}]^{\mathsf T}$ contains no interface multipliers, penalty or stabilization parameters. For the unreduced residual $\bs{r}=\mat{K}_{\mathrm{full}}\mat{G}\hat{\bs{d}}-\tilde{\bs{f}}_{\mathrm{full}}$, the reduced residual is $\mat{G}^{\mathrm H}\bs{r}$. Equation~\eqref{eq:global_system} therefore enforces virtual work over all variations satisfying the interface constraint. Its block equations also give equilibrium in the retained master space, $\bs{f}_{\Gamma}^{\mathrm T}+\boldsymbol{\Pi}^{\mathrm H}\bs{f}_{\Gamma}^{\mathrm Q}=\bs{0}$, where $\bs{f}_{\Gamma}^{\mathrm T}=\mat{R}_{\Gamma,\mathrm T}\bs{r}_{\mathrm T}$ and $\bs{f}_{\Gamma}^{\mathrm Q}=\bs{r}_{\Gamma}^{\mathrm Q}$ are the interface residual forces of the two regions.

The projection is constructed from real basis functions and quadrature data, so $\boldsymbol{\Pi}^{\mathrm H}=\boldsymbol{\Pi}^{\mathsf T}$ and $\mat{G}^{\mathrm H}=\mat{G}^{\mathsf T}$ in this discretization; the identity blocks in \eqref{eq:hybrid_transformation} give $\mat{G}$ full column rank. The congruence in \eqref{eq:global_system} preserves Hermitian or complex-symmetric structure whenever the unreduced operator has the corresponding structure. It also maps the Hermitian and symmetric parts of the unreduced operator to the respective parts of the reduced operator. In contrast, using a different force map $\mat{T}^{\mathrm H}$ with the kinematic map $\boldsymbol{\Pi}$ gives $\mat{G}_{T}^{\mathrm H}\mat{K}_{\mathrm{full}}\mat{G}$ with $\mat{G}_{T}\ne\mat{G}$. Such an unpaired reduction does not in general express the virtual work of the admissible variations or preserve those structures. A collocation map paired with its own adjoint remains work-consistent; the effect of an unpaired transfer is examined in \Cref{tab:cavity_transfer_ablation}.

After solving \eqref{eq:global_system}, the eliminated slave trace is recovered from \eqref{eq:vector_interface_projection}. The same construction applies to disconnected interface components. \Cref{fig:coupled_reduction_workflow} summarizes the regional operators, interface reduction and response recovery.

Three complementary measures characterize trace consistency. The algebraic constraint residual is
\begin{equation}
    \eta_u^\Gamma
    =
    \frac{
    \left\|
    \tilde{\bs{d}}_\Gamma^{\mathrm Q}
    -
    \boldsymbol{\Pi}\mat{R}_{\Gamma,\mathrm T}\tilde{\bs{d}}_{\mathrm T}
    \right\|_2}
    {\max\left(\left\|\tilde{\bs{d}}_\Gamma^{\mathrm Q}\right\|_2,\epsilon_{\mathrm{mach}}\right)},
    \label{eq:interface_displacement_residual}
\end{equation}
where $\epsilon_{\mathrm{mach}}>0$ prevents a vanishing denominator. This coefficient-space residual reaches numerical precision after elimination and recovery. By contrast, evaluating $\mat{N}_{\mathrm Q}\tilde{\bs{d}}_\Gamma^{\mathrm Q}-\mat{N}_{\mathrm T}\tilde{\bs{d}}_\Gamma^{\mathrm T}$ at common physical points gives a trace-jump measure that need not vanish for nonmatching spaces; \eqref{eq:cavity_interface_jump} uses that physical measure. Interface equilibrium in the retained master space is quantified by
\begin{equation}
    \eta_t^\Gamma
    =
    \frac{
    \left\|
    \bs{f}_\Gamma^{\mathrm T}
    +
    \boldsymbol{\Pi}^{\mathrm H}\bs{f}_\Gamma^{\mathrm Q}
    \right\|_2}
    {\max\left(
    \left\|\bs{f}_\Gamma^{\mathrm T}\right\|_2,
    \left\|\boldsymbol{\Pi}^{\mathrm H}\bs{f}_\Gamma^{\mathrm Q}\right\|_2,
    \epsilon_{\mathrm{mach}}
    \right)} .
    \label{eq:interface_force_balance}
\end{equation}
The two internal-force vectors are extracted from the unreduced subdomain equations and carry opposite outward-normal signs, so their projected sum vanishes for an equilibrated coupled solution.

\subsection{Moving-Load Excitation and Response Reconstruction}
\label{subsec:moving_load_spectrum}

The moving load enters the right-hand side of \Cref{eq:global_system}. A wheel-based representation describes a single wheel, dual wheel, or multi-axle vehicle through the wheel positions and forces. For wheel $m$, the vertical contact pressure is written as
\begin{equation}
    p_m(x,z,t)
    =
    -Q_m\,g_x(x-x_m)\,g_z(z-z_m-ct),
    \label{eq:moving_pressure}
\end{equation}
where $Q_m$ is the wheel force, $c$ is the vehicle speed, and the minus sign follows the vertical force convention used in the numerical examples. For a Gaussian contact patch,
\begin{equation}
    g_\alpha(s)
    =
    \frac{1}{\sqrt{2\pi}\sigma_\alpha}
    \exp\left(-\frac{s^2}{2\sigma_\alpha^2}\right),
    \qquad \alpha=x,z ,
    \label{eq:gaussian_patch}
\end{equation}
the longitudinal transform is
\begin{equation}
    \widehat{g}_z(k)
    =
    \exp\left(-\frac{1}{2}\sigma_z^2 k^2\right).
    \label{eq:gaussian_k}
\end{equation}
For a uniform rectangular footprint of half-length $a_z$ in the travelling direction, the corresponding factor is $\operatorname{sinc}(ka_z)$, with $\operatorname{sinc}(s)=\sin(s)/s$ and the limiting value one at $s=0$.

For a pure moving load, the temporal frequency and the longitudinal wavenumber are linked by the moving-load dispersion line
\begin{equation}
    \omega_f=2\pi f,\qquad k_f=\frac{\omega_f}{c}.
    \label{eq:pure_moving_relation}
\end{equation}
The wavenumber-line traction coefficient is
\begin{equation}
    \widehat{t}_y(x;k_f)
    =
    -\sum_{m=1}^{n_w}
    Q_m\,
    g_x(x-x_m)
    \widehat{g}_z(k_f)
    \exp(+\ii k_f z_m).
    \label{eq:moving_load_wavenumber_coeff}
\end{equation}
Applying the forward transform in \eqref{eq:fourier_u} to \eqref{eq:moving_pressure} gives
\begin{equation}
    \tilde{p}_m(x;k,\omega)
    =
    -2\pi Q_m g_x(x-x_m)\widehat{g}_z(k)
    \exp(+\ii k z_m)\,\delta(\omega-kc).
    \label{eq:moving_pressure_transform}
\end{equation}
Thus a positive initial longitudinal offset $z_m$ in \eqref{eq:moving_pressure} produces the positive phase in \eqref{eq:moving_load_wavenumber_coeff}. Integrating the spectral line with respect to $k$ gives $\delta(\omega-kc)=c^{-1}\delta(k-\omega/c)$ for $c>0$. The temporal spectrum on the dispersion line therefore contains the travelling-load Jacobian,
\begin{equation}
    \tilde{t}_y(x;\omega_f)
    =
    \frac{1}{c}\,
    \widehat{t}_y(x;k_f).
    \label{eq:moving_load_spectrum}
\end{equation}
For a harmonic wheel-force modulation with circular frequency $\omega_0$, the same construction is used with $k=(\omega-\omega_0)/c$ and the phase $\exp(+\ii k z_m)$ associated with the definition in \eqref{eq:moving_pressure}. The transformed load vector is assembled by integrating this traction against the THB or quadtree trace basis on the loaded surface and then reduced by the same transformation used in \eqref{eq:global_system}.

For general frequency--wavenumber excitation, any scalar response $\psi$ is recovered by the double inverse transform in \eqref{eq:fourier_u}. For a pure moving load, the delta distribution in \eqref{eq:moving_pressure_transform} reduces the wavenumber integral to the line \eqref{eq:pure_moving_relation}, contributing the single factor $1/c$ in \eqref{eq:moving_load_spectrum}. If $\widehat{\psi}(x,y;k_f,\omega_f)$ denotes the monitored response obtained from the wavenumber-line coefficient \eqref{eq:moving_load_wavenumber_coeff}, the time-domain reconstruction is the one-parameter inverse transform
\begin{equation}
    \psi(x,y,z,t)
    =
    \frac{1}{2\pi}
    \int_{\omega}
    \frac{1}{c}\,
    \widehat{\psi}(x,y;k_f,\omega_f)
    \exp\left[\ii(\omega_f t-k_f z)\right]
    \dd\omega_f .
    \label{eq:moving_load_line_reconstruction}
\end{equation}
This reconstruction retains the phase convention of \Cref{eq:fourier_u} and is used in the numerical verification below.

\section{Numerical Verification and Representative Applications}
\label{sec:numerical}

Homogeneous half-space solutions verify the bounded and exterior SBIGA operators and the moving-load reconstruction. A cylindrical cavity with a closed-form solution then tests the complete nonmatching coupling. The applications combine open and closed interfaces in a layered pavement with a buried conduit and couple two disconnected near-field regions through the surrounding domain for twin modified-D tunnels.

\subsection{Homogeneous Half-Space Verification of the Quadtree-SBIGA and Moving-Load Components}
\label{subsec:num_verification}

The comparisons examine interior spatial resolution, load assembly, and broadband reconstruction.

\paragraph{Analytical reference}
The reference configuration is a homogeneous isotropic half-space with a flat traction-free surface, density $\rho=2000~\mathrm{kg/m^3}$, shear-wave speed $c_s=100~\mathrm{m/s}$, and Poisson ratio $\nu=0.25$ ($c_p=173.2~\mathrm{m/s}$). Material damping is represented by the complex moduli of \Cref{eq:complex_moduli}. The transverse coordinate is $x$, the travelling direction is $z$, and the vertical coordinate is $y$. The road surface is located at $y=0$ and the half-space occupies $y\leq 0$. A downward surface traction is therefore negative in $y$, and the plots report the downward-positive response $-u_y$. In the analytical half-space formula it is convenient to use the positive depth variable $d=-y\geq 0$. The complex Lam\'e constants and wave speeds are
\begin{equation}
    \mu=\frac{E^\ast}{2(1+\nu)},\qquad
    \lambda=\frac{E^\ast\nu}{(1+\nu)(1-2\nu)},\qquad
    c_s=\sqrt{\frac{\mu}{\rho}},\qquad
    c_p=\sqrt{\frac{\lambda+2\mu}{\rho}} .
    \label{eq:halfspace_material}
\end{equation}

The benchmark follows the homogeneous half-space configuration adopted by \citet{WangLinLi2022}, with the semi-analytical viscoelastic moving-load response of \citet{HungYang2001} as reference. Both calculations use $\zeta=0.05$ and the longitudinal and temporal phase convention of \Cref{eq:fourier_u}. For the $j$th harmonic component, $k_j=(\omega-\omega_j)/c$, reducing to $k_j=\omega/c$ for the non-self-oscillating loads considered here. The dilatational and shear square roots are selected with non-negative real parts to enforce decay or outward radiation in the damped half-space. The point-load FRFs of \citet{HungYang2001} are integrated numerically with breakpoints at the branch points and the Rayleigh pole.

\paragraph{Single-element line-load response}
A unit vertical line load at $x_0=0$, moving at $c=90~\mathrm{m/s}$, removes the additional effects of a finite footprint and multiple wheels. The response is compared with the analytical point-load FRF on the spectral line $k=\omega/c$; in transverse wavenumber space, the load is
\begin{equation}
    \tilde{q}^{\mathrm{line}}_y(k_x)
    =
    -Q_0\exp(-\ii k_xx_0),
    \qquad x_0=0 .
    \label{eq:verification_line_load}
\end{equation}
The response plotted below is the downward displacement at $y=-1~\mathrm{m}$ over $0\leq x\leq 4~\mathrm{m}$. The observation line lies below the loaded surface and tests the interior scaled-boundary solution.

\begin{figure}[!htbp]
    \centering
    \includegraphics[width=0.60\linewidth]{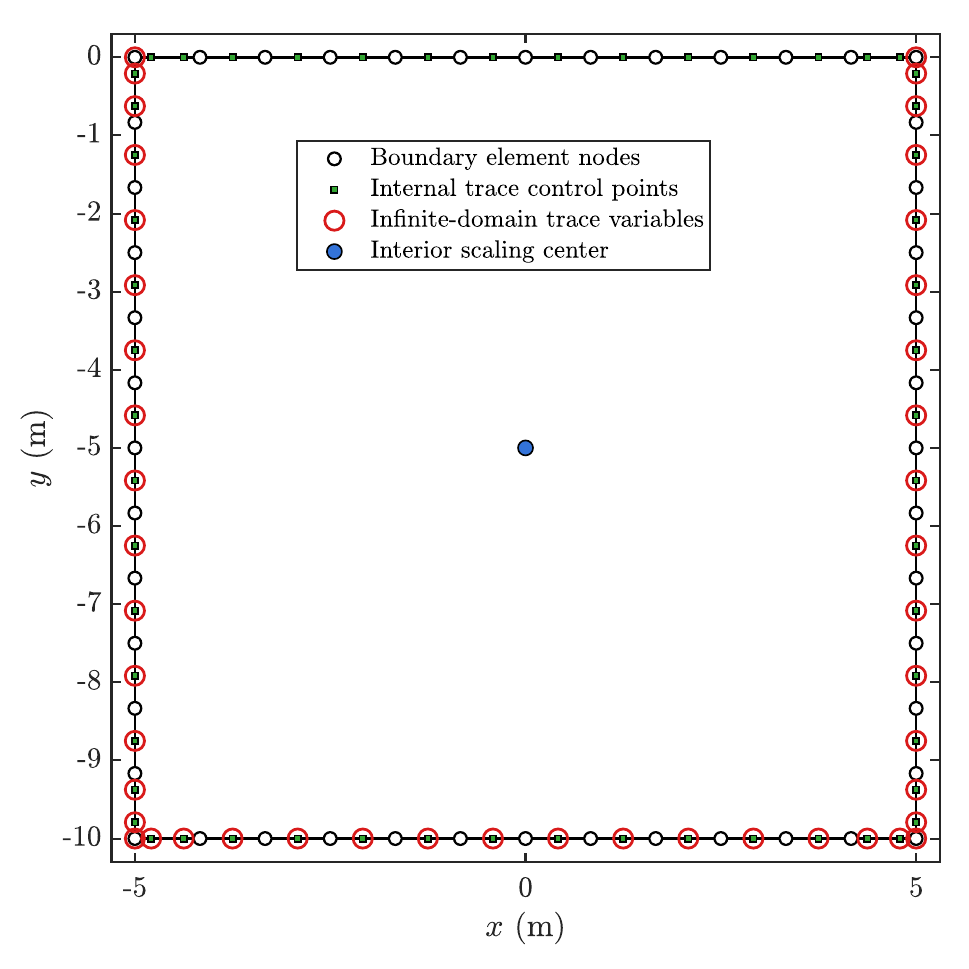}
    \caption{Single-S-element 2.5D SBIGA model: one scaled-boundary element with high-order traces on its four sides, shared $C^0$ corners, and the exterior scaled-boundary stiffness attached to the left, bottom, and right boundaries.}
    \label{fig:single_s_element_model}
\end{figure}

The rectangular S-element in \Cref{fig:single_s_element_model} uses maximally smooth open traces within each side and carries the load on its top side. This model tests the assembled 2.5D coefficient matrices, frequency-dependent inertia, sign and phase convention, and bounded/exterior coupling.

\begin{figure}[!htbp]
    \centering
    \begin{subfigure}[t]{0.47\textwidth}
        \centering
        \includegraphics[width=\linewidth]{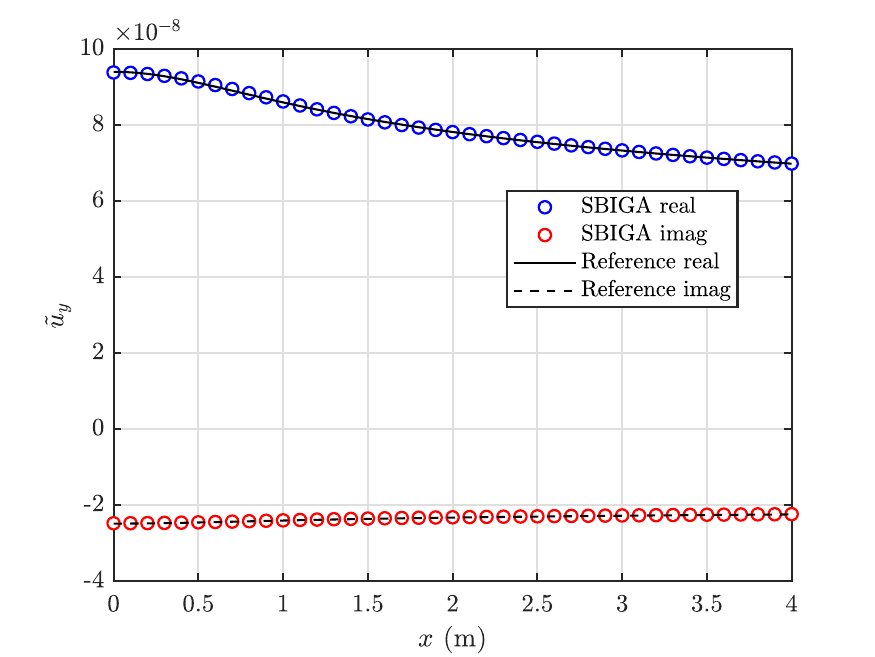}
        \caption{$0.05$ Hz}
        % \label{fig:single_s_response_005hz}
    \end{subfigure}
    \hfill
    \begin{subfigure}[t]{0.47\textwidth}
        \centering
        \includegraphics[width=\linewidth]{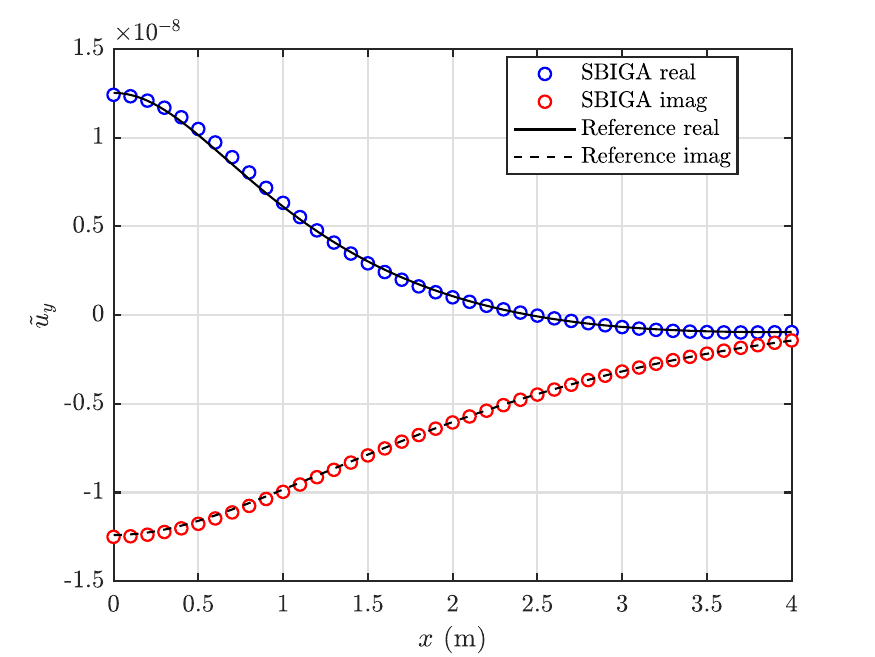}
        \caption{$32$ Hz}
        % \label{fig:single_s_response_32hz}
    \end{subfigure}

    \vspace{0.5em}
    \begin{subfigure}[t]{0.47\textwidth}
        \centering
        \includegraphics[width=\linewidth]{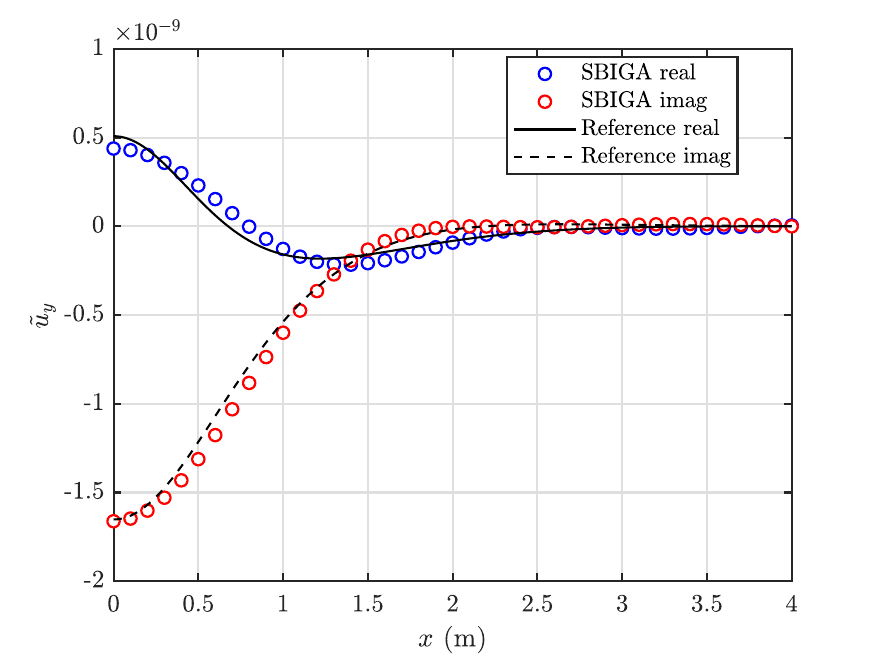}
        \caption{$100$ Hz}
        % \label{fig:single_s_response_100hz}
    \end{subfigure}
    \caption{Single-frequency line-load verification at $y=-1~\mathrm{m}$ along $0\leq x\leq 4~\mathrm{m}$. The single S-element reproduces the reference at low frequency; the deviation at higher frequencies reflects its limited interior resolution.}
    \label{fig:singleresponse}
\end{figure}

At $0.05$ Hz, the numerical and analytical profiles in \Cref{fig:singleresponse} agree closely along the observation line. At $32$ and $100$ Hz, deviations appear because a single scaling centre cannot resolve the shorter wavelengths in the interior. Raising the spline degree or the number of knot spans on the four sides does not remove the high-frequency deviation, so interior resolution must be supplied by partitioning the domain into several S-elements.

\paragraph{Internally partitioned response and wheel-load assembly}
The partitioned model resolves the interior through independent rectangular S-elements, refined towards the surface and the loading zone, whose sides carry maximally smooth open spline/NURBS traces with shared $C^0$ vertices. The same partitioned model is used to verify distributed wheel loads and multi-wheel phase assembly.

Wheel positions, footprints, phases, and the $1/c$ Jacobian follow \Cref{subsec:moving_load_spectrum}. Because $\int_{-\infty}^{\infty}g_{x,m}(s)\,\dd s=1$ and $\widehat g_{z,m}(0)=1$, $Q_m$ remains the wheel resultant. The semi-analytical reference convolves the point-load half-space FRF with the same footprint.

Spectral-point comparisons are reported in terms of the normalized wavenumber-line displacement
\begin{equation}
\tilde{U}_y(x,y;k_f,\omega_f)
=
\frac{2\pi \mu}{c}
\widehat{u}_y(x,y;k_f,\omega_f),
\label{eq:verification_normalized_spectral_disp}
\end{equation}
where $\mu$ is the shear modulus in \eqref{eq:halfspace_material}; time histories apply the Jacobian of \eqref{eq:moving_load_spectrum} before the inverse transform.

Two footprints with the same wheel resultant $Q=10~\mathrm{kN}$ are compared: the Gaussian patch of \Cref{eq:gaussian_patch,eq:gaussian_k} with $\sigma_x=\sigma_z=0.062~\mathrm{m}$, and a uniform rectangular patch of half-width $a_x=0.15~\mathrm{m}$ and half-length $a_z=0.20~\mathrm{m}$,
\begin{equation}
g_x^{\mathrm{R}}(s)
=
\begin{cases}
\dfrac{1}{2a_x},
& |s|\leq a_x,\\[4pt]
0,
& |s|>a_x,
\end{cases}
\qquad
\tilde{g}_z^{\mathrm{R}}(k)
=
\frac{\sin(k a_z)}{k a_z}.
\label{eq:verification_rectangular_footprint}
\end{equation}

\begin{figure}[!htbp]
    \centering
    \includegraphics[width=0.94\linewidth]{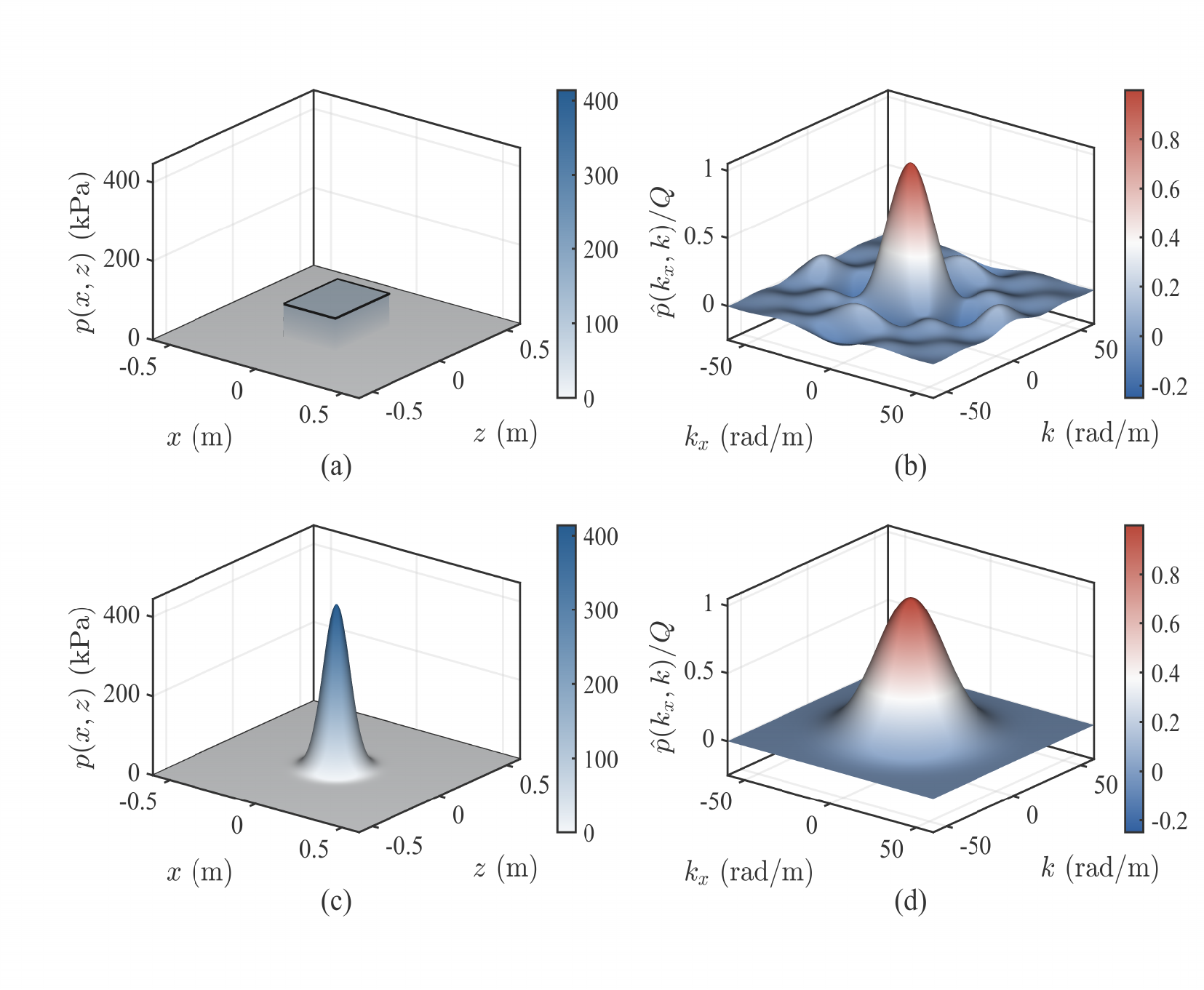}
    \caption{Gaussian and rectangular wheel-load footprints with the same resultant force $Q=10~\mathrm{kN}$; the spectra are normalized by $Q$. Panels (a) and (b) show the rectangular footprint in the spatial and wavenumber domains, and panels (c) and (d) show the corresponding Gaussian footprint.}
    \label{fig:gaussian_rectangular_loads}
\end{figure}

\Cref{fig:gaussian_rectangular_loads} shows both footprints in the spatial and wavenumber domains. The smooth Gaussian and the discontinuous rectangle, with its sinc-type spectrum, test distributed-traction assembly for loads of different regularity and spectral content.

\begin{figure}[!htbp]
\centering
\begin{subfigure}[t]{0.47\linewidth}
\centering
\includegraphics[width=\linewidth]{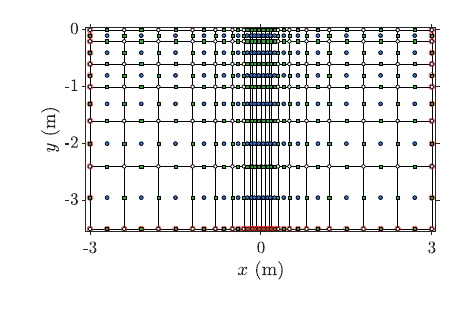}
\caption{Single-wheel partitioned S-element model: $Q=10~\mathrm{kN}$, $x=z=0$.}
\label{fig:single_wheel_partitioned_mesh}
\end{subfigure}
\hfill
\begin{subfigure}[t]{0.47\linewidth}
\centering
\includegraphics[width=\linewidth]{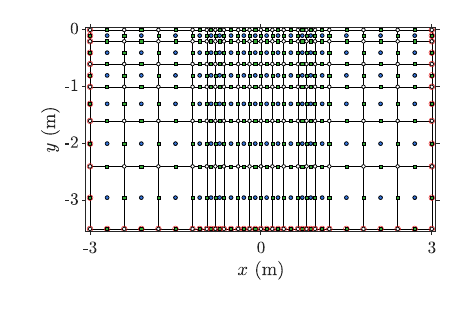}
\caption{Two-axle four-wheel partitioned S-element model: axle loads of $40$ and $60~\mathrm{kN}$ at $z=1.20$ and $-1.20~\mathrm{m}$, track width $1.60~\mathrm{m}$.}
\label{fig:multi_wheel_partitioned_mesh}
\end{subfigure}
\caption{Internally partitioned S-element models used for the homogeneous half-space validation. The white circles denote knot-span vertices on the spline/NURBS traces, the green squares denote additional high-order trace variables, the red outer rings identify trace variables used for the exterior half-space treatment, and the blue circles indicate the interior scaling centres.}
\label{fig:partitioned_s_element_models}
\end{figure}

The partitioned model is applied to a single wheel and to a two-axle, four-wheel vehicle travelling at $c=50~\mathrm{m/s}$ (\Cref{fig:partitioned_s_element_models}). The single-wheel model is refined around the loaded region, and the four-wheel model over a wider zone covering both wheel tracks. The plotted cells are S-elements; each side additionally carries knot-span subdivisions and high-order trace variables.

\begin{figure}[!htbp]
\centering
\begin{subfigure}[t]{0.95\linewidth}
    \centering
    \includegraphics[width=\linewidth]{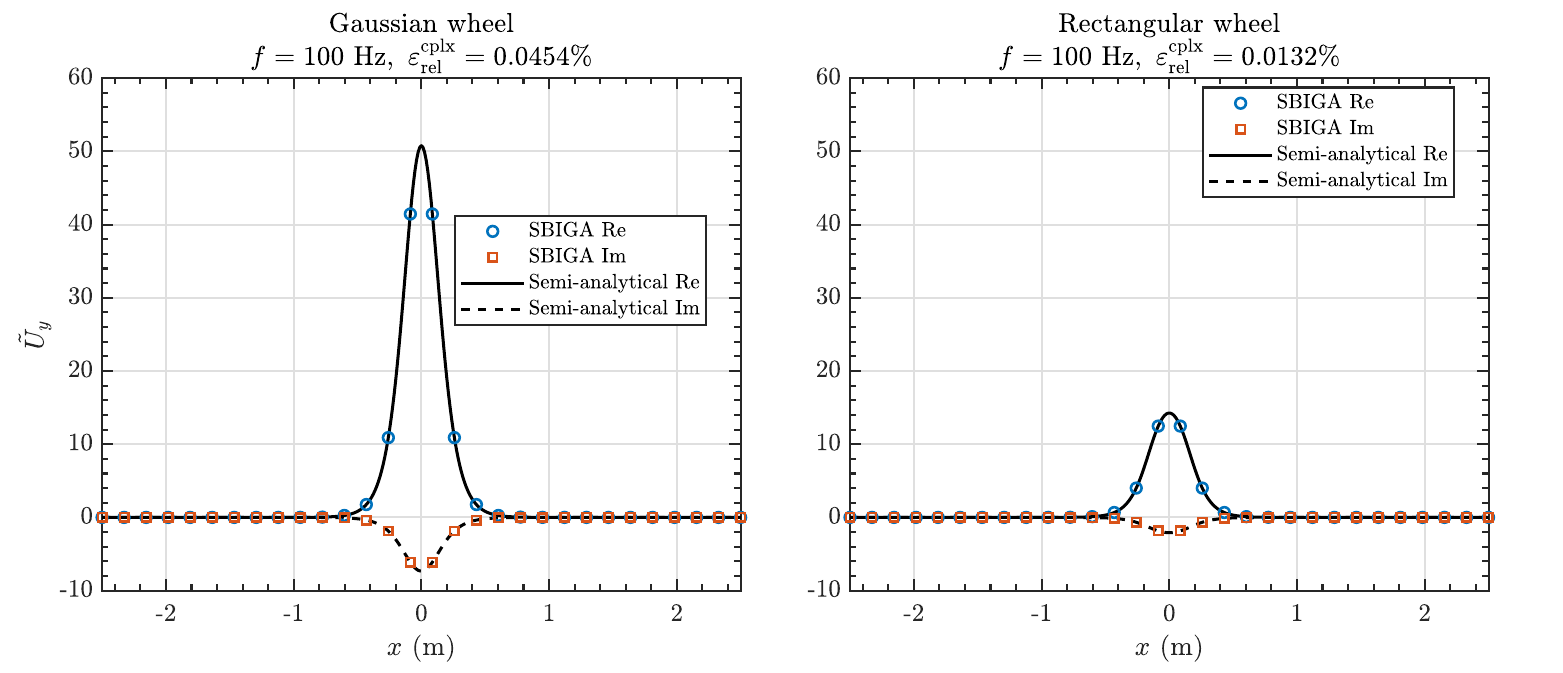}
    \caption{Accuracy comparison at $100~\mathrm{Hz}$.}
    \label{fig:selems-accuracy-100hz}
\end{subfigure}
\vspace{0.8em}
\begin{subfigure}[t]{0.95\linewidth}
    \centering
    \includegraphics[width=\linewidth]{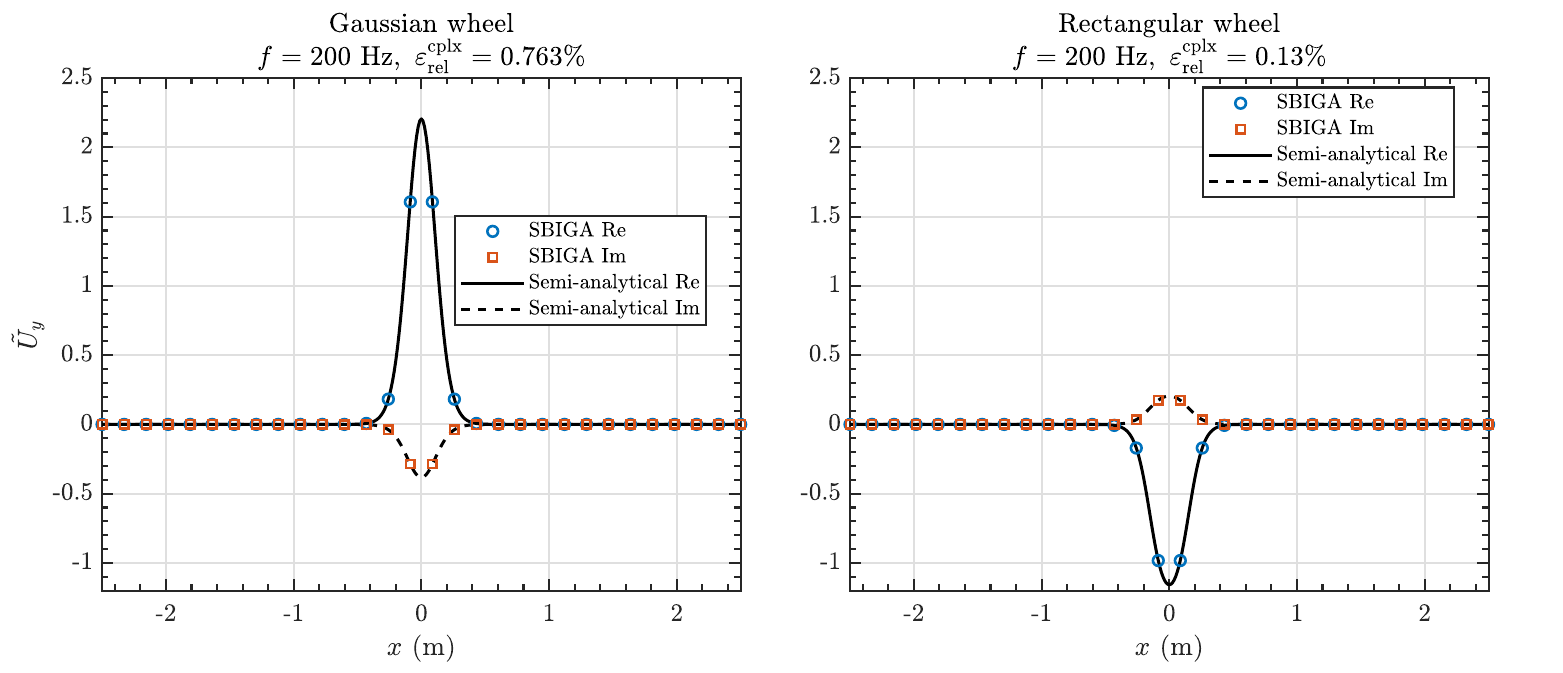}
    \caption{Accuracy comparison at $200~\mathrm{Hz}$.}
    \label{fig:selems-accuracy-200hz}
\end{subfigure}

\caption{Single-wheel response of the internally partitioned SBIGA model for Gaussian and rectangular footprints at two spectral points, compared with the footprint-convolved semi-analytical half-space reference.}
\label{fig:selems-accuracy}

\end{figure}

At representative spectral points of $100$ and $200~\mathrm{Hz}$, the partitioned model agrees with the semi-analytical half-space response in \Cref{fig:selems-accuracy} without visible discontinuities across the internal interfaces.

\begin{figure}[!htbp]
\centering
\begin{subfigure}[t]{0.95\linewidth}
    \centering
    \includegraphics[width=\linewidth]{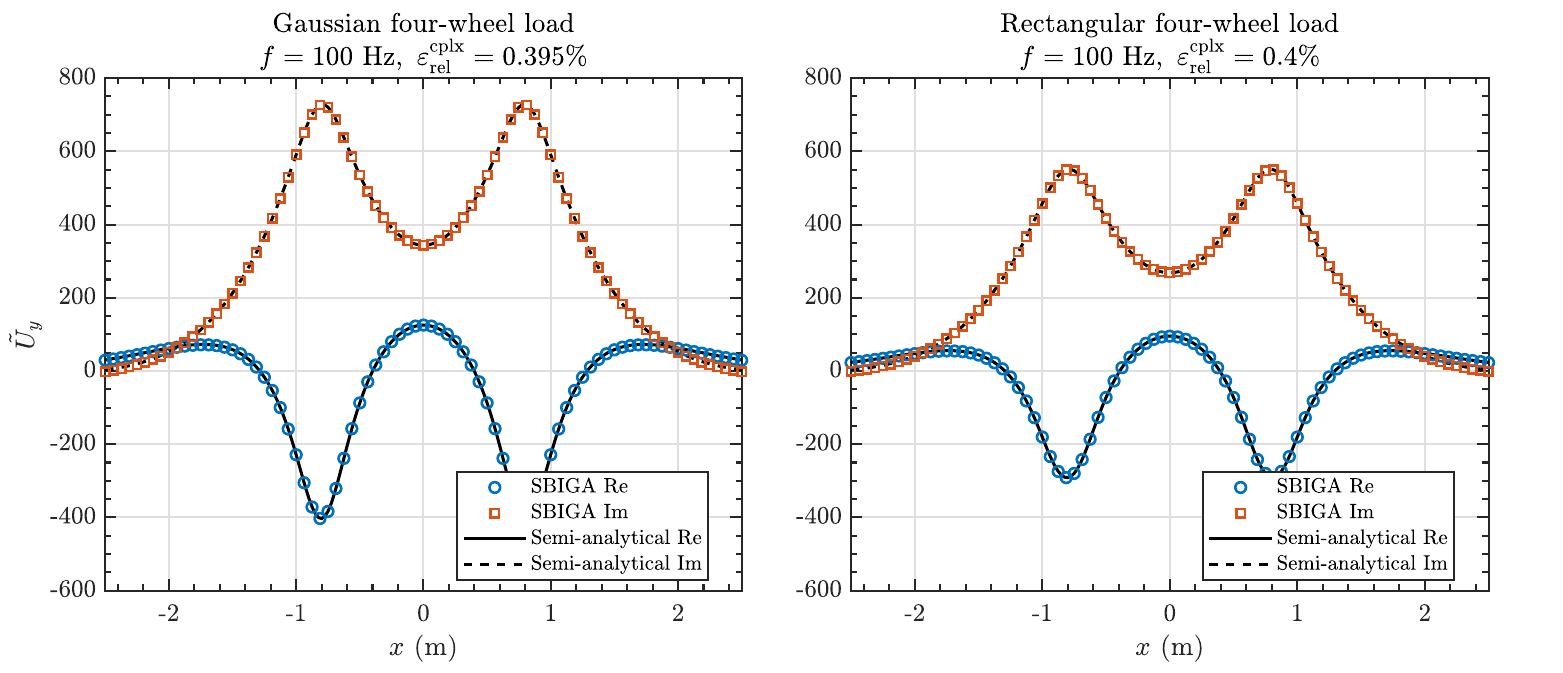}
    \caption{$f=100~\mathrm{Hz}$.}
    \label{fig:selems-wheel-accuracy-100hz}
\end{subfigure}
\vspace{0.8em}
\begin{subfigure}[t]{0.95\linewidth}
    \centering
    \includegraphics[width=\linewidth]{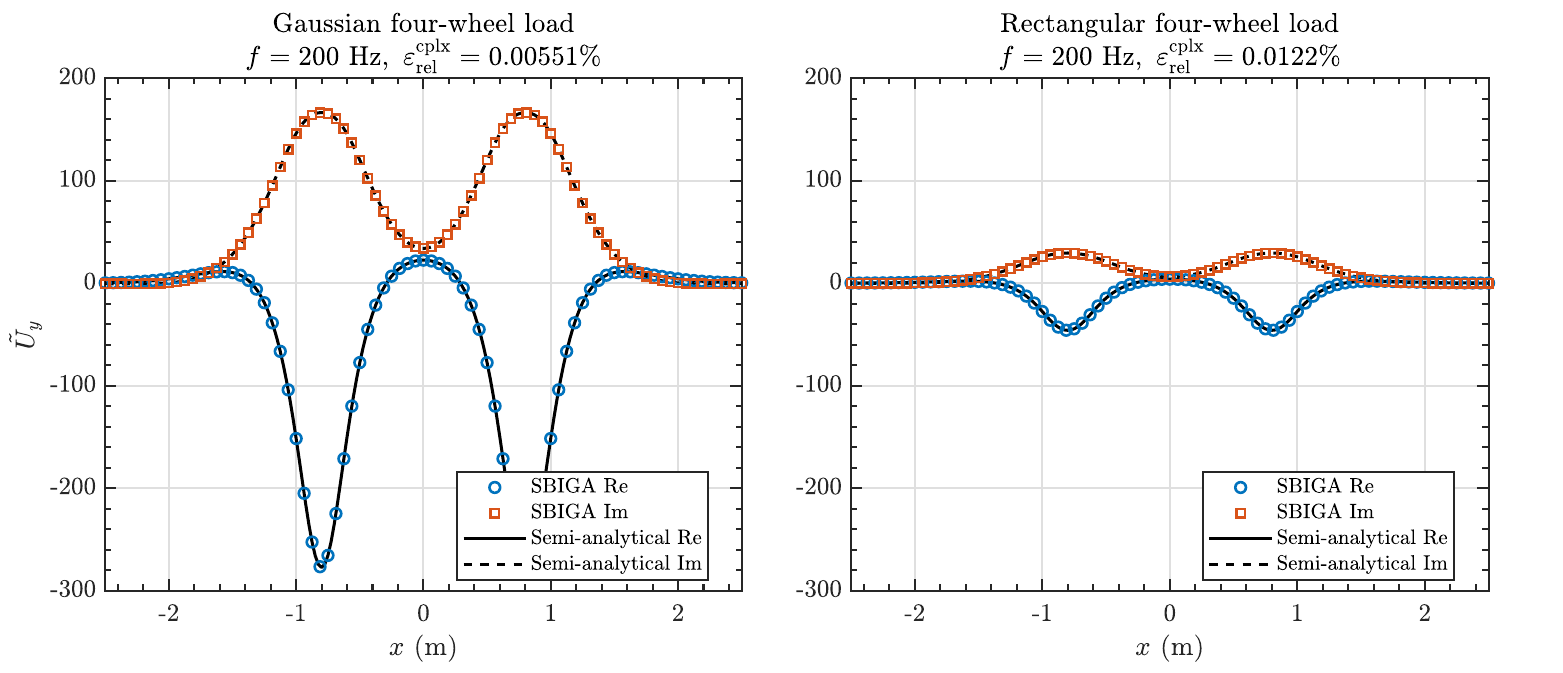}
    \caption{$f=200~\mathrm{Hz}$.}
    \label{fig:selems-wheel-accuracy-200hz}
\end{subfigure}
\caption{Four-wheel vehicle response of the internally partitioned SBIGA model for Gaussian and rectangular footprints, compared with the footprint-convolved semi-analytical half-space reference.}
\label{fig:selems-wheel-accuracy}
\end{figure}

For the four-wheel vehicle, the semi-analytical reference sums four phase-shifted, footprint-convolved point-load responses. The agreement at $100$ and $200~\mathrm{Hz}$ in \Cref{fig:selems-wheel-accuracy} for both footprints verifies the integration of distributed tractions over the high-order top trace together with the wheel-position phases.

The relative complex-valued response error is defined as
\begin{equation}
\varepsilon_{\mathrm{rel}}^{\mathrm{cplx}}
=
\frac{
\left|
\tilde{\bs{u}}_y^{\mathrm{SBIGA}}
-
\tilde{\bs{u}}_y^{\mathrm{ref}}
\right|_2
}{
\left|
\tilde{\bs{u}}_y^{\mathrm{ref}}
\right|_2
},
\label{eq:verification_complex_error}
\end{equation}
where $\tilde{\bs{u}}_y^{\mathrm{SBIGA}}$ and $\tilde{\bs{u}}_y^{\mathrm{ref}}$ collect the complex vertical responses at the observation points.

For the four-wheel vehicle, this error is about $0.4\%$ for both footprints at $100~\mathrm{Hz}$ and does not exceed $1.2\times10^{-4}$ at $200~\mathrm{Hz}$; for the single wheel it remains below $0.8\%$ at both frequencies. In all cases the assembled traction reproduces the prescribed resultant to round-off, and direct four-wheel assembly coincides with the superposition of single-wheel solutions to within $4\times10^{-15}$.

\paragraph{Adaptive quadtree and broadband response}
The regular partition is replaced by a locally refined quadtree, the surrounding-domain discretization used in all subsequent examples. The load is a single $80~\mathrm{kN}$ axle on two Gaussian wheels ($\sigma_x=\sigma_z=0.062~\mathrm{m}$, track width $1.60~\mathrm{m}$) travelling at $c=30~\mathrm{m/s}$. Refinement targets the wheel footprints, shallow response region, and observation lines. Because every leaf is an S-element, this model simultaneously exercises high-order extraction, hanging-interface constraints, global trace assembly, bounded 2.5D dynamic stiffness, and exterior coupling.

\begin{figure}[!htbp]
    \centering
    \begin{subfigure}[t]{0.32\textwidth}
        \centering
        \includegraphics[width=\linewidth]{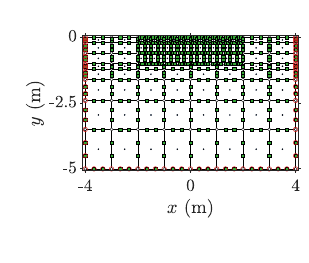}
        \caption{$\ell_{\max}=5$}
        \label{fig:quadtree_mesh_level_5}
    \end{subfigure}
    \hfill
    \begin{subfigure}[t]{0.32\textwidth}
        \centering
        \includegraphics[width=\linewidth]{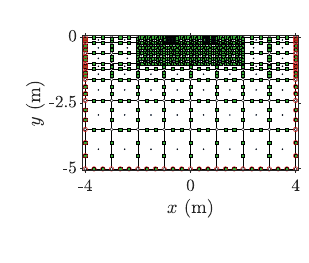}
        \caption{$\ell_{\max}=7$}
        \label{fig:quadtree_mesh_level_7}
    \end{subfigure}
    \hfill
    \begin{subfigure}[t]{0.32\textwidth}
        \centering
        \includegraphics[width=\linewidth]{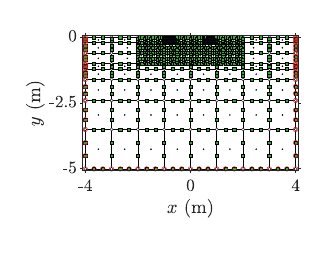}
        \caption{$\ell_{\max}=9$}
        \label{fig:quadtree_mesh_level_9}
    \end{subfigure}
    \caption{Locally refined quadtree S-element partitions generated with increasing maximum refinement levels. Refinement is concentrated near the loading and observation regions, whereas the remote region and the exterior boundary remain comparatively coarse. White circles denote quadtree vertices, green squares denote additional high-order edge-trace variables, red outer rings identify trace variables used for exterior scaled-boundary coupling, and blue circles indicate interior scaling centres.}
    \label{fig:quadtree_mesh_levels}
\end{figure}

\Cref{fig:quadtree_mesh_levels} illustrates the effect of the admissible quadtree depth. Here $\ell_{\max}$ is the maximum local refinement level permitted by the quadtree generator; as it increases, the loading and observation regions are resolved more densely while the far field remains coarse.

\begin{figure}[!htbp]
    \centering
    \includegraphics[width=0.70\linewidth]{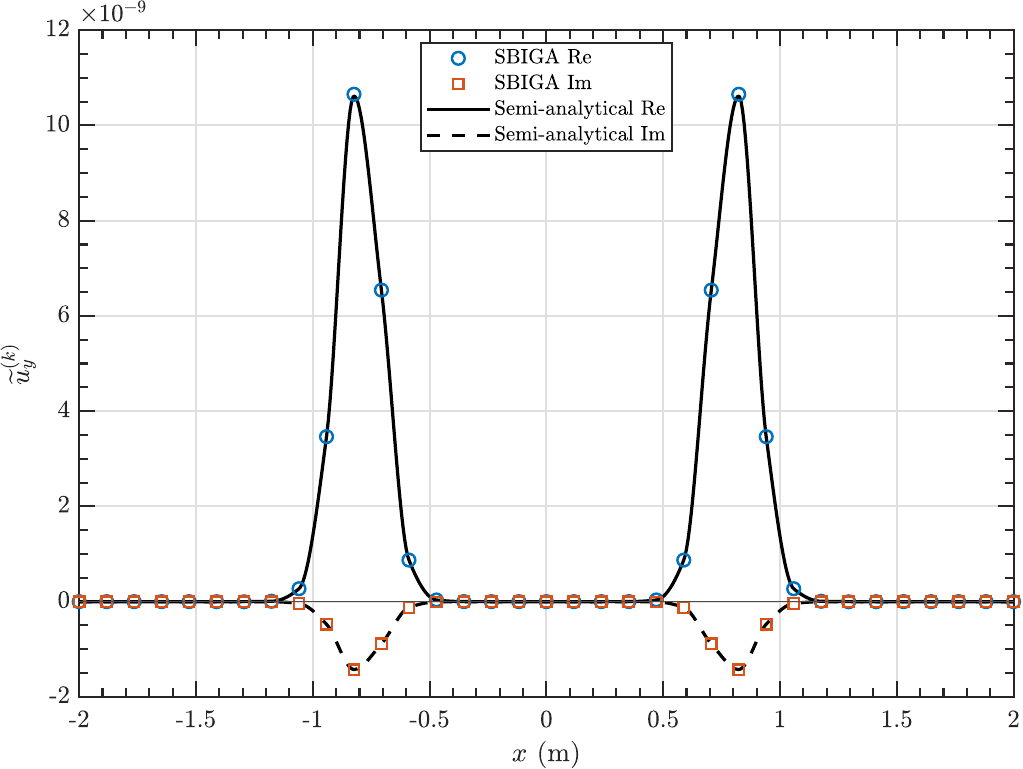}
    \caption{High-frequency quadtree validation at $f=200~\mathrm{Hz}$, $p=5$, and $y=-0.2~\mathrm{m}$. The solid and dashed black curves denote the real and imaginary parts of the semi-analytical half-space response, respectively. The hollow markers denote the corresponding SBIGA results.}
    \label{fig:quadtree_200hz_response}
\end{figure}

The $200~\mathrm{Hz}$ comparison in \Cref{fig:quadtree_200hz_response} probes the quadtree model where the response is sensitive to local spatial resolution and trace richness. The SBIGA markers closely follow the semi-analytical complex response on the shallow observation line.

Trace enrichment is examined on a coarse quadtree skeleton with $\ell_{\max}=3$, comparing degree elevation of the edge trace with knot-span refinement along each edge. Because the $200~\mathrm{Hz}$ response decays rapidly with depth, the error is evaluated on the shallow line $y=-0.2~\mathrm{m}$. \Cref{tab:quadtree_trace_enrichment_200hz} shows that both enrichments reduce the error on the same topology; the extracted trace can therefore be enriched independently of the quadtree skeleton.

\begin{table}[!htbp]
    \centering
    \caption{Trace-space enrichment on the same coarse quadtree skeleton ($\ell_{\max}=3$) at $f=200~\mathrm{Hz}$. The relative complex-valued response error is evaluated only along the shallow observation line $y=-0.2~\mathrm{m}$.}
    \label{tab:quadtree_trace_enrichment_200hz}
    \begin{tabular*}{0.90\linewidth}{@{\extracolsep{\fill}}ccc@{\hspace{1.0cm}}ccc@{}}
        \toprule
        \multicolumn{3}{c}{Degree elevation, $n_{\mathrm{span}}=1$}
        & \multicolumn{3}{c}{Knot-span refinement, $p=3$} \\
        \cmidrule(lr){1-3}\cmidrule(lr){4-6}
        $p$ & DOFs & Error
        & $n_{\mathrm{span}}$ & DOFs & Error \\
        \midrule
        2 & 675  & $1.87$
        & 1 & 1107 & $1.26$ \\
        3 & 1107 & $1.26$
        & 2 & 2403 & $2.06\times10^{-1}$ \\
        4 & 1539 & $4.22\times10^{-1}$
        & 3 & 3699 & $7.93\times10^{-2}$ \\
        5 & 1971 & $3.22\times10^{-1}$
        & 4 & 4995 & $3.25\times10^{-2}$ \\
        \bottomrule
    \end{tabular*}
\end{table}

\begin{figure}[!htbp]
    \centering
    \includegraphics[width=0.70\linewidth]{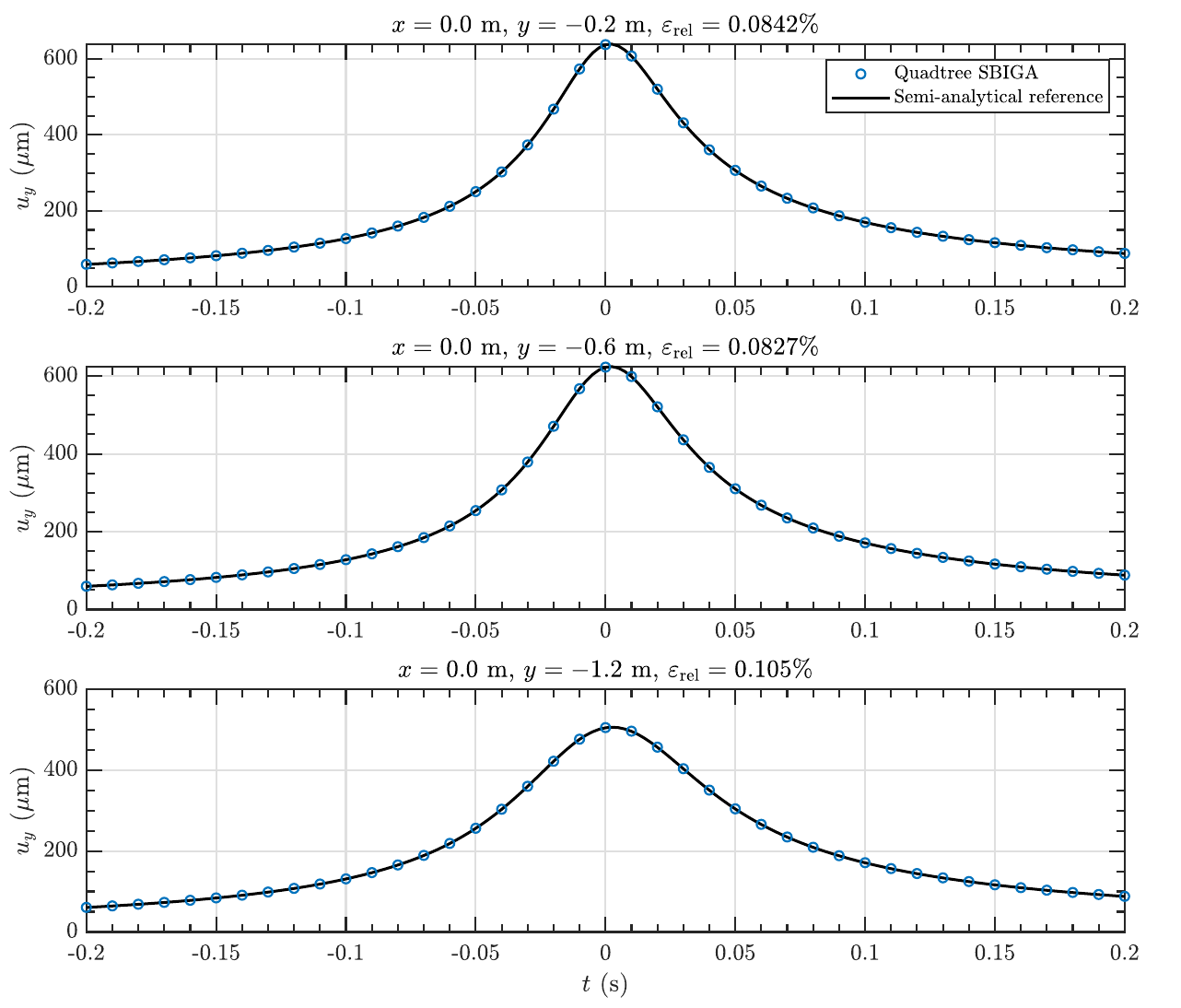}
    \caption{Time-domain reconstruction of the quadtree SBIGA response. The reconstructed vertical displacement histories $u_y(t)$ are compared with the corresponding semi-analytical half-space/IFFT reference responses at selected observation depths.}
    \label{fig:quadtree_time_domain_response}
\end{figure}

Finally, a broadband frequency sweep and inverse Fourier reconstruction produce the vertical displacement history $u_y(t)$. The comparison in \Cref{fig:quadtree_time_domain_response} tests the contact spectrum, wheel phases, 2.5D bounded/exterior solution, and reconstructed physical response.

The half-space comparisons establish the accuracy of the bounded and exterior SBIGA operators and the moving-load reconstruction for the tested frequencies and loads. \Cref{subsec:num_cylindrical_cavity} next evaluates these components in the coupled THB-GIFT/SBIGA formulation.

\subsection{Analytical Validation of the Coupled THB-GIFT/SBIGA Formulation Using a Cylindrical Cavity}
\label{subsec:num_cylindrical_cavity}

At the specified frequency and longitudinal wavenumber, the cylindrical cavity provides a closed-form 2.5D field for testing the coupled THB near field, quadtree-SBIGA surrounding field, recovered interface traces, and scaled-boundary exterior operator.

Consider a cylindrical cavity of radius $a$ in an unbounded homogeneous isotropic viscoelastic medium, with a spatially uniform harmonic pressure of amplitude $p_0$ applied to the cavity wall. The cross-section comprises the four-patch THB-GIFT annulus $\Omega_{\mathrm T}=\{(r,\theta):a\leq r\leq b\}$, the bounded quadtree-SBIGA domain $\Omega_{\mathrm Q}$, and the unbounded exterior $\Omega_\infty$. The circle $r=b$ is the nonmatching interface $\Gamma_{\mathrm{TQ}}$, and the outer trace of $\Omega_{\mathrm Q}$ is $\Gamma_\infty$, where the exterior impedance $\mat{S}_\infty$ acts. NURBS describe the cavity wall and $\Gamma_{\mathrm{TQ}}$ exactly. The THB space supplies local resolution in the cavity neighbourhood, whereas the quadtree-SBIGA discretization connects this near field to the radiating exterior.

\begin{figure}[!htbp]
    \centering
    \includegraphics[width=0.98\linewidth]{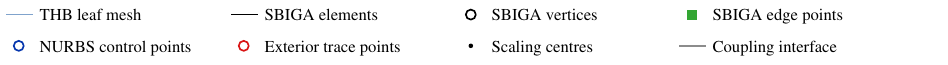}

    \begin{subfigure}[t]{0.495\textwidth}
        \centering
        \includegraphics[
            width=\linewidth, trim=0 0 0 25bp, clip
        ]{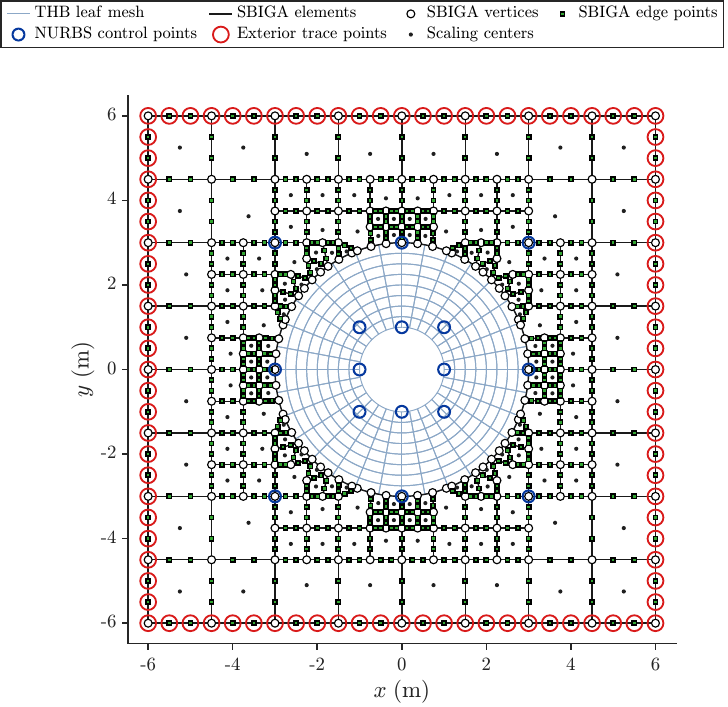}
        \caption{Baseline coupled discretization.}
        \label{fig:cavity_mesh_baseline}
    \end{subfigure}
    \hfill
    \begin{subfigure}[t]{0.425\textwidth}
        \centering
        \includegraphics[
            width=\linewidth, trim=0 0 0 16bp, clip
        ]{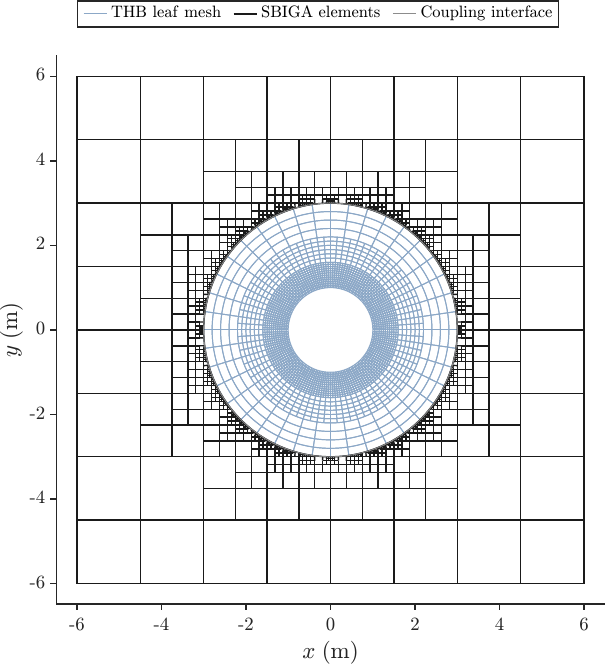}
        \caption{Combined THB and SBIGA refinement.}
        \label{fig:cavity_mesh_refined}
    \end{subfigure}

    \caption{Coupled discretizations for the cylindrical-cavity benchmark: (a) the baseline configuration and (b) an example combining radial THB-GIFT refinement with quadtree-SBIGA levels up to L8, which illustrates that both regions are refined independently. In both panels, the cavity wall is \(r=a\), the outer boundary of \(\Omega_{\mathrm T}\) is the coupling interface \(\Gamma_{\mathrm{TQ}}\), and the outer SBIGA trace represents \(\Gamma_\infty\).}
    \label{fig:cavity_coupled_mesh}
\end{figure}

The baseline parameters are $a=1~\mathrm{m}$, $b=3~\mathrm{m}$, an outer box $[-6,6]~\mathrm{m}\times[-6,6]~\mathrm{m}$, and an observation radius of $6~\mathrm{m}$. The medium has $\rho=1800~\mathrm{kg/m^3}$, $c_p=300~\mathrm{m/s}$, $c_s=150~\mathrm{m/s}$, $\nu=1/3$, and the common hysteretic damping ratio $\zeta=0.05$. The calculation is performed at $\omega=75~\mathrm{rad/s}$ and $k=0.45~\mathrm{m}^{-1}$ with $p_0=1$. The interface projection uses 12-point Gauss--Legendre quadrature per common segment. The bounded Riccati equation is integrated with adaptive Runge--Kutta (relative tolerance $10^{-9}$, absolute tolerance $10^{-12}$); the exterior uses 1800 fourth-order Runge--Kutta steps from $\xi_h=15$. Each annular patch uses degree-three THB functions on an initial $8\times8$ mesh; the SBIGA traces are also degree three, and the surrounding quadtree uses absolute levels L3--L5. \Cref{fig:cavity_coupled_mesh} shows the resulting computational model.

\begin{figure}[!htbp]
    \centering
    \begin{subfigure}[t]{0.31\linewidth}
        \centering
        \includegraphics[width=\linewidth]{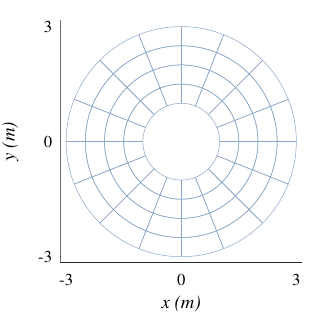}
        \caption{Uniform: $4\times4$ per patch.}
    \end{subfigure}\hfill
    \begin{subfigure}[t]{0.31\linewidth}
        \centering
        \includegraphics[width=\linewidth]{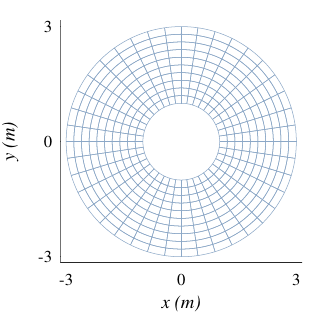}
        \caption{Uniform: $10\times10$ per patch.}
    \end{subfigure}\hfill
    \begin{subfigure}[t]{0.31\linewidth}
        \centering
        \includegraphics[width=\linewidth]{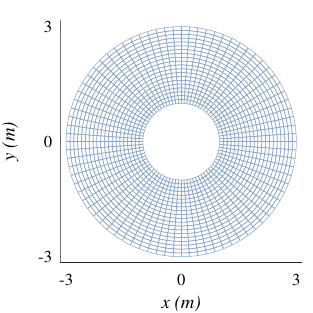}
        \caption{Uniform: $20\times20$ per patch.}
    \end{subfigure}

    \medskip
    \begin{subfigure}[t]{0.31\linewidth}
        \centering
        \includegraphics[width=\linewidth]{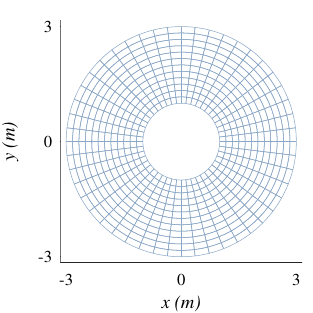}
        \caption{Hierarchical: L0 throughout.}
    \end{subfigure}\hfill
    \begin{subfigure}[t]{0.31\linewidth}
        \centering
        \includegraphics[width=\linewidth]{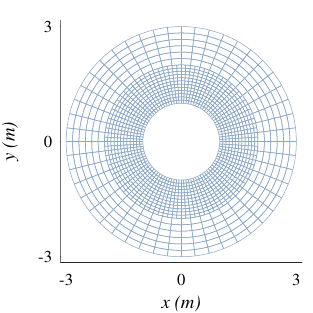}
        \caption{Hierarchical: inner L1, outer L0.}
    \end{subfigure}\hfill
    \begin{subfigure}[t]{0.31\linewidth}
        \centering
        \includegraphics[width=\linewidth]{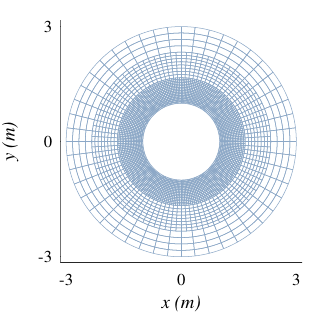}
        \caption{Hierarchical: radial L2/L1/L0 bands.}
    \end{subfigure}
    \caption{THB near-field discretizations used in the cylindrical-cavity studies. The upper row shows uniformly rebuilt annuli of increasing initial mesh density. The lower row shows the three radial hierarchies; refinement moves toward the cavity while the outer THB trace remains at L0.}
    \label{fig:cavity_thb_meshes}
\end{figure}

The two trace meshes on $\Gamma_{\mathrm{TQ}}$ are independent, with potentially different basis functions, endpoints, and partitions. Their combined breakpoints define the common integration partition used by $\boldsymbol{\Pi}$ in \Cref{subsec:trace_projection}, which transfers the THB master trace to the SBIGA slave trace. The constraint is embedded in $\mat{G}$ before solving \eqref{eq:global_system}; the eliminated SBIGA interface variables and both physical subdomain fields are then recovered. \Cref{fig:cavity_thb_meshes} isolates the THB discretizations varied in the two studies. The fixed SBIGA and exterior parts remain those in \Cref{fig:cavity_coupled_mesh}.

Numerical and analytical fields are compared in the frequency--wavenumber domain. Because the outward normal of the material domain on the cavity wall is $\bs{n}=-\bs{e}_r$, the pressure and zero tangential traction impose
\begin{equation}
    \sigma_{rr}(a)=-p_0,
    \qquad
    \sigma_{rz}(a)=0.
    \label{eq:cavity_wall_tractions}
\end{equation}
The reference is the full-space cylindrical-cavity solution of \citet{FrancoisSchevenels-57}. For the axisymmetric $n=0$ P--SV mode, $\tilde{u}_\theta^{\mathrm{a}}=0$, while $\tilde{u}_r^{\mathrm{a}}$ and $\tilde{u}_z^{\mathrm{a}}$ are coupled for $k\neq0$ and contain dilatational- and shear-wave contributions. The $k=0$ limit recovers the corresponding two-dimensional cavity field. The analytical field is evaluated at the numerical observation points and transformed to Cartesian components.

Subdomain accuracy and coupling consistency are measured separately.

Polar sampling grids and quadrature weights define the sampled $L^2$-type relative displacement error
\begin{equation}
    \varepsilon_X
    =
    \left(
    \frac{\sum_{q\in X}w_q
    \lVert\tilde{\bs{u}}_h(\bs{x}_q)-
    \tilde{\bs{u}}^{\mathrm{a}}(\bs{x}_q)\rVert_2^2}
    {\sum_{q\in X}w_q
    \lVert\tilde{\bs{u}}^{\mathrm{a}}(\bs{x}_q)\rVert_2^2}
    \right)^{1/2},
    \qquad X\in\{\Omega_{\mathrm T},\Omega_{\mathrm Q},\Omega_{\mathrm T}\cup\Omega_{\mathrm Q}\},
    \label{eq:cavity_sampled_error}
\end{equation}
where $\tilde{\bs{u}}^{\mathrm{a}}$ is the analytical complex 2.5D solution. Thus $\varepsilon_{\Omega_{\mathrm T}}$, $\varepsilon_{\Omega_{\mathrm Q}}$, and $\varepsilon_\Omega$ denote sampled polar-quadrature relative displacement errors in the THB domain, the bounded SBIGA domain, and their union. The analytical trace errors $\varepsilon_{\Gamma,\mathrm T}$ and $\varepsilon_{\Gamma,\mathrm Q}$ compare the THB-side and SBIGA-side samples, respectively, with $\tilde{\bs{u}}^{\mathrm{a}}$ on $r=b$. In contrast, the normalized displacement jump
\begin{equation}
    \varepsilon_{\Gamma,\mathrm{jump}}
    =
    \frac{\lVert\tilde{\bs{u}}_{\mathrm T}-\tilde{\bs{u}}_{\mathrm Q}\rVert_{W,\Gamma_{\mathrm{TQ}}}}
    {\max\!\left(\lVert\tilde{\bs{u}}_{\mathrm T}\rVert_{W,\Gamma_{\mathrm{TQ}}},
    \lVert\tilde{\bs{u}}_{\mathrm Q}\rVert_{W,\Gamma_{\mathrm{TQ}}}\right)},
    \label{eq:cavity_interface_jump}
\end{equation}
uses the same periodic angular quadrature on both numerical traces and measures their mutual consistency.

\begin{figure}[!htbp]
    \centering
    \begin{subfigure}[t]{0.59\linewidth}
        \centering
        \includegraphics[width=\linewidth]{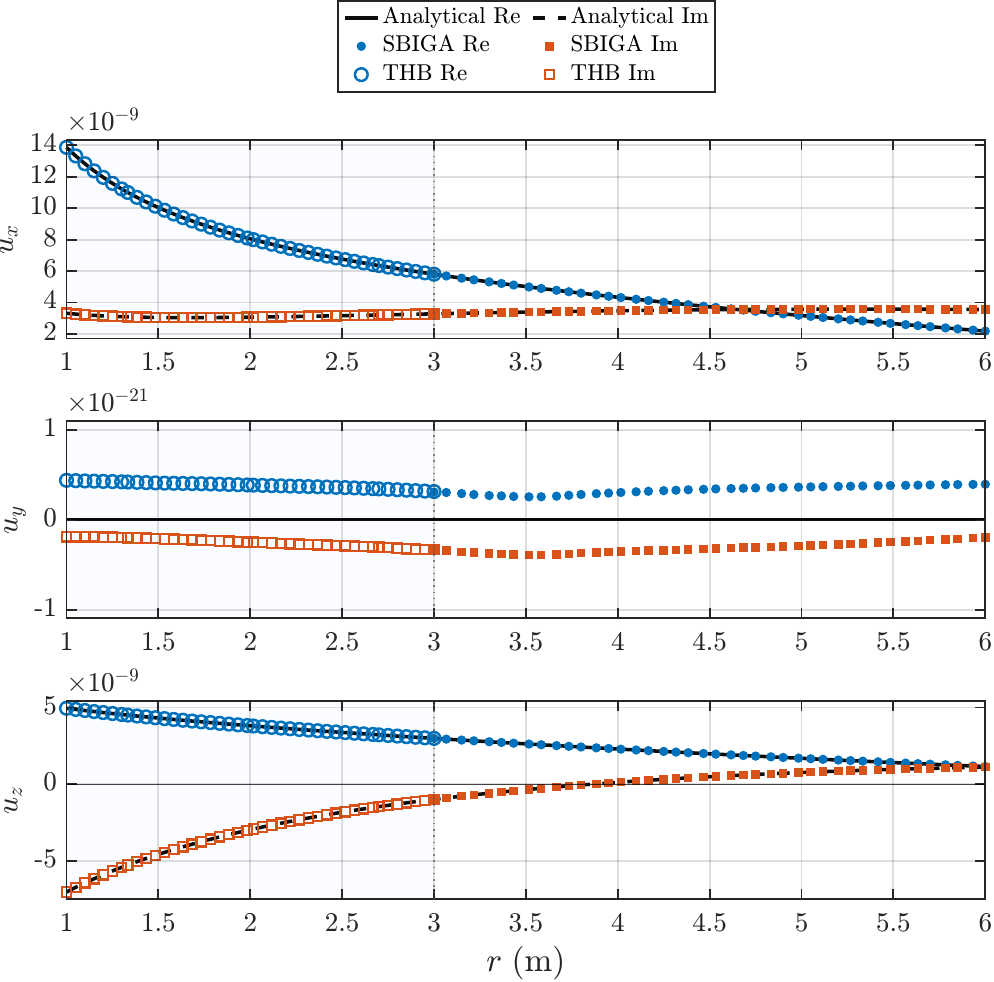}
        \caption{Radial cut.}
    \end{subfigure}

    \vspace{0.5em}
    \begin{subfigure}[t]{0.59\linewidth}
        \centering
        \includegraphics[width=\linewidth]{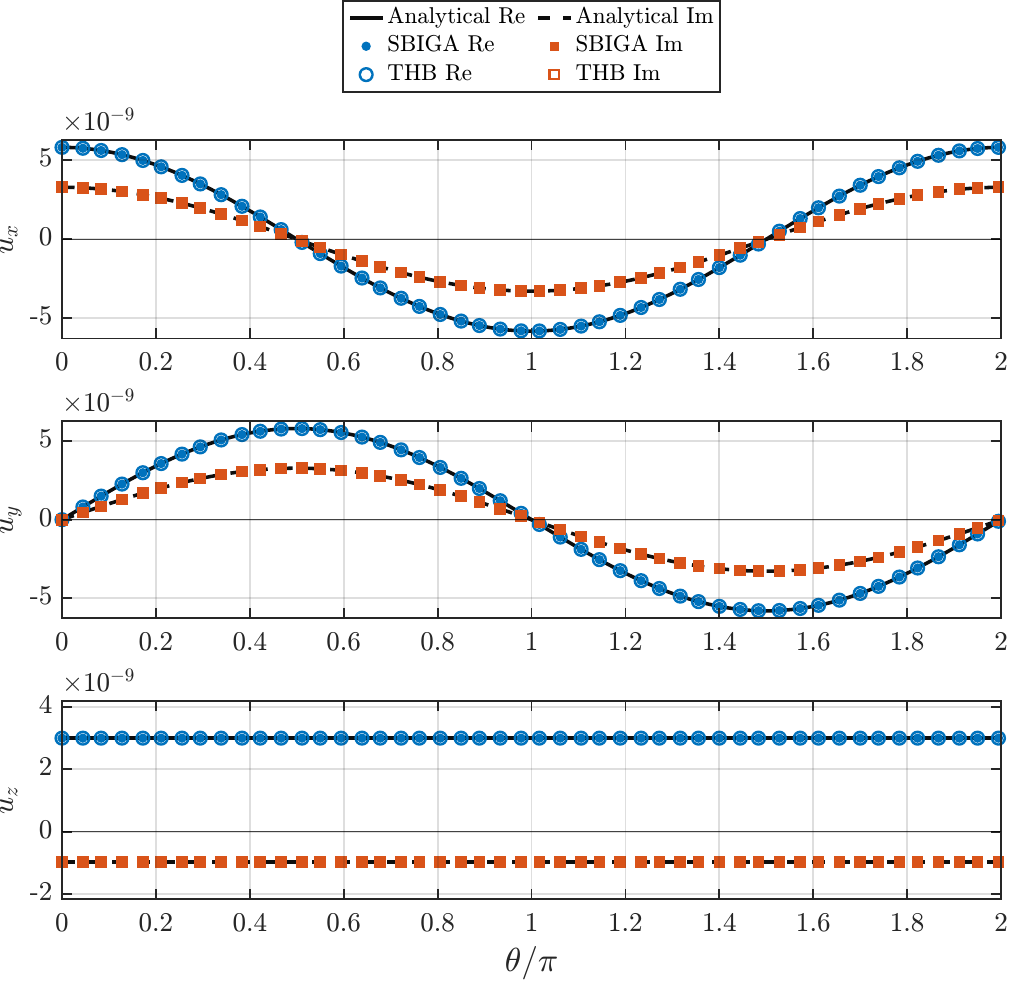}
        \caption{Interface trace on $r=b$.}
    \end{subfigure}
    \caption{Baseline numerical and analytical complex displacement responses in the common Cartesian basis $(u_x,u_y,u_z)$. Continuous curves are the analytical real and imaginary parts; open and filled markers are the THB and SBIGA values. The radial cut follows $\theta=0$ with horizontal coordinate $r$ in metres and crosses $\Gamma_{\mathrm{TQ}}$ at $r=b=3~\mathrm{m}$. The interface comparison is evaluated on $r=b$ with horizontal coordinate $\theta/\pi$.}
    \label{fig:cavity_baseline_response}
\end{figure}

\Cref{fig:cavity_baseline_response} compares the complex displacement components along a radial cut and around the coupling circle. The radial cut is taken at $\theta=0$, starts at the cavity wall, crosses $\Omega_{\mathrm T}$ and $\Gamma_{\mathrm{TQ}}$, and continues through $\Omega_{\mathrm Q}$. All ordinates are the Cartesian components $u_x$, $u_y$, and $u_z$; the analytical cylindrical solution is transformed to this basis, and $u_x=u_r$ only on the selected $\theta=0$ ray. The circumferential plots use $\theta/\pi$ on $r=b$ and compare the two recovered numerical traces with the same analytical trace. The real and imaginary parts are shown separately to assess both amplitude and phase. On the radial cut $u_y$ vanishes analytically, and the computed values are at round-off level (note the $10^{-21}$ scale).

For the baseline discretization, $\varepsilon_{\Omega_{\mathrm T}}=1.32\times10^{-5}$, $\varepsilon_{\Omega_{\mathrm Q}}=8.35\times10^{-6}$, and $\varepsilon_\Omega=1.10\times10^{-5}$. On $\Gamma_{\mathrm{TQ}}$, the two trace errors agree, $\varepsilon_{\Gamma,\mathrm T}=4.47\times10^{-6}$ and $\varepsilon_{\Gamma,\mathrm Q}=4.47\times10^{-6}$, and the jump between the numerical traces is $\varepsilon_{\Gamma,\mathrm{jump}}=2.43\times10^{-7}$, more than an order of magnitude below either trace error. The interface error in this test is therefore governed by trace approximation.

\begin{figure}[!htbp]
    \centering
    \includegraphics[width=0.90\linewidth]{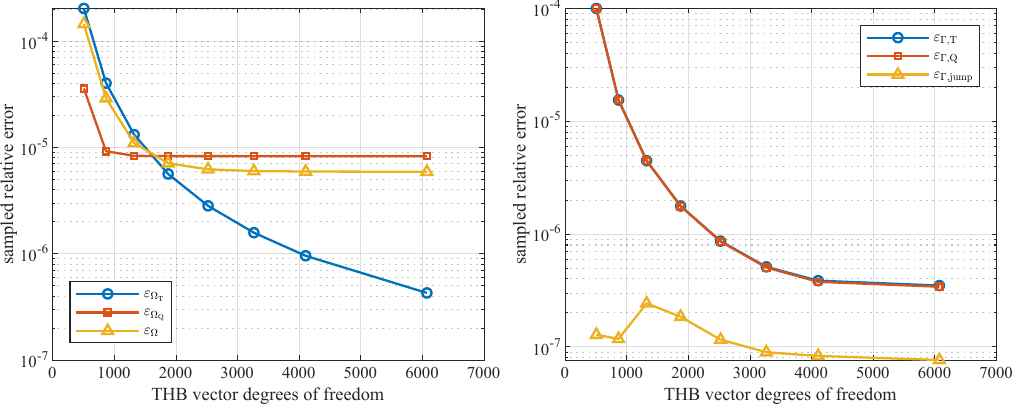}
    \caption{Error trends under uniform THB $h$-refinement from the $4\times4$ to the $20\times20$ annular mesh. The abscissa is the number of THB vector degrees of freedom. The eight points correspond to $4\times4$, $6\times6$, $8\times8$, $10\times10$, $12\times12$, $14\times14$, $16\times16$, and $20\times20$ initial meshes per annular patch. The left panel reports $\varepsilon_{\Omega_{\mathrm T}}$, $\varepsilon_{\Omega_{\mathrm Q}}$, and $\varepsilon_\Omega$; the right panel reports $\varepsilon_{\Gamma,\mathrm T}$, $\varepsilon_{\Gamma,\mathrm Q}$, and $\varepsilon_{\Gamma,\mathrm{jump}}$.}
    \label{fig:cavity_p02_error_trends}
\end{figure}

The uniform THB study refines the four THB-GIFT patches uniformly while all SBIGA, exterior, loading, and observation settings remain fixed. In \Cref{fig:cavity_p02_error_trends}, $N_{\mathrm{THB}}$ denotes displacement unknowns after expansion of the scalar basis to the three-component field. From the $4\times4$ to the $20\times20$ mesh ($504\leq N_{\mathrm{THB}}\leq6072$), the THB-domain error decreases monotonically from $2.03\times10^{-4}$ to $4.28\times10^{-7}$, with observed orders between $3.6$ and $4.0$ with respect to the initial element size, and the THB-side trace error decreases from $1.00\times10^{-4}$ to $3.50\times10^{-7}$. The SBIGA-side trace error follows the THB-side value to within $3\%$, and the jump remains between $7.6\times10^{-8}$ and $2.4\times10^{-7}$.

The SBIGA-domain error decreases from $3.59\times10^{-5}$ on the $4\times4$ mesh to $8.35\times10^{-6}$ on the $8\times8$ mesh and then remains between $8.29\times10^{-6}$ and $8.30\times10^{-6}$. On the coarse meshes this error is concentrated next to $\Gamma_{\mathrm{TQ}}$: in the band $3\leq r\leq3.5~\mathrm{m}$ it falls from $6.65\times10^{-5}$ to $3.20\times10^{-6}$ between the $4\times4$ and $8\times8$ meshes, consistent with the THB trace error entering $\Omega_{\mathrm Q}$ through the projected slave trace. The coupled-domain error decreases from $1.45\times10^{-4}$ to $1.10\times10^{-5}$ at the $8\times8$ mesh. From the $10\times10$ mesh onward, the THB-domain error is smaller than the SBIGA-domain error, and refinement to the $20\times20$ mesh reduces the former by a factor of $13$ but the coupled-domain error only from $7.11\times10^{-6}$ to $5.91\times10^{-6}$, which is then governed by the SBIGA-domain contribution of the fixed L3--L5 discretization.

\Cref{tab:cavity_p02_timing} gives the computational profile of the same refinement sequence. The SBIGA and exterior operators do not change with the THB mesh, and their construction is excluded, so the table isolates the THB-side work: construction of the THB model, assembly of the near-field, coupling and load terms, and the reduced solve. Observation-point postprocessing is also excluded. THB construction dominates and grows from $2.9$ to $254$~s as $N_{\mathrm{THB}}$ increases from $504$ to $6072$, whereas assembly grows from $4.7$ to $21$~s and the reduced solve remains below $0.7$~s, with a relative algebraic residual of order $10^{-14}$. Because $\boldsymbol{\Pi}$ and $\mat{G}$ are frequency independent for the fixed interface discretization (\Cref{subsec:trace_projection}), the construction cost is incurred once per discretization, while assembly and solve are repeated per spectral sample. All timings were obtained on the same machine.

\begin{table}[!htbp]
    \centering
    \scriptsize
    \caption{Computational profile of the uniform THB $h$-refinement runs, excluding the construction of the fixed SBIGA and exterior operators. The initial mesh is given per annular patch; assembly combines the THB, coupling/projection and load assembly.}
    \label{tab:cavity_p02_timing}
    \begin{tabular}{rrrrr}
        \toprule
        Initial mesh & $N_{\mathrm{THB}}$ & THB build (s) & Assembly (s) & Solve (s) \\
        \midrule
        $4\times4$   & 504  & 2.89   & 4.72  & 0.151 \\
        $6\times6$   & 864  & 5.16   & 5.46  & 0.171 \\
        $8\times8$   & 1320 & 10.28  & 6.57  & 0.213 \\
        $10\times10$ & 1872 & 20.32  & 7.67  & 0.248 \\
        $12\times12$ & 2520 & 37.48  & 9.42  & 0.286 \\
        $14\times14$ & 3264 & 62.39  & 10.75 & 0.341 \\
        $16\times16$ & 4104 & 107.80 & 13.88 & 0.428 \\
        $20\times20$ & 6072 & 253.53 & 20.54 & 0.658 \\
        \bottomrule
    \end{tabular}
\end{table}

Conversely, \Cref{tab:cavity_sbiga_interface_resolution} fixes the THB annulus at the $10\times10$ mesh and raises only the maximum SBIGA level, with the minimum level kept at L3. The sampled SBIGA- and coupled-domain errors fall mainly from L4 to L5, with smaller gains at L6. The THB-domain error changes by less than $0.4\%$.

Over fixed radial bands, the step from L5 to L6 reduces the SBIGA-domain error by $35\%$ for $3\leq r\leq3.5~\mathrm{m}$ and by $39\%$ for $3.5<r\leq4~\mathrm{m}$, where the additional level refines the quadtree. In the outer band $5<r\leq6~\mathrm{m}$ the cells remain at levels L3 and L4, the error stays at $1.34\times10^{-5}$, and this band contributes $78\%$ and $82\%$ of $\varepsilon_{\Omega_{\mathrm Q}}^2$ at L5 and L6. These results locate the remaining SBIGA-domain error but do not separate the contributions of the coarse outer cells and of the exterior impedance on $\Gamma_\infty$. By L5, the SBIGA-side trace error approaches that of the fixed THB master trace, $1.77\times10^{-6}$. It remains near that level at L6 as the displacement jump continues to decrease.

\begin{table}[!htbp]
    \centering
    \scriptsize
    \caption{SBIGA interface-resolution sensitivity at fixed THB $10\times10$ discretization. Error columns are scaled as indicated in the table header.}
    \label{tab:cavity_sbiga_interface_resolution}
    \begin{tabular}{lrrrrrr}
        \toprule
        SBIGA levels & Elements & SBIGA vector DOFs & $10^6\varepsilon_{\Omega_{\mathrm Q}}$ & $10^6\varepsilon_{\Omega}$ & $10^6\varepsilon_{\Gamma,\mathrm Q}$ & $10^7\varepsilon_{\Gamma,\mathrm{jump}}$ \\
        \midrule
        L3--L4 & 84  & 1656 & 16.6986 & 12.5322 & 2.71813 & 8.63452 \\
        L3--L5 & 156 & 3060 & 8.29713 & 7.10879 & 1.76433 & 1.84949 \\
        L3--L6 & 284 & 5520 & 8.11389 & 7.00076 & 1.77116 & 0.48495 \\
        \bottomrule
    \end{tabular}
\end{table}

In the hierarchical study, an interface collar keeps the outer THB trace at L0 while the annulus is refined towards the cavity (lower row of \Cref{fig:cavity_thb_meshes}). The collar contains the support cells of the active THB functions with nonzero interface trace and may include additional adjacent leaf layers. These cells remain at or below the prescribed interface level, here L0, and fine-level supports activated in the interior remain disjoint from the interface. The THB trace, common partition and projection are therefore unchanged; the SBIGA and exterior discretizations are also fixed. \Cref{tab:cavity_p03_hierarchy} reports the errors over fixed radial bands, $\bar r=(r-a)/(b-a)$, using \eqref{eq:cavity_sampled_error}. From L0 to L2/L1/L0, the error decreases by a factor of $41$ in the inner band $\bar r\leq1/3$ and by a factor of $5.5$ in the THB domain. The error in the outer band $\bar r\geq2/3$, which remains at L0 next to the coupling trace, changes by less than $1\%$, the interface jump is unchanged to the digits shown, and the coupled-domain error decreases by $5.0\%$ because the SBIGA-domain error, $8.29\times10^{-6}$ in all three configurations, exceeds the THB-domain error. Local refinement therefore reduces the error in the refined region while leaving the interface trace and exterior response unchanged.

\begin{table}[!htbp]
    \centering
    \scriptsize
    \caption{Hierarchical radial-refinement comparison. $N_{\mathrm{THB}}$ is the number of THB vector degrees of freedom. Error columns are scaled as indicated in the table header.}
    \label{tab:cavity_p03_hierarchy}
    \begin{tabular}{lrrrrr}
        \toprule
        Configuration & $N_{\mathrm{THB}}$ & $10^6 \varepsilon_{\mathrm{inner}}$ & $10^6 \varepsilon_{\Omega_{\mathrm T}}$ & $10^6 \varepsilon_{\Omega}$ & $10^7 \varepsilon_{\Gamma,\mathrm{jump}}$ \\
        \midrule
        L0 & 2520 & 4.30226 & 2.82323 & 6.22575 & 1.15804 \\
        L1/L0 & 5256 & 0.32533 & 0.68133 & 5.92027 & 1.15804 \\
        L2/L1/L0 & 13272 & 0.10557 & 0.51233 & 5.91184 & 1.15804 \\
        \bottomrule
    \end{tabular}
\end{table}

At comparable THB vector degrees of freedom, the L1/L0 hierarchy gives an inner-band displacement error of $3.25\times10^{-7}$ with $5256$ degrees of freedom, compared with $6.31\times10^{-7}$ and $6072$ degrees of freedom for the uniformly refined $20\times20$ THB mesh.

The interface transformation is assessed for projection stability, representable-trace reproduction and transfer accuracy on the baseline SBIGA discretization combined with the $4\times4$, $8\times8$, and $16\times16$ THB meshes. Trace errors in this verification are relative $L^2(\Gamma_{\mathrm{TQ}})$ norms evaluated with the 12-point Gauss rule on every segment of $\mathcal{P}_\Gamma$. The fixed slave Gram matrix is symmetric and full rank on all three meshes, with spectral condition number $76.86$ agreeing to thirteen digits, consistent with the projection construction in \Cref{subsec:trace_projection}. A constant master trace is reproduced with $\max_J|(\boldsymbol{\Pi}_s\bs{1})_J-1|\leq9\times10^{-15}$.

The coordinate functions $x$ and $y$ test transfer when the spaces differ: because the rational slave basis carries the weights of the NURBS boundary, both coordinates belong to the slave trace space, whereas the polynomial THB-GIFT trace space contains neither. Their best approximations in the slave space reproduce them to $1.4\times10^{-15}$. The THB-side best approximation $u_{\mathrm T}$ of $x$ has relative errors of $1.04\times10^{-4}$, $4.90\times10^{-6}$, and $2.84\times10^{-7}$ on the three meshes, and its transfer $\boldsymbol{\Pi}_s u_{\mathrm T}$ has errors of $1.04\times10^{-4}$, $4.89\times10^{-6}$, and $2.69\times10^{-7}$; the values for $y$ are identical. Since $\boldsymbol{\Pi}_s$ is the orthogonal projection, in the quadrature inner product of $\mat{M}_{\mathrm{QQ}}$, onto a space that contains $x$, $\|\boldsymbol{\Pi}_s u_{\mathrm T}-x\|\leq\|u_{\mathrm T}-x\|$ holds for every master trace $u_{\mathrm T}$, and the computed values satisfy this bound.

The role of adjoint force transfer is tested by changing only the transfer pair in the baseline reduction. Besides the adjoint $L^2$ pair $(\boldsymbol{\Pi},\boldsymbol{\Pi}^{\mathrm H})$ of \eqref{eq:global_system}, denoted V0, a collocation map $\boldsymbol{\Pi}^{\mathrm c}=\boldsymbol{\Pi}_s^{\mathrm c}\otimes\mat{I}_3$ was formed by interpolating the THB trace at $p_{\mathrm Q}+1$ equally spaced angular points on every slave edge, and $\mat{G}^{\mathrm c}$ denotes $\mat{G}$ with $\boldsymbol{\Pi}$ replaced by $\boldsymbol{\Pi}^{\mathrm c}$. Variant V1 uses the adjoint collocation pair, $\mat{K}_{\mathrm{red}}=(\mat{G}^{\mathrm c})^{\mathrm H}\mat{K}_{\mathrm{full}}\mat{G}^{\mathrm c}$. Variant V2 keeps the $L^2$ kinematics but returns the forces by collocation, $\mat{K}_{\mathrm{red}}=(\mat{G}^{\mathrm c})^{\mathrm H}\mat{K}_{\mathrm{full}}\mat{G}$, giving an unpaired force transfer.

Each variant is assessed against its own kinematic transformation $\mat{G}_k$ and interface map $\boldsymbol{\Pi}_k$: the virtual-work residual $\rho_W=\|\mat{G}_k^{\mathrm H}\bs{r}\|_2/\|\mat{G}_k^{\mathrm H}\tilde{\bs{f}}_{\mathrm{full}}\|_2$ measures virtual-work balance over the admissible variations of that constraint, and $\eta_t^\Gamma$ is evaluated from \eqref{eq:interface_force_balance} with $\boldsymbol{\Pi}^{\mathrm H}$ replaced by $\boldsymbol{\Pi}_k^{\mathrm H}$. For the damped 2.5D operator, $\mat{K}_{\mathrm{full}}$ is neither complex symmetric nor Hermitian: $\|\mat{K}_{\mathrm{full}}-\mat{K}_{\mathrm{full}}^{\mathsf T}\|_F/\|\mat{K}_{\mathrm{full}}\|_F=0.06$--$0.11$ and $\|\mat{K}_{\mathrm{full}}-\mat{K}_{\mathrm{full}}^{\mathrm H}\|_F/\|\mat{K}_{\mathrm{full}}\|_F=0.20$. Structure preservation is therefore measured on the symmetric part $\mat{K}_s=(\mat{K}_{\mathrm{full}}+\mat{K}_{\mathrm{full}}^{\mathsf T})/2$ by $\delta_s=\|\mat{A}_s-\mat{A}_s^{\mathsf T}\|_F/\|\mat{A}_s\|_F$, $\mat{A}_s=\mat{G}_{\mathrm{test}}^{\mathrm H}\mat{K}_s\mat{G}_k$, where $\mat{G}_{\mathrm{test}}$ is the transformation applied from the left, and on the Hermitian part $\mat{K}_h=(\mat{K}_{\mathrm{full}}+\mat{K}_{\mathrm{full}}^{\mathrm H})/2$ by the analogous $\delta_h$. These measures isolate the contribution of the transfer and vanish for an adjoint pair, as described in \Cref{subsec:hybrid_global_system}.

\begin{table}[!htbp]
    \centering
    \scriptsize
    \caption{Interface-transfer ablation on the cylindrical-cavity benchmark with the baseline SBIGA discretization. Each variant is evaluated with its own kinematic map; $\delta_s$ and $\delta_h$ agree to the digits shown. The last column measures the distance of the computed slave trace from the $L^2$ constraint \eqref{eq:interface_displacement_residual}.}
    \label{tab:cavity_transfer_ablation}
    \begin{tabular}{lllcccc}
        \toprule
        Variant & Transfer pair & THB mesh & $\rho_W$ & $\eta_t^\Gamma$ & $\delta_s$, $\delta_h$ & $\eta_u^\Gamma$ \\
        \midrule
        V0 & $(\boldsymbol{\Pi},\boldsymbol{\Pi}^{\mathrm H})$ & $4\times4$ & $8.4\times10^{-15}$ & $1.5\times10^{-14}$ & $1.3\times10^{-16}$ & 0 \\
           & & $8\times8$ & $9.5\times10^{-15}$ & $1.6\times10^{-14}$ & $1.1\times10^{-16}$ & 0 \\
           & & $16\times16$ & $1.7\times10^{-14}$ & $2.6\times10^{-14}$ & $8.8\times10^{-17}$ & 0 \\
        \midrule
        V1 & $(\boldsymbol{\Pi}^{\mathrm c},(\boldsymbol{\Pi}^{\mathrm c})^{\mathrm H})$ & $4\times4$ & $6.6\times10^{-15}$ & $1.2\times10^{-14}$ & $1.0\times10^{-16}$ & $2.6\times10^{-7}$ \\
           & & $8\times8$ & $7.1\times10^{-15}$ & $1.0\times10^{-14}$ & $8.0\times10^{-17}$ & $8.8\times10^{-7}$ \\
           & & $16\times16$ & $1.2\times10^{-14}$ & $1.4\times10^{-14}$ & $6.8\times10^{-17}$ & $3.3\times10^{-5}$ \\
        \midrule
        V2 & $(\boldsymbol{\Pi},(\boldsymbol{\Pi}^{\mathrm c})^{\mathrm H})$ & $4\times4$ & $6.5\times10^{-6}$ & $1.3\times10^{-5}$ & $4.5\times10^{-6}$ & 0 \\
           & & $8\times8$ & $2.6\times10^{-4}$ & $5.2\times10^{-4}$ & $1.6\times10^{-4}$ & 0 \\
           & & $16\times16$ & $1.9\times10^{-3}$ & $3.7\times10^{-3}$ & $1.0\times10^{-3}$ & 0 \\
        \bottomrule
    \end{tabular}
\end{table}

Both adjoint pairs satisfy virtual work, interface equilibrium, and structure preservation to round-off on all three meshes (\Cref{tab:cavity_transfer_ablation}). The non-adjoint pair V2 gives $\rho_W$, $\eta_t^\Gamma$, $\delta_s$, and $\delta_h$ between $4.5\times10^{-6}$ and $3.7\times10^{-3}$, and these residuals grow by more than two orders of magnitude as the THB trace is refined from $4\times4$ to $16\times16$ while the slave trace is fixed. Over the same range, $\eta_u^\Gamma$ for V1, which measures the difference between the collocated and projected slave traces, grows from $2.6\times10^{-7}$ to $3.3\times10^{-5}$. These algebraic measures complement the analytical-response errors reported above.

For the same analytical solution and polar sampling grid, V0 and V1 give nearly identical sampled full-domain displacement errors of approximately $1.45\times10^{-4}$ on the $4\times4$ mesh. On the $8\times8$ mesh, $\varepsilon_\Omega$ is $1.10\times10^{-5}$ for V0 and $1.65\times10^{-5}$ for V1; on the $16\times16$ mesh, the corresponding errors are $5.94\times10^{-6}$ and $2.74\times10^{-5}$. The V1-to-V0 error ratios are therefore approximately $1.5$ and $4.6$, and the V1 error increases between these two meshes. On the same machine with six threads, one-time construction of the V0 projection took $9.9$--$24.9$~s; the V1 collocation-map step took $1.3$--$2.0$~s using interface data already obtained during projection construction. The sums of warm median reduction and solve times were $0.25$--$0.77$~s for V0 and $0.20$--$0.51$~s for V1, while the common SBIGA operator assembly took $148$--$159$~s in these runs.

The cavity problem also compares the coupled discretization with a single-domain quadtree-SBIGA model. Both use the same material, exact NURBS cavity geometry, loading, trace degree, exterior-impedance settings and polar sampling grids. The selected models have the same L4 outer-boundary discretization; the single-domain model also refines the near-field annular band. The errors $\varepsilon_{\mathrm{in}}$ and $\varepsilon_{\mathrm{out}}$ are sampled over $a\leq r\leq b$ and $b\leq r\leq6~\mathrm{m}$, respectively, and $\varepsilon_\Omega$ over their union. This comparison concerns complete regional discretization architectures.

The full time covers model construction, assembly and solution, including projection and reduction for the coupled model. The warm per-sample time repeats the frequency-dependent assembly, load construction and solve in the same process. Both use the same sparse solver and exclude observation-point evaluation; a per-sample time is not the cost of a complete frequency sweep. The reported tests used the same machine with six computational threads.

\begin{table}[!htbp]
    \centering
    \scriptsize
    \caption{Cavity architecture comparisons. The first pair uses outer-boundary levels L3 and L4, whereas the second pair uses L4 for both models. $N$ is the number of system unknowns, $n_e$ the number of S-elements, and nnz the number of matrix nonzeros. Errors are sampled over the inner and outer regions and their union. Times exclude observation-point evaluation.}
    \label{tab:cavity_architecture_comparison}
    \begin{tabular}{llrrrrrrrr}
        \toprule
        Architecture & Discretization & $N$ & $n_e$ & nnz ($10^6$) & $10^6\varepsilon_{\mathrm{in}}$ & $10^6\varepsilon_{\mathrm{out}}$ & $10^6\varepsilon_\Omega$ & Full (s) & Per sample (s) \\
        \midrule
        Coupled & $12\times12$/L3--L6 & 6960 & 284 & 1.3414 & 2.8231 & 8.1065 & 6.1018 & 122.8 & 77.7 \\
        Single-domain & L4--L8 & 11208 & 620 & 1.1028 & 10.785 & 5.1692 & 8.4211 & 275.6 & 270.6 \\
        \midrule
        Coupled & $12\times12$/L4--L6 & 8436 & 380 & 1.6692 & 2.819 & 3.0819 & 2.9552 & 337.3 & 281.3 \\
        Single-domain & L4--L8, annular band & 12864 & 728 & 1.2173 & 3.0414 & 3.8669 & 3.4842 & 288.8 & 280.0 \\
        \bottomrule
    \end{tabular}
\end{table}

The first two rows use different outer-boundary levels and show that the L4 single-domain model has a lower outer-region error than the L3 coupled model. For the matched-L4 comparison, the coupled model uses approximately $34\%$ fewer unknowns and $48\%$ fewer S-elements, while reducing the sampled full-domain error by approximately $15\%$ (\Cref{tab:cavity_architecture_comparison}). Its system contains more nonzero entries. The warm per-sample times are comparable, whereas the coupled full run is longer because it includes construction of the THB region and interface projection.

The cavity benchmark verifies the nonmatching coupled formulation at the specified frequency and wavenumber. Across the tested nonmatching coupled configurations, the interface displacement jump ranges from $4.8\times10^{-8}$ to $8.6\times10^{-7}$ and remains below the approximation errors of both traces. The refinement studies demonstrate local accuracy control and interaction between independently designed regions: near-field refinement improves the local solution and can influence the adjacent SBIGA field. The architecture comparison illustrates the trade-off between near- and outer-region accuracy, system size and computational cost under the stated settings. The next examples examine conduit demand and tunnel interaction.

\subsection{Layered Pavement with a Buried Circular Utility Conduit}
\label{subsec:num_layered_pavement}

The formulation is applied to a finite-width layered pavement containing a buried circular utility conduit. In this configuration, the pavement layers and the conduit wall require local resolution, whereas the foundation soil carries the transmitted wave field to the exterior. The model combines a layered THB-GIFT pavement region, a quadtree-SBIGA foundation, and a scaled-boundary exterior. The CAD-exact circular four-patch THB-GIFT region lies wholly within the homogeneous foundation soil and is coupled directly to the surrounding quadtree domain through a second, disconnected, closed nonmatching interface. For this application, the complete interface in \eqref{eq:general_interface} is $\Gamma_{\mathrm{TQ}}=\Gamma_{\mathrm{PQ}}\cup\Gamma_{\mathrm{CQ}}$, where $\Gamma_{\mathrm{PQ}}$ is the open THB--SBIGA interface located within the homogeneous foundation soil, and $\Gamma_{\mathrm{CQ}}$ is the circular interface. The scaled-boundary exterior treatment acts on $\Gamma_{\infty}$.

The example examines the conduit response to a generic moving multi-axle load, including its circumferential stress distribution and sensitivity to cover and transverse vehicle offset. A reference model without the conduit isolates its effect on the surface response.

The upper THB-GIFT region is $10.0~\mathrm{m}$ wide and $2.50~\mathrm{m}$ deep. The finite layered pavement occupies $-3.75\leq x\leq3.75~\mathrm{m}$; homogeneous foundation-soil collars of width $1.25~\mathrm{m}$ are retained on both sides, and a $0.50~\mathrm{m}$ homogeneous collar is retained below the roadbed. The surface-course, base-course, and roadbed thicknesses are $0.20$, $0.40$, and $1.40~\mathrm{m}$, respectively. Their material interfaces coincide with THB analysis knot lines, every integration leaf is assigned exactly one material, and the complete open interface $\Gamma_{\mathrm{PQ}}$ lies in foundation soil.

\begin{figure}[!htbp]
    \centering
    \begin{subfigure}[t]{0.55\linewidth}
        \centering
        \includegraphics[width=\linewidth]{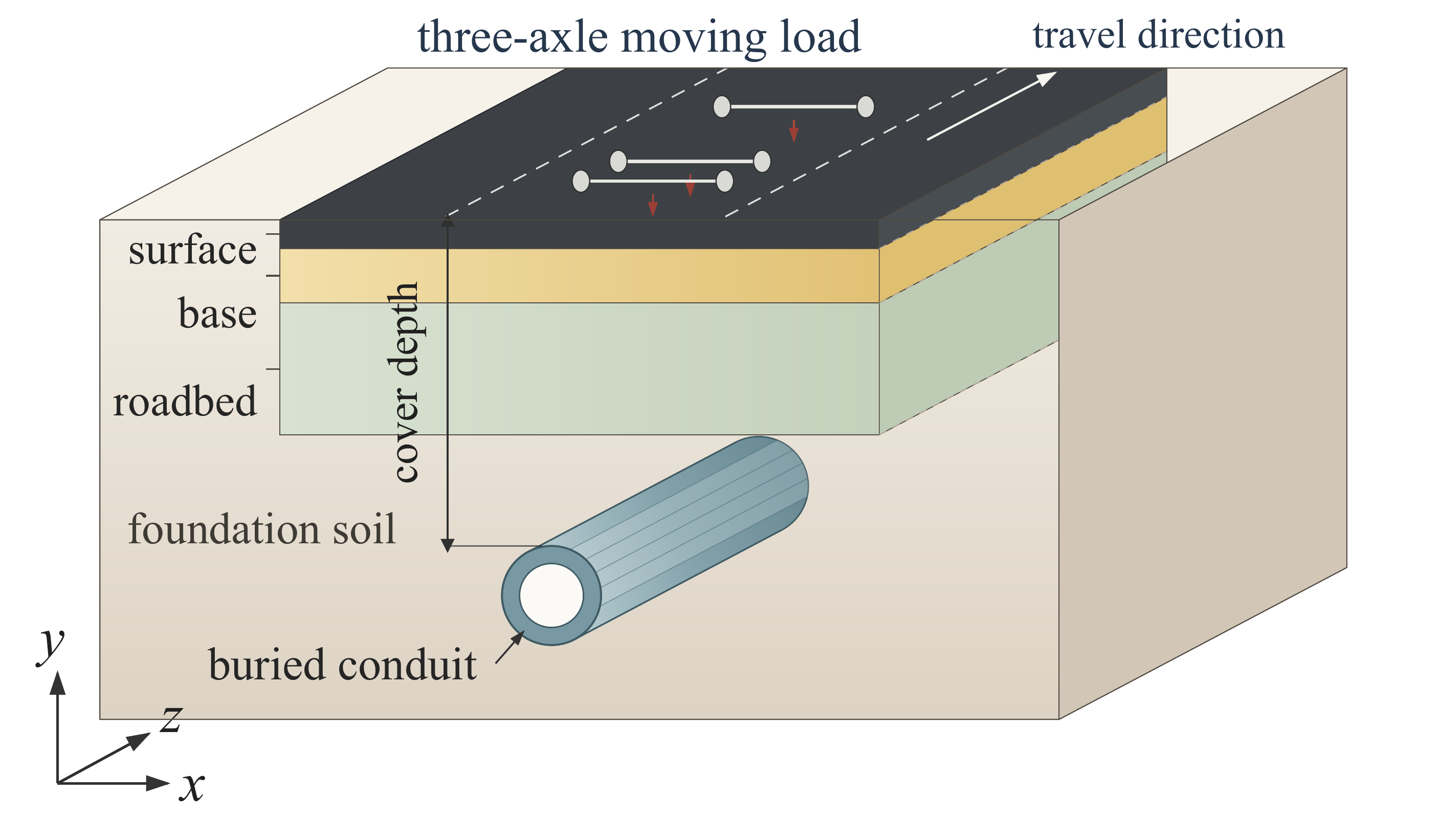}
        \caption{Physical configuration and three-axle moving-load arrangement.}
    \end{subfigure}

    \vspace{0.45em}
    \begin{subfigure}[t]{0.95\linewidth}
        \centering
        \includegraphics[width=\linewidth]{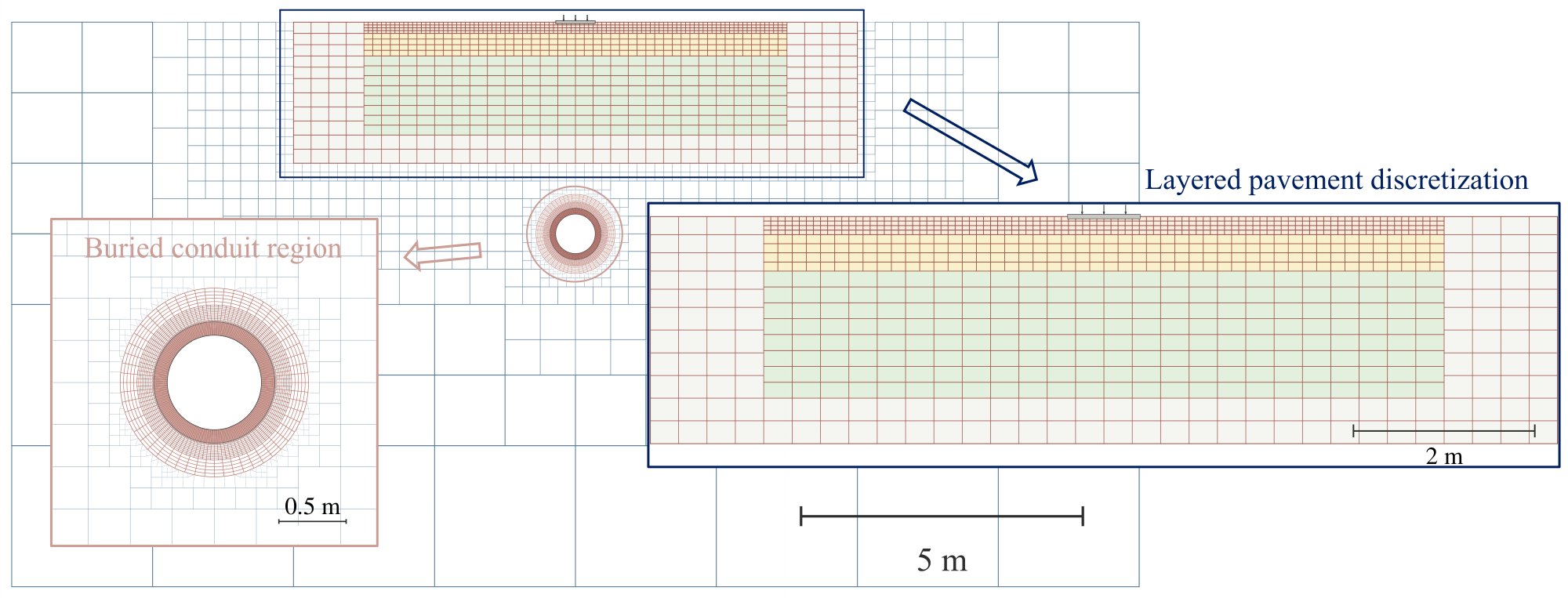}
        \caption{Layered and circular THB-GIFT near fields within the quadtree-SBIGA exterior soil.}
    \end{subfigure}
    \caption{Physical and computational configuration of the layered pavement with a buried conduit. Panel (a) shows the finite-width layered road, conduit placement, coordinate convention, and three-axle moving-load sequence along $z$. Panel (b) distinguishes the layer-aware THB-GIFT pavement and circular THB-GIFT conduit region from the surrounding quadtree-SBIGA soil, including local refinement near the conduit and the coarse-to-fine exterior transition. Grid lines in the upper THB region of panel (b) are drawn at reduced density for legibility.}
    \label{fig:case1_configuration}
\end{figure}

The conduit axis is parallel to $z$ and its centre is $(x_c,y_c)=(0,-3.75)~\mathrm{m}$. The inner and outer radii are $R_i=0.35~\mathrm{m}$ and $R_o=0.45~\mathrm{m}$, giving an outside diameter $D=0.90~\mathrm{m}$, wall thickness $0.10~\mathrm{m}$, and surface-to-crown cover $C=3.30~\mathrm{m}$ ($C/D=3.67$). The coupling radius $b=0.70~\mathrm{m}$ leaves $0.25~\mathrm{m}$ of homogeneous soil between the conduit wall and $\Gamma_{\mathrm{CQ}}$ and $0.55~\mathrm{m}$ between the top of $\Gamma_{\mathrm{CQ}}$ and $\Gamma_{\mathrm{PQ}}$. The physical conduit--soil material interface is at $r=R_o$. The annulus $R_o<r<b$ and the quadtree soil immediately outside $r=b$ have the same foundation-soil properties. The numerical interface $\Gamma_{\mathrm{CQ}}$ therefore lies within one homogeneous material, where the locally refined quadtree couples directly to the circular THB-GIFT region without an intermediate subdomain (\Cref{fig:case1_configuration}).

The baseline linear isotropic properties are listed in \Cref{tab:case1_materials}. The buried structure is a generic hollow concrete-like utility conduit. All regions use the equivalent hysteretic material model in \Cref{eq:complex_moduli}, with $\zeta=0.05$.

\begin{table}[!htbp]
    \centering
    \caption{Baseline linear isotropic material parameters. The conduit centre is $(0,-3.75)~\mathrm{m}$, with $R_i=0.35~\mathrm{m}$, $R_o=0.45~\mathrm{m}$, and coupling radius $b=0.70~\mathrm{m}$.}
    \label{tab:case1_materials}
    \begin{tabular}{lrrrr}
        \toprule
        Region & Thickness (m) & $E$ (MPa) & $\nu$ & $\rho$ (kg/m$^3$) \\
        \midrule
        Surface course & 0.20 & 3450 & 0.25 & 2000 \\
        Base course & 0.40 & 1000 & 0.25 & 2000 \\
        Roadbed & 1.40 & 60 & 0.40 & 1500 \\
        Foundation soil & -- & 35 & 0.25 & 1400 \\
        Conduit wall & 0.10 & 30000 & 0.20 & 2400 \\
        \bottomrule
    \end{tabular}
\end{table}

The open-interface projection uses six Gauss points on each segment of the common physical partition formed by the THB and SBIGA trace breaks, with the slave trace oriented consistently with its B\'ezier extraction. The closed-interface projection uses 28 points per common segment. The discretization has 19,110 reduced vector degrees of freedom, and the time-domain responses are reconstructed over $0.01\leq f\leq200~\mathrm{Hz}$.

\begin{samepage}
To account for the corrected open-interface projection, we interpolate the complex response difference between the corrected and original solutions from 12 frequencies onto the production grid. We set the correction to zero above $20~\mathrm{Hz}$ and add it to the original spectra before time reconstruction.
\par
\end{samepage}

\begin{samepage}
For the baseline, cover and offset cases, comparisons with references based on more correction frequencies, supported by independent frequency checks, give differences of about $0.013\%$ in peak magnitude, below $0.05\%$ in the time histories, and approximately $0.10\%$ in the responses at critical circumferential locations.
\par
\end{samepage}

The exterior impedance $\mat{S}_\infty$ is applied below $80~\mathrm{Hz}$. At $80$ Hz and above, the subsonic moving-load field decays spatially towards the truncation boundary, and $\mat{S}_\infty$ is omitted, leaving a traction-free artificial boundary. At $0.35$ and $6$ Hz, the RMS displacement jump sampled at common physical locations on $\Gamma_{\mathrm{PQ}}$ is at most $3.1\times10^{-6}$ of the interface trace RMS. At $40$, $80$ and $120$ Hz, the jump RMS is at most $2.4\times10^{-7}$ of the maximum surface response. The sampled closed-interface jump on $\Gamma_{\mathrm{CQ}}$ remains below $1.40\times10^{-9}$ at the independently checked frequencies $0.35$, $6$, $40$ and $80$ Hz. Comparing the corrected solution with a 37,821-degree-of-freedom discretization at 15 frequencies gives maximum time-history differences of $0.084\%$, $0.022\%$ and $0.008\%$ of the finer-mesh peak for surface $-u_y$, crown $u_r$ and crown $\sigma_\theta$, respectively, and $1.12\%$ at any monitored conduit location.

\begin{figure}[!htbp]
    \centering
    \begin{subfigure}[t]{0.55\linewidth}
        \centering
        \includegraphics[width=\linewidth]{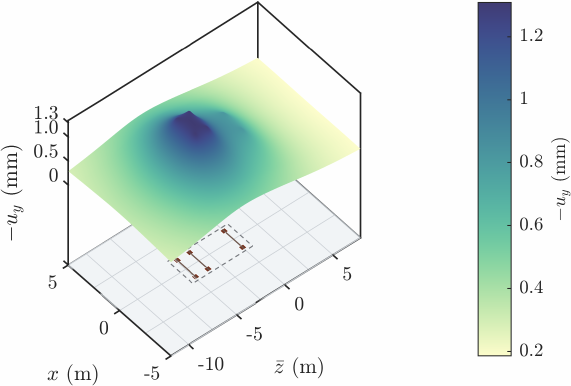}
        \caption{Downward-positive travelling surface-response field in the $x$--$\bar z$ frame. The response relief is displayed at $6\times$ vertical magnification; the ordinate and colour scale remain dimensional.}
    \end{subfigure}

    \vspace{0.5em}

    \begin{subfigure}[t]{0.48\linewidth}
        \centering
        \includegraphics[width=\linewidth]{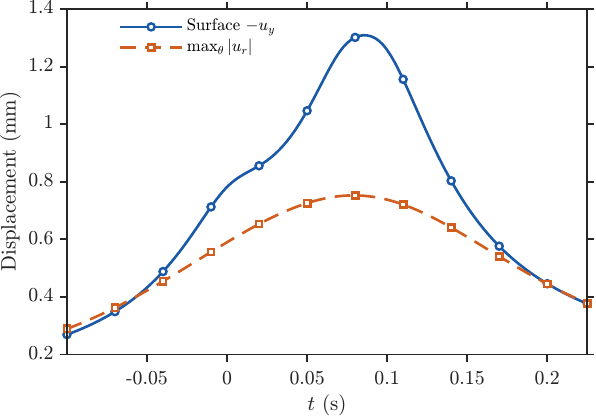}
        \caption{Downward-positive surface and conduit displacement histories.}
    \end{subfigure}\hfill
    \begin{subfigure}[t]{0.48\linewidth}
        \centering
        \includegraphics[width=\linewidth]{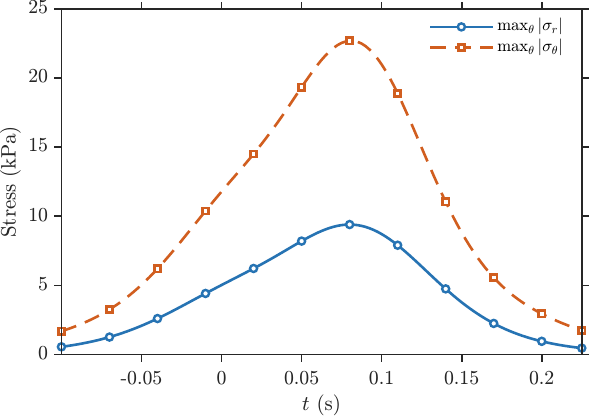}
        \caption{Conduit radial- and hoop-stress envelopes.}
    \end{subfigure}
    \caption{Baseline response to the generic three-axle moving load. In panel (a), the lower grid is a projection plane marking the wheel positions in the travelling frame.}
    \label{fig:case1_baseline_response}
\end{figure}

The generic moving-load configuration consists of three axles represented by six equivalent wheel groups. The front pair carries $35~\mathrm{kN}$ per side at $z_m=0$, and the two rear pairs carry $55~\mathrm{kN}$ per side at $z_m=-3.60$ and $-4.90~\mathrm{m}$. Every axle uses $x_m=\pm0.90~\mathrm{m}$, the vehicle speed is $50~\mathrm{m/s}$, and the Gaussian footprint parameters are $\sigma_x=0.10~\mathrm{m}$ and $\sigma_z=0.12~\mathrm{m}$. For a pure moving load, the dispersion line $k=\omega/c$ makes the response stationary in the travelling coordinate $\bar z=z-ct$. \Cref{fig:case1_baseline_response}(a) therefore presents the reconstructed surface field in the $x$--$\bar z$ frame. Consistent with \Cref{subsec:num_verification}, downward surface displacement is reported as $-u_y$ throughout this subsection.

The complete moving load produces a downward surface-displacement peak of $1.309~\mathrm{mm}$ at $t=0.085~\mathrm{s}$. The maximum conduit radial displacement is $0.7526~\mathrm{mm}$, and the maximum radial- and hoop-stress magnitudes are $9.404$ and $22.70~\mathrm{kPa}$, respectively; all three conduit maxima occur at $t=0.080~\mathrm{s}$.

The circumferential distributions locate the conduit-wall sectors associated with these peak demands. On the monitored conduit wall, $u_r(\theta,t)$ denotes radial displacement, $\sigma_r(\theta,t)$ radial normal stress, and $\sigma_\theta(\theta,t)$ circumferential (hoop) normal stress. The angle is measured counterclockwise from the right springline: $\theta=0^\circ$ is the right springline, $90^\circ$ the crown, $180^\circ$ the left springline, and $270^\circ$ the invert. These signed quantities are recovered at 144 circumferential locations ($2.5^\circ$ spacing). The baseline maxima occur near the crown ($88.75^\circ/91.25^\circ$) for $u_r$, at the springlines ($1.25^\circ/178.75^\circ$) for $\sigma_r$, and in sectors approximately $11^\circ$--$14^\circ$ above the springlines for $\sigma_\theta$, where the neighbouring sampled peaks are nearly equal.

\begin{figure}[!htbp]
    \centering
    \begin{subfigure}[t]{0.53\linewidth}
        \centering
        \includegraphics[width=\linewidth]{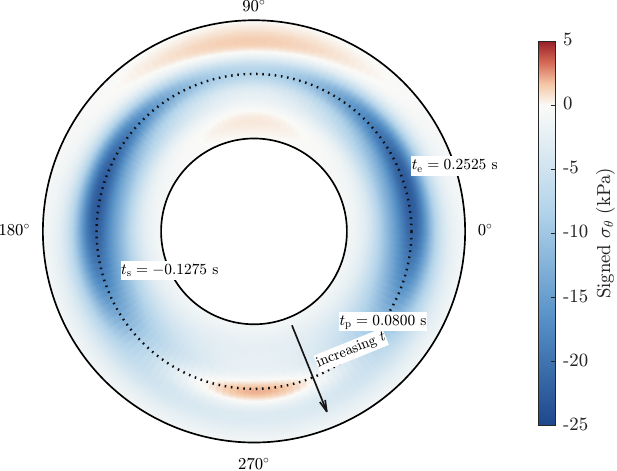}
        \caption{Signed annular $\sigma_\theta(\theta,t)$ history over the displayed active-response interval. Each concentric ring represents one time instant and time increases radially outward.}
    \end{subfigure}\hfill
    \begin{subfigure}[t]{0.43\linewidth}
        \centering
        \includegraphics[width=\linewidth]{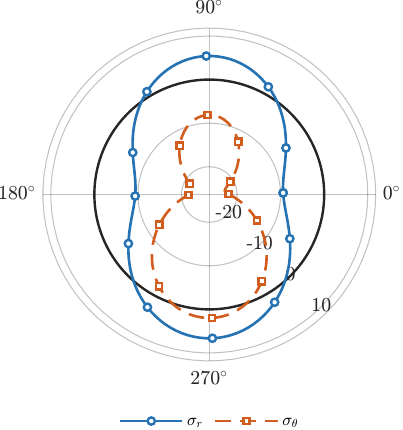}
        \caption{Signed $\sigma_r(\theta)$ and $\sigma_\theta(\theta)$ at $t_p=0.080~\mathrm{s}$ on a common kPa scale.}
    \end{subfigure}
    \caption{Full-circumference conduit stress mechanics using the angular convention defined in the text. Panel (a) shows the evolution of the governing hoop stress and marks the peak-envelope time with a dotted ring; panel (b) shows that radial and hoop stresses govern in different angular sectors and that the hoop stress is larger.}
    \label{fig:case1_circumferential_stress}
\end{figure}

The hoop-stress envelope reaches its maximum at $t_p=0.080~\mathrm{s}$. \Cref{fig:case1_circumferential_stress}(a) maps the active response interval around the vehicle passage radially outward in time, while panel (b) compares the signed radial and hoop stresses at the same governing instant. In the offset-polar representation, the reference circle denotes zero stress, compression is plotted inward, tension outward, and both stress components use the same kPa scale.

The conduit-induced surface change is isolated with a reference model in which the conduit region is replaced by foundation soil, all other settings being identical. The conduit and reference configurations give downward-positive surface peaks of $1.309$ and $1.353~\mathrm{mm}$, respectively. The maximum difference between the two downward-positive surface histories is $43.9~\mu\mathrm{m}$, or 3.25\% of the reference peak. The difference field in the $x$--$t$ plane is broad and nearly symmetric, so the conduit redistributes the surface motion rather than producing a localized scattering pattern.

The cover and transverse-offset cases are summarized in \Cref{tab:case1_parameters}. Relative to the baseline $C/D=3.67$, shallower cover increases the conduit radial displacement and both stress peaks. Deeper cover reduces all three peaks. Across the complete tested cover range, radial and hoop stress decrease by 25.16\% and 26.24\%, while the downward surface peak changes by 1.36\%. The normalized circumferential profiles remain nearly unchanged, and their governing locations shift by no more than one $2.5^\circ$ angular sample. Over the investigated range, cover changes the amplitude of conduit demand while the governing circumferential pattern remains essentially unchanged.

\begin{figure}[!htbp]
    \centering
    \begin{subfigure}[t]{0.485\linewidth}
        \centering
        \includegraphics[width=\linewidth]{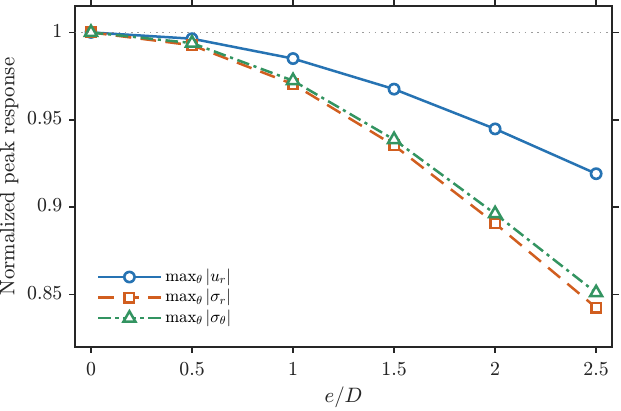}
        \caption{Normalized conduit-demand measures.}
    \end{subfigure}\hfill
    \begin{subfigure}[t]{0.485\linewidth}
        \centering
        \includegraphics[width=\linewidth]{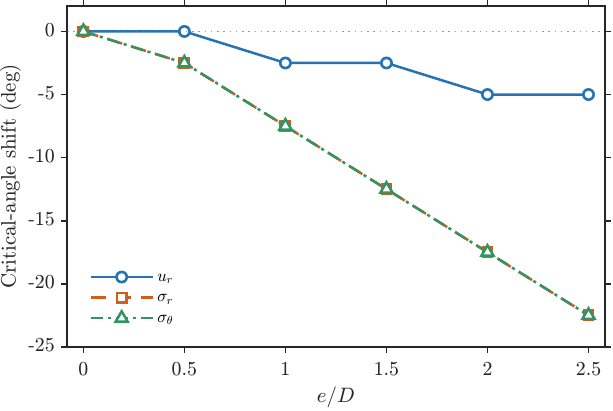}
        \caption{Migration of the governing response locations.}
    \end{subfigure}
    \caption{Effect of vehicle-to-conduit transverse offset. Panel (a) shows each peak conduit response normalized by its corresponding value at $e/D=0$; panel (b) uses the shortest circular shift relative to the centred reference branch. The line and marker identities for $u_r$, $\sigma_r$, and $\sigma_\theta$ are shared by both panels. Markers denote the six computed cases.}
    \label{fig:case1_offset}
\end{figure}

The transverse offset $e$ is measured from the conduit axis to the vehicle centreline. \Cref{fig:case1_offset}(a) normalizes the conduit peak magnitudes of $u_r$, $\sigma_r$, and $\sigma_\theta$ by their centred-case values at $e/D=0$. \Cref{tab:case1_parameters} separately reports the peak surface displacement at the fixed observation point. At $e/D=2.5$, the three conduit peaks are 8.09\%, 15.76\%, and 14.90\% below their centred values, respectively. The finer-mesh comparison at 16 frequencies gives a maximum difference of 0.91\% at any monitored conduit location.

\begin{table}[!htbp]
    \centering
    \caption{Cover-ratio and transverse-offset cases with the corresponding peak responses. Surface displacement is reported using the downward-positive convention, and cover changes are calculated from unrounded peaks relative to the baseline $C/D=3.67$.}
    \label{tab:case1_parameters}
    \small
    \setlength{\tabcolsep}{4.2pt}
    \begin{tabular}{rrrrrr}
        \toprule
        \multicolumn{6}{l}{\textit{Block A: cover ratio}} \\
        $C/D$ & $C$ (m) & peak $-u_y$ (mm) & $\max|u_r|$ (mm) &
        $\max|\sigma_r|$ (kPa) & $\max|\sigma_\theta|$ (kPa) \\
        \midrule
        3.35 & 3.015 & 1.302 & 0.7845 & 10.38 & 25.21 \\
        \multicolumn{2}{r}{\emph{Change (\%)}} & $-0.53$ & $+4.23$ & $+10.35$ & $+11.04$ \\
        3.67 & 3.300 & 1.309 & 0.7526 & 9.404 & 22.70 \\
        \multicolumn{2}{r}{\emph{Change (\%)}} & 0.00 & 0.00 & 0.00 & 0.00 \\
        4.35 & 3.915 & 1.320 & 0.6918 & 7.766 & 18.60 \\
        \multicolumn{2}{r}{\emph{Change (\%)}} & $+0.83$ & $-8.08$ & $-17.42$ & $-18.09$ \\
        \midrule
        \multicolumn{6}{l}{\textit{Block B: transverse offset}} \\
        $e/D$ & $e$ (m) & peak $-u_y$ (mm) & $\max|u_r|$ (mm) &
        $\max|\sigma_r|$ (kPa) & $\max|\sigma_\theta|$ (kPa) \\
        \midrule
        0.0 & 0.00 & 1.309 & 0.7526 & 9.404 & 22.70 \\
        0.5 & 0.45 & 1.306 & 0.7499 & 9.335 & 22.57 \\
        1.0 & 0.90 & 1.304 & 0.7414 & 9.125 & 22.08 \\
        1.5 & 1.35 & 1.184 & 0.7282 & 8.795 & 21.31 \\
        2.0 & 1.80 & 1.067 & 0.7110 & 8.375 & 20.35 \\
        2.5 & 2.25 & 0.9536 & 0.6917 & 7.922 & 19.32 \\
        \bottomrule
    \end{tabular}
\end{table}

For the centred configuration, the radial-displacement maximum has the mirror-equivalent locations $88.75^\circ$ and $91.25^\circ$. Angular migration follows the branch continuous with the positive-offset cases, which moves from $88.75^\circ$ to $83.75^\circ$ over the investigated range. At $e/D=2.5$, the governing radial- and hoop-stress sectors have each shifted by approximately $-22.5^\circ$, whereas the radial-displacement sector shifts by $-5.0^\circ$. Increasing offset therefore reduces overall conduit demand while moving the governing stress sectors; fixed crown, invert, and springline monitoring locations do not necessarily track the critical response.

\subsection{Dynamic Interaction of Twin Modified-D Tunnels under Moving Excitation}
\label{subsec:num_twin_tunnels}

\begin{figure}[!htbp]
    \centering
    \begin{subfigure}[t]{0.3794\linewidth}
        \centering
        \includegraphics[width=\linewidth]{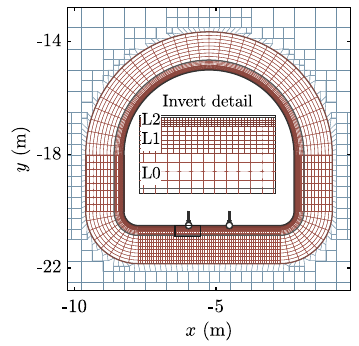}
        \caption{Tunnel detail with the refined lining.}
    \end{subfigure}\hfill
    \begin{subfigure}[t]{0.6018\linewidth}
        \centering
        \includegraphics[width=\linewidth]{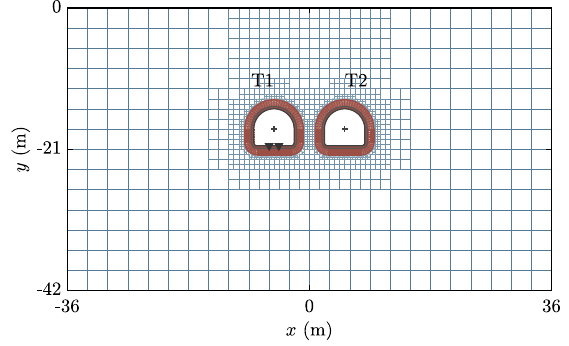}
        \caption{Complete computational domain.}
    \end{subfigure}
    \caption{Twin-tunnel discretization at $S/B=1.75$. In panel (a), the lining cells are L2 at the cavity side, L1 transitional and L0 outer, surrounded by an eight-layer soil collar; THB-GIFT (warm) and quadtree-SBIGA (cool) edges meet on the exact modified-D interface $\Gamma_{\mathrm{TQ}}$. In panel (b), $x\in[-36,36]$ m and $y\in[-42,0]$ m; the upper edge is the free surface, and the quadtree coarsens towards the far field.}
    \label{fig:case2_mesh}
\end{figure}

This application examines the interaction between a loaded and an unloaded tunnel, for which both linings require local resolution and the soil between them transmits the dynamic response. Two side-by-side modified-D tunnels are embedded in a homogeneous half-space at centre spacings $S/B=1.75$, $2.00$, $2.50$ and $3.00$. The excavation width is $B=6$~m, the roof-circle centre depth is $18$~m and the crown cover is $15$~m. Each tunnel has a $0.35$ m lining and a $1$ m soil collar; the two disconnected THB-GIFT regions couple through the common quadtree-SBIGA half-space (\Cref{fig:case2_mesh}). The soil properties are $E_s=108$ MPa, $\nu_s=1/3$, $\rho_s=1800$ kg\,m$^{-3}$ and $c_s=150$ m\,s$^{-1}$; the concrete-like lining has $E_l=30$ GPa, $\nu_l=0.20$ and $\rho_l=2400$ kg\,m$^{-3}$. Equivalent hysteretic damping uses $E^*=E(1+2\ii\zeta)$, with $\zeta_s=0.05$ and $\zeta_l=0.02$.

Tunnel~1 carries a moving railway load, represented as an equivalent traction transmitted to its invert, whereas Tunnel~2 is unloaded. Four $140$ kN axles travel along $+z$ at $v=11.5$ m\,s$^{-1}$ and apply vertical traction along $-y$. The axle offsets are $z=\pm6.15,\pm4.25$ m, the rail gauge is $1.435$ m, and the transverse and longitudinal Gaussian widths are $0.12$ and $0.15$ m. Vehicle--track interaction is not modelled. Mean cross-sectional motion, residual cross-sectional motion and local lining stress on the lining midline are compared to assess how spacing affects transmitted motion and local demand.

The interface projection uses 40-point Gauss--Legendre quadrature per common segment. The load speed is well below the soil shear-wave speed ($v/c_s\approx0.08$), and the subsonic moving-load field decays spatially towards the truncation boundary. Above $0.5$ Hz the exterior impedance $\mat{S}_\infty$ is omitted, leaving a traction-free artificial boundary: including it changes the response by a relative amount of at most $2\times10^{-4}$ at $0.5$ Hz and by less than $10^{-8}$ above $1$ Hz, and enlarging the domain to $[-48,48]\times[-54,0]$ m has an effect of the same order. Below $0.5$ Hz, $\mat{S}_\infty$ is retained.

For time reconstruction, the response is computed at $0.25$ Hz intervals from $0.25$ to $20$ Hz and at $0.5$ Hz intervals up to $40$ Hz. The integral from zero to $0.25$ Hz uses an endpoint expansion fitted at four Gauss--Legendre nodes, so that no solution at $f=0$ is required. Additional responses at $50$, $60$ and $80$ Hz check the high-frequency tail, which contributes negligibly to the time histories. The broadband results use the discretization of \Cref{fig:case2_mesh}(b); the hierarchically refined lining in \Cref{fig:case2_mesh}(a) changes the response by at most $0.1\%$ between $10$ and $40$ Hz.

\begin{figure}[!htbp]
    \centering
    \begin{subfigure}[t]{1.0\linewidth}
        \centering
        \includegraphics[width=\linewidth]{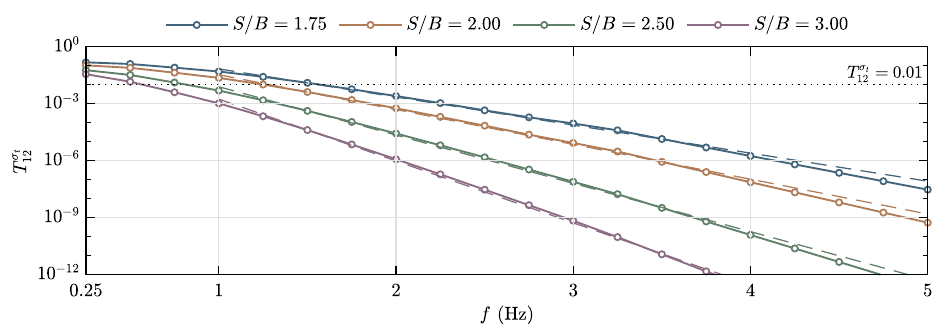}
        \caption{Tangential-stress transfer $T_{12}^{\sigma_t}$; dashed lines are exponential fits over $1$--$4$ Hz, and the dotted line marks $T_{12}^{\sigma_t}=0.01$.}
    \end{subfigure}

    \vspace{0.5em}
    \begin{subfigure}[t]{0.795\linewidth}
        \centering
        \includegraphics[width=\linewidth]{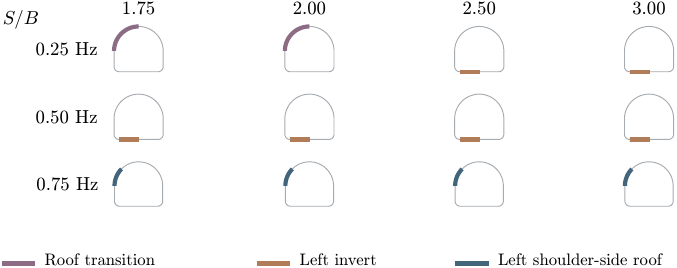}
        \caption{Governing sectors on the passive lining.}
    \end{subfigure}
    \caption{Passive tangential-stress transfer and governing lining sectors. In panel (b), both left roof arcs are marked where they compete for the maximum.}
    \label{fig:case2_transfer}
\end{figure}

For response $R$, let $P_j^R(f)=\max_q|\widehat R_j(q,f)|$, where $q=s/L$ is normalized inner-contour arclength and $j=1,2$ denotes the loaded and unloaded tunnels. Their same-frequency transfer ratio is
\begin{equation}
 T_{12}^R(f)=\frac{P_2^R(f)}{P_1^R(f)}.
 \label{eq:case2_interaction}
\end{equation}
The circumferential amplitude maxima in this ratio may occur at different locations. The tangential stress is $\sigma_t=\boldsymbol t^{\mathsf T}\boldsymbol\sigma_{xy}\boldsymbol t$. The effective low-frequency range of tangential-stress transfer narrows as spacing increases. At $0.25$ Hz, $T_{12}^{\sigma_t}$ decreases from $0.1517$ to $0.0361$ as $S/B$ increases from $1.75$ to $3.00$; at $1$ Hz it decreases from $0.0498$ to $0.00104$ (\Cref{fig:case2_transfer}). Over the same spacing interval, the interpolated frequency at which $T_{12}^{\sigma_t}$ falls to $10\%$ of its own sampled $0.25$--$40$ Hz maximum decreases from approximately $1.45$ to $0.79$ Hz. For the absolute threshold $T_{12}^{\sigma_t}=0.01$, the corresponding frequency decreases from $1.60$ to $0.61$ Hz. On a logarithmic scale, $T_{12}^{\sigma_t}$ decays almost exponentially with frequency above about $1$ Hz.

This decay follows from the moving-load line. In the Fourier-reduced Navier equation, the 2.5D fundamental solution of a homogeneous medium depends on the radial wavenumbers $k_\alpha=(\omega^2/c_\alpha^2-k^2)^{1/2}$, $\alpha=p,s$ \citep{TadeuKausel2000}. On $k_f=\omega_f/v$ with $v<c_s$, both are imaginary with $|k_\alpha|=k_f(1-v^2/c_\alpha^2)^{1/2}\approx k_f$, so the field transmitted through the soil is evanescent and decays approximately as $\exp(-k_f r)$ with distance $r$, as for the quasi-static contribution of subcritical axle loads \citep{LombaertDegrande2009}. A fit $T_{12}^{\sigma_t}\propto\exp(-k_f\Delta)$ over $1$--$4$ Hz gives effective decay distances $\Delta=6.2$, $7.7$, $10.7$ and $13.7$ m for the four spacings. The difference in fitted decay rate between any two spacings agrees with $2\pi/v$ times the spacing difference to within $0.3\%$, with $\Delta\approx S-4.3~\mathrm{m}$.

At $0.25$ Hz, the closer spacings have nearly tied left roof sectors, whereas the wider spacings are governed by the left invert. At $0.50$ Hz the left invert controls all four cases; at $0.75$ Hz the left shoulder-side roof controls. Where two sectors compete, both are marked in \Cref{fig:case2_transfer}(b). A fixed crown or invert receiver therefore cannot represent the governing region across frequencies and spacings.

\begin{figure}[!htbp]
\centering\includegraphics[width=0.65\linewidth]{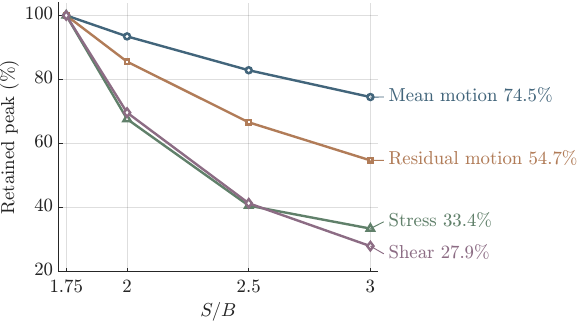}
\caption{Passive peak retention relative to $S/B=1.75$ for the four computed spacings; the maxima of different quantities need not be simultaneous.}
\label{fig:case2_retention}
\end{figure}

Mean motion and residual cross-sectional motion on the midline $\Gamma_m$, of length $L_m$, are defined by
\begin{equation}
 \overline{\boldsymbol u}(t)=\frac{1}{L_m}\int_{\Gamma_m}
 \boldsymbol u\,\mathrm{d}s,\quad
 U(t)=\|\overline{\boldsymbol u}(t)\|,\quad
 D(t)=\left[\frac{1}{L_m}\int_{\Gamma_m}
 \|\boldsymbol u-\overline{\boldsymbol u}\|^2\,\mathrm{d}s\right]^{1/2}.
\label{eq:case2_motion_decomp}
\end{equation}
All three displacement components are included; $D(t)$ measures the residual after mean translation, including any rotation. The local demand envelopes are $\Sigma_t(t)=\max_q|\sigma_t(q,t)|$ and $\mathcal{T}_{nt}(t)=\max_q|\tau_{nt}(q,t)|$. \Cref{fig:case2_retention} compares their peaks over $\mathcal W=[-1.5,1.5]$ s relative to $S/B=1.75$.

At $S/B=3$, the whole-window norms of $U$, $D$, $\Sigma_t$ and $\mathcal{T}_{nt}$ retain $76.7\%$, $57.9\%$, $39.7\%$ and $37.3\%$ of their values at $S/B=1.75$, respectively. The norm is $\|X\|_{\mathcal W}=[\int_{\mathcal W}X(t)^2\,\mathrm{d}t]^{1/2}$. Peak and whole-window measures give the same ordering: transmitted mean cross-sectional motion attenuates more slowly with spacing than residual cross-sectional motion and local lining demand. The different decay rates support tracking mean motion and lining demand separately. \Cref{tab:case2_peaks} lists the absolute peaks for the unloaded tunnel. The time-domain peak tangential stress in the loaded tunnel varies by less than $0.6\%$ across the four spacings.

\begin{table}[!htbp]
\centering\small
\caption{Absolute passive-tunnel peaks over $[-1.5,1.5]$ s; $D_{\max}$ denotes peak residual cross-sectional motion.}
\label{tab:case2_peaks}
\begin{tabular}{@{}crrrr@{}}\toprule
$S/B$ & $U_{\max}$ ($\mu$m) & $D_{\max}$ ($\mu$m) & $\Sigma_{t,\max}$ (kPa) & $\mathcal{T}_{nt,\max}$ (kPa)\\\midrule
1.75 &92.28&11.04&4.030&2.241\\
2.00 &86.23&9.45&2.725&1.560\\
2.50 &76.48&7.35&1.634&0.925\\
3.00 &68.75&6.04&1.345&0.625\\\bottomrule
\end{tabular}
\end{table}

\begin{figure}[H]
    \centering
    \includegraphics[width=0.661\linewidth]{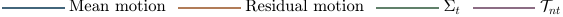}

    \vspace{0.3em}
    \begin{subfigure}[t]{0.42\linewidth}
        \centering
        \includegraphics[width=\linewidth]{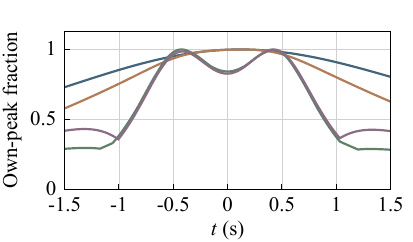}
        \caption{$S/B=1.75$.}
    \end{subfigure}\hspace{0.02\linewidth}
    \begin{subfigure}[t]{0.42\linewidth}
        \centering
        \includegraphics[width=\linewidth]{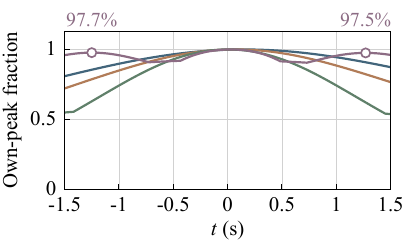}
        \caption{$S/B=3.00$.}
    \end{subfigure}
    \caption{Passive histories over $[-1.5,1.5]$ s, normalized by each quantity's own peak. In panel (a), each stress envelope has two nearly equal local peaks, both offset from the residual motion peak. In panel (b), the shear envelope retains side lobes near $t=-1.25$ and $+1.27$~s.}
    \label{fig:case2_histories}
\end{figure}

The peak-normalized histories in \Cref{fig:case2_histories} show how the timing and shape of the response vary. Tangential stress follows residual motion more closely than mean motion, although the stress and residual-motion peaks occur at different times. At $S/B=3$, the shear side lobes reach $97.7\%$ and $97.5\%$ of the shear peak. The different peak times motivate separate temporal envelopes of residual motion and local lining demand.

\begin{figure}[!htbp]
    \centering
    \begin{minipage}[c]{0.46\linewidth}
        \centering
        \begin{subfigure}[t]{\linewidth}
            \centering
            \includegraphics[width=\linewidth]{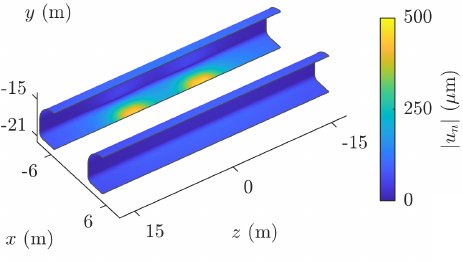}
            \caption{Normal displacement, common scale for both tunnels.}
        \end{subfigure}
        \par\vspace{0.6em}
        \begin{subfigure}[t]{\linewidth}
            \centering
            \includegraphics[width=\linewidth]{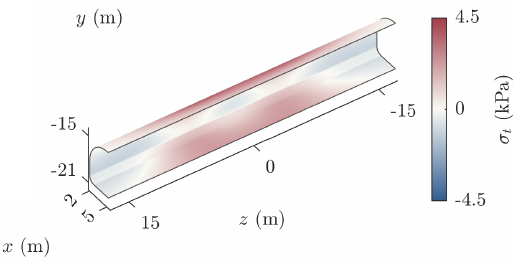}
            \caption{Passive-lining midline stress.}
        \end{subfigure}
    \end{minipage}\hfill
    \begin{subfigure}[c]{0.51\linewidth}
        \centering
        \includegraphics[width=\linewidth]{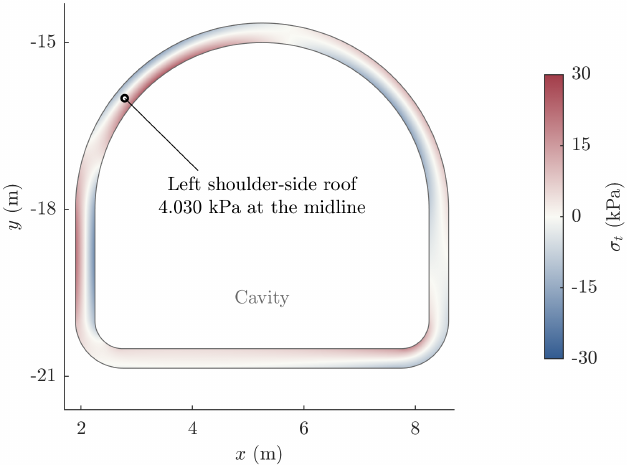}
        \caption{Passive full-thickness stress at $z=4.8875$ m.}
    \end{subfigure}
    \caption{Physical fields at $S/B=1.75$, $t_\star=0$. Panels (a) and (b) show the open left halves of the lining midline surfaces; panel (c) marks the $4.030$ kPa midline maximum. All spatial axes retain equal physical scale.}
    \label{fig:case2_spatial}
\end{figure}

For constant-speed excitation in this longitudinally invariant model, the field along $z$ follows the travelling relation
\[
R(x,y,z,t_\star)=R(x,y,0,t_\star-z/v).
\]
\Cref{fig:case2_spatial} shows the longitudinal variation at $t_\star=0$ and $S/B=1.75$, with a common displacement scale for both tunnels and a separate passive stress scale. The full-thickness section at $z=4.8875$ m locates the critical region on the left shoulder-side roof. Opposite stress signs across parts of the roof lining indicate a bending contribution.

% Flush pending floats before the concluding section.
\FloatBarrier
\section{Conclusions}
\label{sec:conclusions}

This study develops a projection-coupled 2.5D formulation for the dynamic response of tunnels and buried conduits under moving loads, for linear elastic or viscoelastic configurations that are invariant along the travelling direction. An $L^2$ trace projection on a common physical partition connects independently discretized THB-GIFT near-field regions around the structures to a quadtree-SBIGA discretization of the surrounding soil. Its Hermitian-adjoint force transfer and the corresponding congruence yield a work-consistent reduction. Unified B\'ezier extraction accommodates polynomial, hanging-edge and rational CAD traces, and all regional operators and the moving-load reconstruction follow one Fourier convention.

Homogeneous half-space benchmarks verify the bounded and exterior SBIGA operators and moving-load reconstruction over the tested frequencies and loads. The projection reproduces traces representable in both interface spaces, including uniform translations (\Cref{subsec:trace_projection}). In the cylindrical-cavity benchmark, the interface displacement jump stays below the approximation errors of both traces, and local THB refinement improves near-field accuracy with the interface trace held fixed. The transfer tests show discrete virtual-work balance to round-off for both $L^2$ projection and collocation when each displacement map is paired with its own adjoint force transfer.

With matched outer-boundary discretizations at the tested frequency and wavenumber, the coupled cavity model gives approximately $15\%$ lower sampled full-domain displacement error with $34\%$ fewer system unknowns than the single-domain quadtree-SBIGA model (\Cref{tab:cavity_architecture_comparison}). Warm per-sample times are comparable. The coupled model takes longer for the full run, which includes construction of the THB near field and the interface projection.

The pavement--conduit application combines a layered THB pavement region and a circular conduit region with an open and a closed numerical coupling interface, both located in homogeneous foundation soil. Over the tested range, cover mainly changes the amplitude of the conduit demand, whereas a transverse load offset reduces the demand and also shifts the governing circumferential stress sectors, so fixed crown, invert and springline locations do not necessarily capture the critical response. In the twin-tunnel application, two disconnected near-field regions are coupled through the same surrounding soil. Increasing the tested spacing narrows the effective low-frequency transfer range, and the mean cross-sectional motion of the unloaded tunnel attenuates more slowly than its residual motion and local lining stress and shear demand. The time-domain peak tangential stress in the loaded tunnel varies by less than $0.6\%$ across the tested spacings.

The near-field and surrounding-soil discretizations can therefore be refined independently in moving-load analyses of tunnels and buried conduits.

\section*{Acknowledgements}

This research was supported by the National Natural Science Foundation of China (Grant No. 52478477 to Bei Zhang and Grant No. 52578540 to Yanhui Zhong), the China Postdoctoral Science Foundation (Grant No. 2024M752937 to Quansheng Zang), the Natural Science Foundation of Henan Province (Grant No. 252600421827 to Yanhui Zhong), the Henan Provincial Science and Technology Research Project (Grant No. 262102221015 to Quansheng Zang), the Key Research Project of Higher Education Institutions of Henan Province (Grant No. 25A560006 to Quansheng Zang), and the Zhengzhou University Young Student Basic Research Project for PhD Students (Grant No. ZDBJ2026065 to Hao Hong).

\bibliographystyle{elsarticle-num-names}
\bibliography{myRef}

\begin{thebibliography}{44}
\expandafter\ifx\csname natexlab\endcsname\relax\def\natexlab#1{#1}\fi
\providecommand{\url}[1]{\texttt{#1}}
\providecommand{\href}[2]{#2}
\providecommand{\path}[1]{#1}
\providecommand{\DOIprefix}{doi:}
\providecommand{\ArXivprefix}{arXiv:}
\providecommand{\URLprefix}{URL: }
\providecommand{\Pubmedprefix}{pmid:}
\providecommand{\doi}[1]{\href{http://dx.doi.org/#1}{\path{#1}}}
\providecommand{\Pubmed}[1]{\href{pmid:#1}{\path{#1}}}
\providecommand{\bibinfo}[2]{#2}
\ifx\xfnm\relax \def\xfnm[#1]{\unskip,\space#1}\fi
%Type = Article
\bibitem[{Zhang et~al.(2023)Zhang, Liu, Pan, and Wang}]{ZhangLiu-66}
\bibinfo{author}{Z.~Zhang}, \bibinfo{author}{S.~Liu}, \bibinfo{author}{E.~Pan},
  \bibinfo{author}{Q.~Wang},
\newblock \bibinfo{title}{Dynamic loading in a transversely isotropic and
  layered elastic half-space},
\newblock \bibinfo{journal}{International Journal of Mechanical Sciences}
  \bibinfo{volume}{260} (\bibinfo{year}{2023}) \bibinfo{pages}{108626}.
  \DOIprefix\doi{10.1016/j.ijmecsci.2023.108626}.
%Type = Article
\bibitem[{Wu et~al.(2020)Wu, Wang, Zhao, Jiang, Qiu, and
  Yusupov}]{Wu2020PavementMovingLoad}
\bibinfo{author}{C.~Wu}, \bibinfo{author}{H.~Wang}, \bibinfo{author}{J.~Zhao},
  \bibinfo{author}{X.~Jiang}, \bibinfo{author}{Y.~Qiu},
  \bibinfo{author}{B.~Yusupov},
\newblock \bibinfo{title}{Prediction of viscoelastic pavement responses under
  moving load and nonuniform tire contact stresses using {2.5-D} finite element
  method},
\newblock \bibinfo{journal}{Mathematical Problems in Engineering}
  \bibinfo{volume}{2020} (\bibinfo{year}{2020}) \bibinfo{pages}{1029089}.
  \DOIprefix\doi{10.1155/2020/1029089}.
%Type = Article
\bibitem[{Zhang and Wong(2024)}]{ZhangWong2024BuriedPipeTruck}
\bibinfo{author}{Y.~Zhang}, \bibinfo{author}{R.~C.-K. Wong},
\newblock \bibinfo{title}{Static and dynamic responses of buried steel pipe
  under truck load: An experimental study},
\newblock \bibinfo{journal}{Engineering Structures} \bibinfo{volume}{312}
  (\bibinfo{year}{2024}) \bibinfo{pages}{118253}.
  \DOIprefix\doi{10.1016/j.engstruct.2024.118253}.
%Type = Article
\bibitem[{Xu et~al.(2017)Xu, Shen, and Rakitin}]{Xu2017BuriedPipelineTraffic}
\bibinfo{author}{M.~Xu}, \bibinfo{author}{D.~Shen},
  \bibinfo{author}{B.~Rakitin},
\newblock \bibinfo{title}{The longitudinal response of buried large-diameter
  reinforced concrete pipeline with gasketed bell-and-spigot joints subjected
  to traffic loading},
\newblock \bibinfo{journal}{Tunnelling and Underground Space Technology}
  \bibinfo{volume}{64} (\bibinfo{year}{2017}) \bibinfo{pages}{117--132}.
  \DOIprefix\doi{10.1016/j.tust.2016.12.020}.
%Type = Article
\bibitem[{Kuo et~al.(2011)Kuo, Hunt, and Hussein}]{Kuo2011TwinTunnel}
\bibinfo{author}{K.~A. Kuo}, \bibinfo{author}{H.~E.~M. Hunt},
  \bibinfo{author}{M.~F.~M. Hussein},
\newblock \bibinfo{title}{The effect of a twin tunnel on the propagation of
  ground-borne vibration from an underground railway},
\newblock \bibinfo{journal}{Journal of Sound and Vibration}
  \bibinfo{volume}{330} (\bibinfo{year}{2011}) \bibinfo{pages}{6203--6222}.
  \DOIprefix\doi{10.1016/j.jsv.2011.07.035}.
%Type = Article
\bibitem[{He et~al.(2019)He, Zhou, Guo, Di, Zhang, and
  Yu}]{He2019TwinTunnelInteraction}
\bibinfo{author}{C.~He}, \bibinfo{author}{S.~Zhou}, \bibinfo{author}{P.~Guo},
  \bibinfo{author}{H.~Di}, \bibinfo{author}{X.~Zhang}, \bibinfo{author}{F.~Yu},
\newblock \bibinfo{title}{Theoretical modelling of the dynamic interaction
  between twin tunnels in a multi-layered half-space},
\newblock \bibinfo{journal}{Journal of Sound and Vibration}
  \bibinfo{volume}{456} (\bibinfo{year}{2019}) \bibinfo{pages}{65--85}.
  \DOIprefix\doi{10.1016/j.jsv.2019.05.025}.
%Type = Article
\bibitem[{Real et~al.(2015)Real, Zamorano, Ribes, and
  Real}]{Real2015TunnelFEM2D3D}
\bibinfo{author}{T.~Real}, \bibinfo{author}{C.~Zamorano},
  \bibinfo{author}{F.~Ribes}, \bibinfo{author}{J.~I. Real},
\newblock \bibinfo{title}{Train-induced vibration prediction in tunnels using
  {2D} and {3D} {FEM} models in time domain},
\newblock \bibinfo{journal}{Tunnelling and Underground Space Technology}
  \bibinfo{volume}{49} (\bibinfo{year}{2015}) \bibinfo{pages}{376--383}.
  \DOIprefix\doi{10.1016/j.tust.2015.05.004}.
%Type = Article
\bibitem[{Yang and Hung(2001)}]{Yang2003MovingLoadFEMIEM}
\bibinfo{author}{Y.~B. Yang}, \bibinfo{author}{H.~H. Hung},
\newblock \bibinfo{title}{A {2.5D} finite/infinite element approach for
  modelling visco-elastic bodies subjected to moving loads},
\newblock \bibinfo{journal}{International Journal for Numerical Methods in
  Engineering} \bibinfo{volume}{51} (\bibinfo{year}{2001})
  \bibinfo{pages}{1317--1336}. \DOIprefix\doi{10.1002/nme.208}.
%Type = Article
\bibitem[{Jin et~al.(2018)Jin, Thompson, Lurcock, Toward, and
  Ntotsios}]{JinThompson2018}
\bibinfo{author}{Q.~Jin}, \bibinfo{author}{D.~J. Thompson},
  \bibinfo{author}{D.~E.~J. Lurcock}, \bibinfo{author}{M.~G.~R. Toward},
  \bibinfo{author}{E.~Ntotsios},
\newblock \bibinfo{title}{A {2.5D} finite element and boundary element model
  for the ground vibration from trains in tunnels and validation using
  measurement data},
\newblock \bibinfo{journal}{Journal of Sound and Vibration}
  \bibinfo{volume}{422} (\bibinfo{year}{2018}) \bibinfo{pages}{373--389}.
  \DOIprefix\doi{10.1016/j.jsv.2018.02.019}.
%Type = Article
\bibitem[{He et~al.(2018)He, Zhou, Guo, Di, and
  Zhang}]{He2018PoroelasticTunnel25D}
\bibinfo{author}{C.~He}, \bibinfo{author}{S.~Zhou}, \bibinfo{author}{P.~Guo},
  \bibinfo{author}{H.~Di}, \bibinfo{author}{X.~Zhang},
\newblock \bibinfo{title}{Modelling of ground vibration from tunnels in a
  poroelastic half-space using a {2.5-D} {FE-BE} formulation},
\newblock \bibinfo{journal}{Tunnelling and Underground Space Technology}
  \bibinfo{volume}{82} (\bibinfo{year}{2018}) \bibinfo{pages}{211--221}.
  \DOIprefix\doi{10.1016/j.tust.2018.08.043}.
%Type = Article
\bibitem[{Gupta et~al.(2007)Gupta, Hussein, Degrande, Hunt, and
  Clouteau}]{Gupta2007UndergroundRailwayModels}
\bibinfo{author}{S.~Gupta}, \bibinfo{author}{M.~F.~M. Hussein},
  \bibinfo{author}{G.~Degrande}, \bibinfo{author}{H.~E.~M. Hunt},
  \bibinfo{author}{D.~Clouteau},
\newblock \bibinfo{title}{A comparison of two numerical models for the
  prediction of vibrations from underground railway traffic},
\newblock \bibinfo{journal}{Soil Dynamics and Earthquake Engineering}
  \bibinfo{volume}{27} (\bibinfo{year}{2007}) \bibinfo{pages}{608--624}.
  \DOIprefix\doi{10.1016/j.soildyn.2006.12.007}.
%Type = Article
\bibitem[{François et~al.(2010)François, Schevenels, Galvín, Lombaert, and
  Degrande}]{FrancoisSchevenels-57}
\bibinfo{author}{S.~François}, \bibinfo{author}{M.~Schevenels},
  \bibinfo{author}{P.~Galvín}, \bibinfo{author}{G.~Lombaert},
  \bibinfo{author}{G.~Degrande},
\newblock \bibinfo{title}{A {2.5D} coupled {FE--BE} methodology for the dynamic
  interaction between longitudinally invariant structures and a layered
  halfspace},
\newblock \bibinfo{journal}{Computer Methods in Applied Mechanics and
  Engineering} \bibinfo{volume}{199} (\bibinfo{year}{2010})
  \bibinfo{pages}{1536--1548}. \DOIprefix\doi{10.1016/j.cma.2010.01.001}.
%Type = Article
\bibitem[{Huang et~al.(2025)Huang, Wei, Ding, and
  Xu}]{Huang2025Improved25DFEBE}
\bibinfo{author}{J.~Huang}, \bibinfo{author}{X.~Wei},
  \bibinfo{author}{Z.~Ding}, \bibinfo{author}{W.~Xu},
\newblock \bibinfo{title}{Research on the dynamic behavior of transversely
  isotropic saturated media due to meeting subway loads using an improved
  {2.5D} {FE-BE} method},
\newblock \bibinfo{journal}{Soil Dynamics and Earthquake Engineering}
  \bibinfo{volume}{191} (\bibinfo{year}{2025}) \bibinfo{pages}{109244}.
  \DOIprefix\doi{10.1016/j.soildyn.2025.109244}.
%Type = Article
\bibitem[{Wolf and Song(2001)}]{WolfSong2001SBFEM}
\bibinfo{author}{J.~P. Wolf}, \bibinfo{author}{C.~Song},
\newblock \bibinfo{title}{The scaled boundary finite-element method--a
  fundamental-solution-less boundary-element method},
\newblock \bibinfo{journal}{Computer Methods in Applied Mechanics and
  Engineering} \bibinfo{volume}{190} (\bibinfo{year}{2001})
  \bibinfo{pages}{5551--5568}. \DOIprefix\doi{10.1016/S0045-7825(01)00183-9}.
%Type = Article
\bibitem[{Yaseri et~al.(2018)Yaseri, Bazyar, and
  Javady}]{YaseriBazyarJavady-XX}
\bibinfo{author}{A.~Yaseri}, \bibinfo{author}{M.~H. Bazyar},
  \bibinfo{author}{S.~Javady},
\newblock \bibinfo{title}{{2.5D} coupled {FEM-SBFEM} analysis of ground
  vibrations induced by train movement},
\newblock \bibinfo{journal}{Soil Dynamics and Earthquake Engineering}
  \bibinfo{volume}{104} (\bibinfo{year}{2018}) \bibinfo{pages}{307--318}.
  \DOIprefix\doi{10.1016/j.soildyn.2017.10.021}.
%Type = Article
\bibitem[{Yang and Li(2022)}]{YangLi2021IGASBIGA}
\bibinfo{author}{Y.~B. Yang}, \bibinfo{author}{J.~Li},
\newblock \bibinfo{title}{{2.5D} prediction of soil vibrations due to railway
  loads by the isogeometric analysis with scaled boundary},
\newblock \bibinfo{journal}{Engineering Analysis with Boundary Elements}
  \bibinfo{volume}{134} (\bibinfo{year}{2022}) \bibinfo{pages}{341--359}.
  \DOIprefix\doi{10.1016/j.enganabound.2021.10.012}.
%Type = Article
\bibitem[{Atroshchenko et~al.(2018)Atroshchenko, Tomar, Xu, and
  Bordas}]{Atroshchenko2018GIFT}
\bibinfo{author}{E.~Atroshchenko}, \bibinfo{author}{S.~Tomar},
  \bibinfo{author}{G.~Xu}, \bibinfo{author}{S.~P.~A. Bordas},
\newblock \bibinfo{title}{Weakening the tight coupling between geometry and
  simulation in isogeometric analysis: from sub- and super-geometric analysis
  to geometry-independent field approximation ({GIFT})},
\newblock \bibinfo{journal}{International Journal for Numerical Methods in
  Engineering} \bibinfo{volume}{114} (\bibinfo{year}{2018})
  \bibinfo{pages}{1131--1159}. \DOIprefix\doi{10.1002/nme.5778}.
%Type = Article
\bibitem[{Videla et~al.(2019)Videla, Anitescu, Khajah, Bordas, and
  Atroshchenko}]{Videla2019AdaptiveGIFT}
\bibinfo{author}{J.~Videla}, \bibinfo{author}{C.~Anitescu},
  \bibinfo{author}{T.~Khajah}, \bibinfo{author}{S.~P.~A. Bordas},
  \bibinfo{author}{E.~Atroshchenko},
\newblock \bibinfo{title}{{h}- and {p}-adaptivity driven by recovery and
  residual-based error estimators for {PHT}-splines applied to time-harmonic
  acoustics},
\newblock \bibinfo{journal}{Computers \& Mathematics with Applications}
  \bibinfo{volume}{77} (\bibinfo{year}{2019}) \bibinfo{pages}{2369--2395}.
  \DOIprefix\doi{10.1016/j.camwa.2018.12.026}.
%Type = Article
\bibitem[{Jansari et~al.(2022)Jansari, Videla, Natarajan, Bordas, and
  Atroshchenko}]{Jansari2022AdaptiveGIFT}
\bibinfo{author}{C.~Jansari}, \bibinfo{author}{J.~Videla},
  \bibinfo{author}{S.~Natarajan}, \bibinfo{author}{S.~P. Bordas},
  \bibinfo{author}{E.~Atroshchenko},
\newblock \bibinfo{title}{Adaptive enriched geometry independent field
  approximation for {2D} time-harmonic acoustics},
\newblock \bibinfo{journal}{Computers \& Structures} \bibinfo{volume}{263}
  (\bibinfo{year}{2022}) \bibinfo{pages}{106728}.
  \DOIprefix\doi{10.1016/j.compstruc.2021.106728}.
%Type = Article
\bibitem[{Giannelli et~al.(2012)Giannelli, Jüttler, and
  Speleers}]{Giannelli2012THB}
\bibinfo{author}{C.~Giannelli}, \bibinfo{author}{B.~Jüttler},
  \bibinfo{author}{H.~Speleers},
\newblock \bibinfo{title}{{THB}-splines: The truncated basis for hierarchical
  splines},
\newblock \bibinfo{journal}{Computer Aided Geometric Design}
  \bibinfo{volume}{29} (\bibinfo{year}{2012}) \bibinfo{pages}{485--498}.
  \DOIprefix\doi{10.1016/j.cagd.2012.03.025}.
%Type = Article
\bibitem[{Giannelli et~al.(2016)Giannelli, Jüttler, Kleiss, Mantzaflaris,
  Simeon, and Špeh}]{Giannelli2016THBIGA}
\bibinfo{author}{C.~Giannelli}, \bibinfo{author}{B.~Jüttler},
  \bibinfo{author}{S.~K. Kleiss}, \bibinfo{author}{A.~Mantzaflaris},
  \bibinfo{author}{B.~Simeon}, \bibinfo{author}{J.~Špeh},
\newblock \bibinfo{title}{{THB}-splines: An effective mathematical technology
  for adaptive refinement in geometric design and isogeometric analysis},
\newblock \bibinfo{journal}{Computer Methods in Applied Mechanics and
  Engineering} \bibinfo{volume}{299} (\bibinfo{year}{2016})
  \bibinfo{pages}{337--365}. \DOIprefix\doi{10.1016/j.cma.2015.11.002}.
%Type = Article
\bibitem[{Dijkstra et~al.(2026)Dijkstra, Giannelli, and
  Toshniwal}]{Dijkstra2026MacroTHB}
\bibinfo{author}{K.~Dijkstra}, \bibinfo{author}{C.~Giannelli},
  \bibinfo{author}{D.~Toshniwal},
\newblock \bibinfo{title}{Macro-element refinement schemes for {THB}-splines:
  Applications to {B\'ezier} projection and structure-preserving
  discretizations},
\newblock \bibinfo{journal}{Computer Methods in Applied Mechanics and
  Engineering} \bibinfo{volume}{452} (\bibinfo{year}{2026})
  \bibinfo{pages}{118707}. \DOIprefix\doi{10.1016/j.cma.2025.118707}.
%Type = Article
\bibitem[{Borden et~al.(2011)Borden, Scott, Evans, and
  Hughes}]{Borden2011BezierExtraction}
\bibinfo{author}{M.~J. Borden}, \bibinfo{author}{M.~A. Scott},
  \bibinfo{author}{J.~A. Evans}, \bibinfo{author}{T.~J.~R. Hughes},
\newblock \bibinfo{title}{Isogeometric finite element data structures based on
  {B}ézier extraction of {NURBS}},
\newblock \bibinfo{journal}{International Journal for Numerical Methods in
  Engineering} \bibinfo{volume}{87} (\bibinfo{year}{2011})
  \bibinfo{pages}{15--47}. \DOIprefix\doi{10.1002/nme.2968}.
%Type = Article
\bibitem[{Liu et~al.(2017)Liu, Saputra, Wang, Tin-Loi, and
  Song}]{Liu2017AutomaticPolyhedralSBFEM}
\bibinfo{author}{Y.~Liu}, \bibinfo{author}{A.~A. Saputra},
  \bibinfo{author}{J.~Wang}, \bibinfo{author}{F.~Tin-Loi},
  \bibinfo{author}{C.~Song},
\newblock \bibinfo{title}{Automatic polyhedral mesh generation and scaled
  boundary finite element analysis of {STL} models},
\newblock \bibinfo{journal}{Computer Methods in Applied Mechanics and
  Engineering} \bibinfo{volume}{313} (\bibinfo{year}{2017})
  \bibinfo{pages}{106--132}. \DOIprefix\doi{10.1016/j.cma.2016.09.038}.
%Type = Article
\bibitem[{Li et~al.(2024)Li, Su, Yang, Gao, and Wang}]{Li2024NQSBFEM}
\bibinfo{author}{X.~Li}, \bibinfo{author}{H.~Su}, \bibinfo{author}{J.~Yang},
  \bibinfo{author}{G.~Gao}, \bibinfo{author}{Y.~Wang},
\newblock \bibinfo{title}{{NURBS}-boundary-based quadtree scaled boundary
  finite element method study for irregular design domain},
\newblock \bibinfo{journal}{Engineering Analysis with Boundary Elements}
  \bibinfo{volume}{159} (\bibinfo{year}{2024}) \bibinfo{pages}{418--433}.
  \DOIprefix\doi{10.1016/j.enganabound.2023.12.007}.
%Type = Article
\bibitem[{Lozano et~al.(2025)Lozano, Gravenkamp, and
  Birk}]{Lozano2025GuidedWaveSBFEM}
\bibinfo{author}{D.~Lozano}, \bibinfo{author}{H.~Gravenkamp},
  \bibinfo{author}{C.~Birk},
\newblock \bibinfo{title}{Simulations of ultrasonic guided wave scattering
  using the scaled boundary finite element method},
\newblock \bibinfo{journal}{Computer Methods in Applied Mechanics and
  Engineering} \bibinfo{volume}{445} (\bibinfo{year}{2025})
  \bibinfo{pages}{118204}. \DOIprefix\doi{10.1016/j.cma.2025.118204}.
%Type = Article
\bibitem[{Brivadis et~al.(2015)Brivadis, Buffa, Wohlmuth, and
  Wunderlich}]{Brivadis2015IGAMortar}
\bibinfo{author}{E.~Brivadis}, \bibinfo{author}{A.~Buffa},
  \bibinfo{author}{B.~Wohlmuth}, \bibinfo{author}{L.~Wunderlich},
\newblock \bibinfo{title}{Isogeometric mortar methods},
\newblock \bibinfo{journal}{Computer Methods in Applied Mechanics and
  Engineering} \bibinfo{volume}{284} (\bibinfo{year}{2015})
  \bibinfo{pages}{292--319}. \DOIprefix\doi{10.1016/j.cma.2014.09.012}.
%Type = Article
\bibitem[{Dornisch et~al.(2015)Dornisch, Vitucci, and
  Klinkel}]{Dornisch2015WeakSubstitution}
\bibinfo{author}{W.~Dornisch}, \bibinfo{author}{G.~Vitucci},
  \bibinfo{author}{S.~Klinkel},
\newblock \bibinfo{title}{The weak substitution method -- an application of the
  mortar method for patch coupling in {NURBS}-based isogeometric analysis},
\newblock \bibinfo{journal}{International Journal for Numerical Methods in
  Engineering} \bibinfo{volume}{103} (\bibinfo{year}{2015})
  \bibinfo{pages}{205--234}. \DOIprefix\doi{10.1002/nme.4918}.
%Type = Article
\bibitem[{Chasapi et~al.(2020)Chasapi, Dornisch, and
  Klinkel}]{Chasapi2020Mortar}
\bibinfo{author}{M.~Chasapi}, \bibinfo{author}{W.~Dornisch},
  \bibinfo{author}{S.~Klinkel},
\newblock \bibinfo{title}{Patch coupling in isogeometric analysis of solids in
  boundary representation using a mortar approach},
\newblock \bibinfo{journal}{International Journal for Numerical Methods in
  Engineering} \bibinfo{volume}{121} (\bibinfo{year}{2020})
  \bibinfo{pages}{3206--3226}. \DOIprefix\doi{10.1002/nme.6354}.
%Type = Article
\bibitem[{Nguyen et~al.(2014)Nguyen, Kerfriden, Brino, Bordas, and
  Bonisoli}]{Nguyen2014NitscheNURBS}
\bibinfo{author}{V.~P. Nguyen}, \bibinfo{author}{P.~Kerfriden},
  \bibinfo{author}{M.~Brino}, \bibinfo{author}{S.~P.~A. Bordas},
  \bibinfo{author}{E.~Bonisoli},
\newblock \bibinfo{title}{Nitsche's method for two and three dimensional
  {NURBS} patch coupling},
\newblock \bibinfo{journal}{Computational Mechanics} \bibinfo{volume}{53}
  (\bibinfo{year}{2014}) \bibinfo{pages}{1163--1182}.
  \DOIprefix\doi{10.1007/s00466-013-0955-3}.
%Type = Article
\bibitem[{Si et~al.(2023)Si, Yu, Li, and
  Natarajan}]{Si2023AdaptiveMultipatchNitsche}
\bibinfo{author}{Z.~Si}, \bibinfo{author}{T.~Yu}, \bibinfo{author}{Y.~Li},
  \bibinfo{author}{S.~Natarajan},
\newblock \bibinfo{title}{Adaptive multi-patch isogeometric phase-field method
  for quasi-static brittle fracture based on {Nitsche's} method},
\newblock \bibinfo{journal}{Computer Methods in Applied Mechanics and
  Engineering} \bibinfo{volume}{414} (\bibinfo{year}{2023})
  \bibinfo{pages}{116154}. \DOIprefix\doi{10.1016/j.cma.2023.116154}.
%Type = Article
\bibitem[{Antonelli et~al.(2026)Antonelli, Gorgi, Zorrilla, and
  Rossi}]{Antonelli2026GapShiftedIGA}
\bibinfo{author}{N.~Antonelli}, \bibinfo{author}{A.~Gorgi},
  \bibinfo{author}{R.~Zorrilla}, \bibinfo{author}{R.~Rossi},
\newblock \bibinfo{title}{Isogeometric multipatch coupling with arbitrary
  refinement and parametrization using the {Gap--Shifted Boundary Method}},
\newblock \bibinfo{journal}{Computer Methods in Applied Mechanics and
  Engineering} \bibinfo{volume}{456} (\bibinfo{year}{2026})
  \bibinfo{pages}{118913}. \DOIprefix\doi{10.1016/j.cma.2026.118913}.
%Type = Article
\bibitem[{Dvořák and González(2024)}]{Dvorak2024InterfaceOperators}
\bibinfo{author}{R.~Dvořák}, \bibinfo{author}{J.~A. González},
\newblock \bibinfo{title}{On the automatic construction of interface coupling
  operators for non-matching meshes by optimization methods},
\newblock \bibinfo{journal}{Computer Methods in Applied Mechanics and
  Engineering} \bibinfo{volume}{432} (\bibinfo{year}{2024})
  \bibinfo{pages}{117336}. \DOIprefix\doi{10.1016/j.cma.2024.117336}.
%Type = Article
\bibitem[{Bai et~al.(2025)Bai, Yang, Huang, Huang, Li, Damil, and
  Hu}]{Bai2025NonmatchingDDMD}
\bibinfo{author}{X.~Bai}, \bibinfo{author}{J.~Yang},
  \bibinfo{author}{Q.~Huang}, \bibinfo{author}{W.~Huang},
  \bibinfo{author}{H.~Li}, \bibinfo{author}{N.~Damil}, \bibinfo{author}{H.~Hu},
\newblock \bibinfo{title}{Coupling of data-driven and model-driven computing
  within non-matching meshes},
\newblock \bibinfo{journal}{Computer Methods in Applied Mechanics and
  Engineering} \bibinfo{volume}{441} (\bibinfo{year}{2025})
  \bibinfo{pages}{117959}. \DOIprefix\doi{10.1016/j.cma.2025.117959}.
%Type = Article
\bibitem[{Yang et~al.(2021)Yang, Dong, and Wu}]{Yang2021NonconformingIGAFEBE}
\bibinfo{author}{H.~S. Yang}, \bibinfo{author}{C.~Y. Dong},
  \bibinfo{author}{Y.~H. Wu},
\newblock \bibinfo{title}{Non-conforming interface coupling and symmetric
  iterative solution in isogeometric {FE--BE} analysis},
\newblock \bibinfo{journal}{Computer Methods in Applied Mechanics and
  Engineering} \bibinfo{volume}{373} (\bibinfo{year}{2021})
  \bibinfo{pages}{113561}. \DOIprefix\doi{10.1016/j.cma.2020.113561}.
%Type = Article
\bibitem[{Zhong et~al.(2027)Zhong, Hong, Zhang, Zang, Rappel, and
  Bordas}]{Zhong2027NURBSTraceInfinite}
\bibinfo{author}{Y.~Zhong}, \bibinfo{author}{H.~Hong},
  \bibinfo{author}{B.~Zhang}, \bibinfo{author}{Q.~Zang},
  \bibinfo{author}{H.~Rappel}, \bibinfo{author}{S.~P. Bordas},
\newblock \bibinfo{title}{A {2.5D} {NURBS}-trace infinite-element method for
  moving-load wave propagation and soil–structure interaction in
  semi-infinite ground},
\newblock \bibinfo{journal}{Computers and Geotechnics} \bibinfo{volume}{201}
  (\bibinfo{year}{2027}) \bibinfo{pages}{108515}.
  \DOIprefix\doi{10.1016/j.compgeo.2026.108515}.
%Type = Article
\bibitem[{Venta-Viñuela et~al.(2026)Venta-Viñuela, Verhelst, Mantzaflaris,
  Giannelli, and Reali}]{VentaVinuela2026THBPhaseField}
\bibinfo{author}{L.~Venta-Viñuela}, \bibinfo{author}{H.~M. Verhelst},
  \bibinfo{author}{A.~Mantzaflaris}, \bibinfo{author}{C.~Giannelli},
  \bibinfo{author}{A.~Reali},
\newblock \bibinfo{title}{Effective phase-field modeling of the
  {Cahn--Hilliard} equation in {2D} and {3D} using a quasi-interpolation
  approach to adaptive isogeometric analysis with {THB}-splines},
\newblock \bibinfo{journal}{Computer Methods in Applied Mechanics and
  Engineering} \bibinfo{volume}{455} (\bibinfo{year}{2026})
  \bibinfo{pages}{118855}. \DOIprefix\doi{10.1016/j.cma.2026.118855}.
%Type = Book
\bibitem[{Song(2018)}]{Song2018SBFEMBook}
\bibinfo{author}{C.~Song}, \bibinfo{title}{The Scaled Boundary Finite Element
  Method: Introduction to Theory and Implementation}, \bibinfo{publisher}{John
  Wiley \& Sons}, \bibinfo{year}{2018}. \DOIprefix\doi{10.1002/9781119388487}.
%Type = Article
\bibitem[{Jiao and Heath(2004)}]{JiaoHeath2004CommonRefinement}
\bibinfo{author}{X.~Jiao}, \bibinfo{author}{M.~T. Heath},
\newblock \bibinfo{title}{Common-refinement-based data transfer between
  non-matching meshes in multiphysics simulations},
\newblock \bibinfo{journal}{International Journal for Numerical Methods in
  Engineering} \bibinfo{volume}{61} (\bibinfo{year}{2004})
  \bibinfo{pages}{2402--2427}. \DOIprefix\doi{10.1002/nme.1147}.
%Type = Article
\bibitem[{Jaiman et~al.(2006)Jaiman, Jiao, Geubelle, and
  Loth}]{Jaiman2006ConservativeTransfer}
\bibinfo{author}{R.~K. Jaiman}, \bibinfo{author}{X.~Jiao},
  \bibinfo{author}{P.~H. Geubelle}, \bibinfo{author}{E.~Loth},
\newblock \bibinfo{title}{Conservative load transfer along curved fluid--solid
  interface with non-matching meshes},
\newblock \bibinfo{journal}{Journal of Computational Physics}
  \bibinfo{volume}{218} (\bibinfo{year}{2006}) \bibinfo{pages}{372--397}.
  \DOIprefix\doi{10.1016/j.jcp.2006.02.016}.
%Type = Article
\bibitem[{Wang et~al.(2022)Wang, Lin, and Li}]{WangLinLi2022}
\bibinfo{author}{Y.~Wang}, \bibinfo{author}{G.~Lin}, \bibinfo{author}{Z.~Li},
\newblock \bibinfo{title}{Dynamic response of anisotropic multilayered road
  structures induced by moving loads based on a novel spectral element method},
\newblock \bibinfo{journal}{Buildings} \bibinfo{volume}{12}
  (\bibinfo{year}{2022}) \bibinfo{pages}{1354}.
  \DOIprefix\doi{10.3390/buildings12091354}.
%Type = Article
\bibitem[{Hung and Yang(2001)}]{HungYang2001}
\bibinfo{author}{H.-H. Hung}, \bibinfo{author}{Y.~B. Yang},
\newblock \bibinfo{title}{Elastic waves in visco-elastic half-space generated
  by various vehicle loads},
\newblock \bibinfo{journal}{Soil Dynamics and Earthquake Engineering}
  \bibinfo{volume}{21} (\bibinfo{year}{2001}) \bibinfo{pages}{1--17}.
  \DOIprefix\doi{10.1016/S0267-7261(00)00078-6}.
%Type = Article
\bibitem[{Tadeu and Kausel(2000)}]{TadeuKausel2000}
\bibinfo{author}{A.~J.~B. Tadeu}, \bibinfo{author}{E.~Kausel},
\newblock \bibinfo{title}{Green's functions for two-and-a-half-dimensional
  elastodynamic problems},
\newblock \bibinfo{journal}{Journal of Engineering Mechanics}
  \bibinfo{volume}{126} (\bibinfo{year}{2000}) \bibinfo{pages}{1093--1097}.
  \DOIprefix\doi{10.1061/(ASCE)0733-9399(2000)126:10(1093)}.
%Type = Article
\bibitem[{Lombaert and Degrande(2009)}]{LombaertDegrande2009}
\bibinfo{author}{G.~Lombaert}, \bibinfo{author}{G.~Degrande},
\newblock \bibinfo{title}{Ground-borne vibration due to static and dynamic axle
  loads of {InterCity} and high-speed trains},
\newblock \bibinfo{journal}{Journal of Sound and Vibration}
  \bibinfo{volume}{319} (\bibinfo{year}{2009}) \bibinfo{pages}{1036--1066}.
  \DOIprefix\doi{10.1016/j.jsv.2008.07.003}.

\end{thebibliography}

\end{document}